\documentstyle[pre,aps,super]{revtex}

\input epsf

\begin{document}

\title{ Miscibility behavior and single chain properties in polymer blends:\\
	a bond fluctuation model study.} 

\author{
M.\ M{\"u}ller
\\
{\small Institut f{\"u}r Physik, WA 331, Johannes Gutenberg Universit{\"a}t}
\\
{\small D-55099 Mainz, Germany}
}
\date{\today}
\maketitle

\begin{abstract}
Computer simulation studies on the miscibility behavior and single chain properties in binary polymer blends are reviewed.
We consider blends of various architectures in order to identify important architectural parameters on a coarse grained level
and study their qualitative consequences for the miscibility behavior. The phase diagram, the relation between the exchange chemical
potential and the composition, and the intermolecular paircorrelation functions for symmetric blends of linear chains, blends of cyclic 
polymers, blends with an asymmetry in cohesive energies, blends with different chain lengths, blends with distinct monomer shapes, and 
blends with a stiffness disparity between the components are discussed.

For strictly symmetric blends the Flory Huggins theory becomes quantitatively correct in the long chain 
length limit, when the $\chi$ parameter is identified via the intermolecular pair correlation function. For small
chain lengths composition fluctuations are important. They manifest themselves in 3D Ising behavior at the critical point 
and an upward parabolic curvature of the SANS $\chi$ parameter close to the critical point. The ratio between the mean
field estimate and the true critical temperature decreases like $\sqrt{\chi}/\rho b^3$ for long chain lengths. 
The chain conformations in the minority phase of a symmetric blend shrink as to reduce the number of energetic unfavorable
interactions. Scaling arguments, detailed SCF calculations and Monte Carlo simulations of chains with up to $512$ effective 
segments agree that the conformational changes decrease around the critical point like $1/\sqrt{N}$. Other mechanisms for a 
composition dependence of the single chain conformations in asymmetric blends are discussed.

If the constituents of the blends have non-additive monomer shapes, one has a large positive chain length independent entropic 
contribution to the $\chi$ parameter. In this case the blend phase separates upon heating at a lower critical solution temperature.
Upon increasing the chain lengths the critical temperatures approach a finite value from above. For blends with a stiffness disparity
an entropic contribution of the $\chi$ parameter of the order $10^{-3}$ is measured with high accuracy. Also the enthalpic contribution 
increases, because a back folding of the stiffer component is suppressed and the stiffer chains possess more intermolecular contacts.

Two aspects of the single chain dynamics in blends are discussed: ({\bf a}) The dynamics of short non--entangled chains in a binary blend is 
studied via dynamic Monte Carlo simulations. There is hardly any coupling between the chain dynamics and the thermodynamic state 
of the mixture. Above the critical temperatures both the translational diffusion and the relaxation of the chain conformations 
are independent of the temperature. 
({\bf b}) Irreversible reactions of a small fraction of reactive polymers at a strongly segregated interface in a symmetric 
binary polymer blend are investigated. End-functionalized homopolymers of different species interact at the interface instantaneously 
and irreversibly to form diblock copolymers. The initial reaction rate for small reactant concentrations is time dependent and
and larger than expected from theory. At later times there is a depletion of the reactive chains at the interface and the reaction is determined
by the flux of the chains to interface. 

Pertinent off-lattice simulations and analytical theories are briefly discussed.

\end{abstract}

\section{Introduction.}
Melt blending of polymers is of both practical as well as fundamental importance: Polymeric materials in daily life
are generally multicomponent systems. Chemically different polymers are ``alloyed'' as to design a material which
combines the favorable characteristics of the individual components\cite{APPLICATION}. Clearly the miscibility behavior of the blend is crucial
for understanding and tailoring properties relevant for practical applications. Miscibility on a microscopic length scale is desirable for 
a high tensile strength of the material. However, when the components are only partially miscible, the system separates into 
regions in which one of the components is enriched. The degree of segregation, the interfacial tension between 
the coexisting phases, and the morphology of the material on a mesoscopic length scale depend sensitively on the
monomeric incompatibility.

The monomeric incompatibility is parameterized in terms of the Flory-Huggins parameter $\chi$. When $\chi$ is used as an adjustable
parameter, the mean field theory\cite{FH} of Flory and Huggins is quite successful in describing many experimental observations.
Most notably the theory rationalizes the fact that long macromolecules tend to demix and provides simple analytical expressions
for the free energy of mixing. The simple form of the bulk free energy lies, e.g., at the basis of self-consistent field calculations\cite{SCF1,SCF2,SCF3,SCF4,SCF5}
for spatially inhomogeneous polymer systems in the framework of the Gaussian chain model. Taking the Flory Huggins free energy of mixing
as a prerequisite, the theory has proven valuable in studying more complex structures (e.g., interfaces between partially miscible polymers\cite{SCF3,FREDDI} and
self-assembly of copolymers\cite{SCF5}).

Using small angle neutron scattering (SANS) the experimentalists measure the incompatibility parameter in the one phase of the blend\cite{BATES}. This yields
valuable information about the thermodynamics of mixing. Also single chain conformations can be extracted from the scattering data. Wide angle
X-ray scattering\cite{WAXS} has been applied to determine the total pair correlation function -- a quantity which characterizes the packing structure
of the monomer fluid. However, many blends of practical relevance are characterized by strong asymmetries. The  complicated interplay between
various asymmetries makes an interpretation rather difficult. Moreover, the parameter range accessible to experiments
might be restricted by chemical degradation at high temperatures or the glass transition at low temperatures\cite{GEHLEN}.

Despite the large success in parameterizing the experimental observations and studying complex structures, a fundamental understanding how to 
relate the measured $\chi$ parameter to the detailed chemical structure of the constituents is still largely missing\cite{SCHWEIZERR,FREED_REV}. Intriguingly, 
one of the fundamental predictions of the Flory Huggins theory -- namely the scaling behavior of the critical temperature in a binary 
homopolymer blends -- was only verified experimentally\cite{GEHLEN} 50 years after the seminal papers\cite{FH} by Flory and Huggins. 
According to the original lattice model of Flory and Huggins the $\chi$ parameter describes the difference in enthalpic interactions
between monomer species:
\begin{equation}
\chi_{\rm FH} = \frac{z}{k_B T} \left( \epsilon_{AB} - \frac{\epsilon_{AA}+\epsilon_{BB}}{2} \right)
\label{eqn:org}
\end{equation}
$z$ denotes the number of neighbors on the lattice with which a monomer interacts, and $\epsilon_{IJ}$ denotes the
interaction energy between species $I$ and $J$. In simple cases of non--polar subunits these interactions are due to van der Waals forces between
the chemical units and their energy scale is on the order $k_BT$. Typical values of $\chi$ parameters for chemically
similar constituents range from 0.01 to $10^{-5}$. Hence, the different enthalpic contributions in the equation above 
cancel to a large extent and a quantitative prediction of the $\chi$ parameter from the atomic structure is extremely 
difficult. Moreover, the original theory cannot rationalize the following key observations: 

({\bf i}) The temperature dependence of the measured values of the $\chi$ parameter often takes the form $\chi = A + B/T$.
      Following common convention, $B$ is denoted as the enthalpic contribution, whereas $A$ is refered to as an additional entropic contribution.
      In the framework of the Flory Huggins theory\cite{FH}, the only contribution to the entropy of mixing stems from the translation 
      of the molecules as a whole. Hence, the entropy of mixing per monomer is of the order $1/N$. However, other entropic contributions 
      on smaller length scales, e.g., due to the dependence of the packing arrangement of the segments on the local environment, may become important. 
      In experiments, however,  the temperature dependence of $\chi$ is only accessible over a rather small temperature range
      and there is a strong interplay between entropic and enthalpic 
      contributions, which makes the standard decomposition of $\chi$ into an entropic and an enthalpic part difficult: e.g., the temperature dependence 
      of the chain conformations or equation of state effects might yield a nonlinear relation between $\chi$ and the inverse temperature. 
      If entropic effects  become dominant the blend might phase separate upon heating at a lower critical solution temperature. Such behavior 
      is observed in binary polymer solutions and blends\cite{SCHWAHN2}. Specific interactions (e.g., hydrogen bonds\cite{HB1,HB2,HB3,HB4}) might also 
      give rise to a phase separation upon heating.
      Moreover, the $\chi$ parameter often depends on composition and chain length as well. More recently the effect of pressure on 
      the incompatibility $\chi$ of homopolymer blends and the ordering in diblock copolymers has attracted attention\cite{PRESSURE}.

({\bf ii}) According to the original mean field theory, the demixing temperature is independent of the molecular architecture.
      Hence, blends of two homopolymers and two ring polymers of the same monomeric units are predicted to have the same
      miscibility behavior. Similarly, the theory does not capture the dependence of the miscibility behavior of the chain stiffness
      or the degree of branching\cite{SCHWEIZERR,FREED_REV}.

({\bf iii}) Being a mean field theory, the Flory Huggins theory neglects fluctuations of the local composition and invokes a random mixing
      approximation. Thus, the behavior in the vicinity of the critical point is described by mean field exponents and the binodals
      have a parabolic shape. In the ultimate vicinity of the critical point, the correlation length grows very large and 
      the polymeric properties become irrelevant. In this critical region the behavior is characterized by the 3D Ising universality 
      class\cite{SARIBAN} which applies to all binary mixtures with short range interactions. The latter behavior manifests itself in a 
      much flatter binodal at the critical point and a stronger divergence of composition fluctuations. This has been observed in neutron scattering
      experiments extremely close to the critical point\cite{BATES,SCHWAHN}.

({\bf iv}) The same intermolecular forces which determine the miscibility behavior alter the conformations of the extended flexible
      macromolecules. Monte Carlo simulations by Sariban and Binder\cite{SARIBAN} for rather short chain lengths reveal a pronounced contraction of the 
      polymer coils in the minority phase. A qualitatively similar behavior has been observed in simulations of various other computational 
      models\cite{CIFRA1,HPD,DYN,MURAT,CLUSTER}. Experiments in highly incompatible poly(methyl methacrylate) (PMMA) and poly(vinyl acetate) (PVAc) blends of rather low molecular
      weight indicate a relative contraction of isolated PMMA chain extensions by 13-15\%\cite{FAYER}. These observations indicate a possible
      coupling between the single chain conformations and the thermodynamic state (i.e., temperature and composition of the mixture).
      SANS experiments\cite{BRIBER} on dilute deuterated PS in a PS or a PVME matrix show an increase in the chain extension upon mixing by about $7\%$
      (The blend shows a LCST, i.e. it phase separates upon heating, and therefore the PS is more extended in the PVME matrix).

Notwithstanding the experimental and theoretical efforts directed towards a quantitative understanding of the above observations,
today's predictive abilities of monomeric incompatibility on an atomistic basis are, by and large, poor. In applications a multitude 
of deviations from the simple Flory Huggins theory is observed. Nevertheless, significant progress has been made in specific areas:
Experiments\cite{BATES,GRAESSLEY,WALSH,RUSSELL}, theory\cite{FREED_REV,SCHWEIZER3,LIU}
and computer simulations\cite{M0,STIFF1,WEINHOLD,GREST,SMILE} on carefully characterized model systems have lead to advances in understanding the qualitative ingredients into 
the $\chi$ parameter. Experiments on pairs of homopolymers and their partially deuterated counterparts have clarified the scaling of 
the critical temperature with the molecular weight\cite{GEHLEN}. Extensive studies on saturated hydrocarbons with various degree of branching have aimed at 
correlating the miscibility behavior and the microstructure\cite{GRAESSLEY} in terms of Hildebrand's\cite{HILDEBRAND} solubility parameters. 
Analytical approaches like the 
P-RISM theory of Schweizer and co-workers\cite{SCHWEIZERR} or the Lattice Cluster theory of Freed and collaborators\cite{FREED_REV} calculate the local structure of the polymer 
fluid and their implications on the miscibility.

In this article we shall review some recent Monte Carlo simulations targeted at identifying the influence
of architectural parameters on the miscibility behavior. We compare simulations of various degree of microscopic detail with theoretical predictions.
Giving a complete overview over this rich and interesting field is of course difficult. Hence, 
the selection of topics we discuss is strongly biased by our own personal research interests.
The paper is arranged as follows: In the next section we provide some simple mean field 
views on the phase behavior and single chain properties in a binary polymer blend. Results of more sophisticated theories are briefly
discussed. The following section describes our computational model and simulation techniques. Subsequently, we present results for the miscibility
in structurally symmetric polymer blends\cite{M0}. The influence of asymmetries is exemplified with the investigation of disparities in the enthalpic interaction\cite{HPD2},
chain length\cite{M0}, and in the chain stiffness\cite{STIFF2}. In section V we highlight the dependence of the chain conformations on the thermodynamic state\cite{CLUSTER}, while we
discuss some findings on the dynamics of single chains\cite{DYN,REAC} in a blend in section VI. We close with a discussion and an outlook on interesting future topics.

\section{Background.}
In this section we give a brief phenomenological introduction into the salient features of the mean field theory of spatially 
homogeneous polymer blends. The derivation serves to reveal the assumptions which underlay the Flory Huggins theory\cite{FH} and to motivate the analysis of the
Monte Carlo data. The starting point is the partition function of a binary polymer mixture of $n_A$ molecules of species $A$
and $n_B$ molecules of species $B$ in the volume $V$ at temperature $T$.
\begin{equation}
{\cal Z}(n_A,n_B) = \frac{1}{2^{n_A}2^{n_B}n_A! n_B!} \int_{\{n_A,n_B\}} {\cal D}[{\bf r}] \exp \left(-{\cal H}[{\bf r}]/k_BT\right)
\end{equation}
the factorials take account of the indistinguishability of polymers of the same species, and we assume the chain ends to be indistinguishable.
The integral ${\cal D}[{\bf r}]$ sums over all allowed conformations 
of the polymers and ${\cal H}$ denotes the thermal interactions. In the following we assume pairwise interactions of the form
$v_{IJ}({\bf r})$ between monomers of type $I$ and $J$ a distance ${\bf r}$ apart. Using this expression we can formally calculate the 
chemical potential difference per monomer between the species:
\begin{eqnarray}
\frac{\Delta \mu}{k_BT} &\equiv&  -\frac{\partial \ln {\cal Z}}{N_A\partial n_A} + \frac{\partial \ln {\cal Z}}{N_B\partial n_B} \nonumber \\
&=& - \frac{1}{N_A} \ln \left( \frac{{\cal Z}(n_A+1,n_B)}{{\cal Z}(n_A,n_B)}\right) + \frac{1}{N_B} \ln \left( \frac{{\cal Z}(n_A,n_B+1)}{{\cal Z}(n_A,n_B)}\right)\nonumber \\
&=& - \frac{1}{N_A} \ln  \left( \frac{ \int_{\{n_A+1,n_B\}} {\cal D}[{\bf r}] \exp \left(-{\cal H}[{\bf r}]/k_BT\right)}
		       {2(n_A+1) \int_{\{n_A,n_B\}} {\cal D}[{\bf r}] \exp \left(-{\cal H}[{\bf r}]/k_BT\right)}\right) 
    + \frac{1}{N_B} \ln  \left( \frac{ \int_{\{n_A,n_B+1\}} {\cal D}[{\bf r}] \exp \left(-{\cal H}[{\bf r}]/k_BT\right)}
		       {2(n_B+1) \int_{\{n_A,n_B\}} {\cal D}[{\bf r}] \exp \left(-{\cal H}[{\bf r}]/k_BT\right)}\right)
\end{eqnarray}
We employ the ratio of the partition functions which differ by a single $A$ or $B$ polymer to define a single chain partition function\cite{STIFF1} of
an $A$ or $B$ polymer in a mixture of composition $\phi=N_A n_A/(N_A n_A+N_B n_B)$.
\begin{equation}
\exp\left(-N_A\frac{e_A-k_BT s_A}{k_BT}\right) \equiv \frac{1}{V} \frac{ \int_{\{n_A+1,n_B\}} {\cal D}[{\bf r}] \exp \left(-{\cal H}[{\bf r}]/k_BT\right)}{\int_{\{n_A,n_B\}} {\cal D}[{\bf r}] \exp \left(-{\cal H}[{\bf r}]/k_BT\right)} = \frac{1}{V} \Big\langle  {\cal D}_A[{\bf r}] \exp \left(-\delta E [{\bf r}]/k_BT\right) \Big\rangle_{\{n_A,n_B\}}
\label{eqn:single}
\end{equation}
The thermal average corresponds to the canonical ensemble of $n_A$ $A$ polymers and $n_B$ $B$ polymers, ${\cal D}_A[{\bf r}]$ sums over all
conformations of the additional $A$ polymer and $\delta E$ denotes the energy associated with inserting this additional $A$ polymer into the 
system. This single chain partition function in a blend can be decomposed into an enthalpic contribution $e_A$ and an entropic one $s_A$.
The first represents the energy associated with inserting a { segment} of an $A$ polymer into a mixture, whereas $s_A$ denotes the conformational
entropy { per segment} in a mixture of composition $\phi$. Both quantities depend in general on the temperature and the 
composition of the mixture. A similar expression is employed for the $B$ species. Using these definitions and rewriting the number of $A$ polymers
$n_A = \rho V \phi/N_A$ in terms of the monomer number density $\rho$, the volume $V$ and the composition $\phi$, we obtain the following exact expression 
for the chemical potential difference:
\begin{equation}
\frac{\Delta \mu}{k_BT} = \frac{1}{N_A} \ln \frac{2\rho \phi}{N_A} - \frac{1}{N_B} \ln \frac{2\rho(1-\phi)}{N_B} + \frac{e_A-e_B}{k_BT} - (s_A-s_B) 
\label{eqn:eos}
\end{equation}
The first two terms are the translational entropy of the polymers. This contribution is reduced by a factor $N_A$ and $N_B$ compared to
mixtures of small molecules. Hence, a small enthalpic or entropic mismatch is generally sufficient to bring about phase separation. The
mean field theory rationalizes why most polymers are only partially miscible. If the enthalpic and entropic contributions $e_A,e_B,s_A,s_B$ were 
known as a function of temperature and composition, the phase behavior of the blend could be readily calculated. The critical point is determined 
via the condition $\partial \Delta \mu/\partial \phi=0$ and
$\partial^2 \Delta \mu/\partial^2 \phi=0$. The spinodals are calculated via $\partial \Delta \mu/\partial \phi=0$ and the binodals
are given via the Maxwell construction. Comparing the above equation with the expression for the experimentally accessible, inverse collective structure 
$S_{\rm coll}({\bf q} \to 0)$ factor, we extract a Flory Huggins parameter $\chi_{\rm SANS}$ according to:
\begin{eqnarray}
\frac{1}{S_{\rm coll}({\bf q} \to 0)} &=& \frac{1}{V\rho (\langle \phi^2\rangle - \langle \phi\rangle^2)} = \frac{\partial}{\partial \phi} \frac{\Delta \mu}{k_BT}
= \frac{1}{N_A\phi} + \frac{1}{N_B (1-\phi)} - 2\chi_{\rm SANS} \label{eqn:sans} \\
\Rightarrow \chi_{\rm SANS} &=&  \frac{1}{2} \frac{\partial}{\partial \phi}  \left( \frac{e_A-e_B}{k_BT} - (s_A-s_B) \right)
\end{eqnarray}
This formal expression suggests  that the entropic portion of the Flory Huggins parameter is associated with the composition dependence of the
segmental conformational entropy, i.e., if the number of conformations a polymer can adopt depends on the composition of the blend we expect an
entropic contribution to the $\chi$ parameter\cite{STIFF1}. Unfortunately, the configurational integrals in Eq.(\ref{eqn:single}) cannot be evaluated and one 
has to resort to approximations. We discuss them in turn:

{\bf (A) Composition independence of single chain conformations:} The segmental entropy $s_A$ denotes the average number of possibilities to add an
      additional $A$ segment to a partially inserted $A$ polymer. Assuming that the single chain conformations in a blend are independent 
      of temperature and composition, we can approximate the conformational entropies $s_A$ and $s_B$ by constants. Within this approximation 
      the last term in Eq.(\ref{eqn:eos}) does not influence the miscibility behavior. By the same token we take the intramolecular energy 
      $e_A^{\rm intra}$ to be independent of composition.  Hence, the intramolecular energy is also immaterial for the miscibility 
      behavior.

{\bf (B) Neglect of fluctuations and random mixing approximation:} The energy $\delta E$ of adding an $A$ polymer to a mixture of composition 
      $\phi$ can be partitioned into intramolecular contributions and intermolecular contributions. Neglecting fluctuations we can evaluate the 
      intermolecular energy per monomer:
      \begin{equation}
      e_A^{\rm inter} \approx \rho \phi \int {\rm d}^3{\bf r}\; g_{AA}^{\rm inter}({\bf r}) v_{AA}({\bf r}) 
		       + \rho (1-\phi) \int {\rm d}^3{\bf r}\; g_{AB}^{\rm inter}({\bf r}) v_{AB}({\bf r})
      \end{equation}
      where the intermolecular pair correlation function $g^{\rm inter}_{AA}({\bf r})$ denotes the probability of finding an $A$ monomer of a different polymer at a distance
      ${\bf r}$ from the reference $A$ monomer. The functions are normalized such that they approach unity for large distances.
      Using a similar expression for the intermolecular energy associated with inserting a $B$ segment, one
      obtains:
\begin{equation}
e^{\rm inter}_A - e^{\rm inter}_B \approx 2 \chi \phi - \rho \phi  \int {\rm d}^3{\bf r}\; \left[ g^{\rm inter}_{AB}({\bf r}) v_{AB}({\bf r}) 
- g^{\rm inter}_{BB}({\bf r}) v_{BB}({\bf r}) \right]
\end{equation}
where we have identified the  Flory-Huggins parameter $\chi$ via the energy of mixing:
\begin{equation}
\chi =  \rho \int {\rm d}^3{\bf r}\; \left[ g^{\rm inter}_{AB}({\bf r}) v_{AB}({\bf r}) - \frac{g^{\rm inter}_{AA}({\bf r}) v_{AA}({\bf r})+g^{\rm inter}_{BB}({\bf r})v_{BB}({\bf r})}{2} \right]
\label{eqn:chis}
\end{equation}
This identification demonstrates that not the total number of interactions determine the miscibility behavior, but that the intramolecular
interactions do not contribute. Even in this simple form it is evident that the chain architecture has pronounced effects on the intermolecular 
paircorrelation function and the miscibility behavior: the more open the chains are, the larger is the number of intermolecular contacts, and the
smaller is the miscibility. Moreover, we shall often approximate the intermolecular pair correlation functions by their athermal values.

For later reference, we briefly summarize the predictions of the mean field theory for the miscibility behavior\cite{BREV1}.
The inverse critical temperature and composition scale like
\begin{equation}
\chi_c= \frac{1}{2}\left(\frac{1}{\sqrt{N_A}} + \frac{1}{\sqrt{N_A}}\right)^2 \qquad \mbox{and} \qquad \phi_c = \frac{1}{1+\sqrt{N_A/N_B}}
\label{eqn:crit}
\end{equation}
For symmetric blends ($N_A=N_B \equiv N$) the relation between the composition of the mixture and the exchange chemical potential $\Delta \mu$ takes the form:
\begin{equation}
\beta \Delta \mu = \frac{1}{N} \ln \left(\frac{\phi}{1-\phi} \right) - \chi (2\phi-1)  \qquad \mbox{SG--EOS}
\end{equation}
where $\beta=1/k_BT$ denotes the inverse temperature. In the following we refer to this relation as semi-grandcanonical equation of state (SG--EOS).
For symmetric blends ($N_A=N_B \equiv N$) the exchange chemical potential $\Delta \mu_{\rm coex}$ at phase coexistence vanishes by virtue of the symmetry with 
respect to the interchange of $A$ and $B$, and the equation above implicitly determines the binodals.
In the vicinity of the critical point the binodals take the form $\phi_{A/B} = 1/2 \pm \sqrt{3(\chi-\chi_c)/4\chi_c}$.
We obtain the strength of composition fluctuations via the second derivative of the free energy of mixing:
\begin{equation}
V\rho (\langle \phi^2\rangle - \langle \phi\rangle^2) = \frac{1}{\frac{1}{N\phi}+\frac{1}{N(1-\phi)}-2\chi}
\end{equation}
In the framework of the random phase approximation, the correlation length $\xi$ of composition fluctuations can be calculated to:
\begin{equation}
\xi = \frac{R_g}{\sqrt{3[1-2\chi N \phi(1-\phi)]}}
\label{eqn:xi}
\end{equation}
where $R_g^2=b^2 N/6$ denotes the radius of gyration and $b$ the statistical segment length.

Using these results and simple scaling arguments, we can partially assess the validity of the above assumptions in deriving the Flory Huggins theory.
The importance of composition fluctuations is expressed in terms of the Ginzburg criterion\cite{GINZBURG}: The neglect of fluctuations
is justified when the concentration fluctuations in one ``correlation volume'' of size $\xi^3$ are small compared to the composition difference
between the two coexisting phases. Using the above expressions, one obtains:
\begin{equation}
(\langle \phi^2\rangle - \langle \phi\rangle^2)_{|\xi^3} \sim \frac{\sqrt{\chi/\chi_c-1}}{\rho b^3/\sqrt{\chi_c}}  \stackrel{!}{\ll} (\phi_A - \phi_B)^2 \sim (\chi/\chi_c-1) 
\qquad \Rightarrow \qquad \frac{\chi - \chi_c}{\chi_c} \stackrel{!}{\gg}  \frac{\chi_c}{\rho^2\bar{b}^6} \sim \frac{1}{N}
\label{eqn:ginz}
\end{equation}
where $\bar{b}^2=(b_A^2/4\phi_c+b_B^2/4(1-\phi_c))$ denotes the square of the statistical segment length.
This Ginzburg criterion\cite{GINZBURG} states that composition fluctuations are important at the critical point, where simulations\cite{SARIBAN} and experiments\cite{GEHLEN,SCHWAHN} 
find 3D Ising critical behavior.
However, unlike the situation in mixtures of small molecules, the temperature interval in which composition fluctuations are dominant is restricted to the ultimate 
vicinity of the critical point $\chi-\chi_c < \chi_c/N$.
General arguments\cite{MON} rationalize that fluctuations lead to an overestimation of the ``true'' critical temperature $T_c$ in the mean field theory $T_c^{\rm MF}$ by an 
amount $(T_c^{\rm MF}-T_c)/T_c \sim 1/\sqrt{N}$. 

The vanishing of composition fluctuations can be exemplified by estimating the scaling of non--random mixing effects with growing chain length.
The strength of composition fluctuations in a volume $V$ is of the order $N/\rho V$. Expressing the composition fluctuations via the correlation functions, we obtain:
\begin{eqnarray}
1 \sim \frac{\rho V}{N} (\langle \phi^2\rangle - \langle \phi\rangle^2) &\sim& 
\frac{\rho}{N} \int {\rm d}^3{\bf r}\; \left[ \langle \phi\rangle g^{\rm inter}_{AA}({\bf r})+(1-\langle \phi\rangle)g^{\rm inter}_{BB}({\bf r}) - g^{\rm inter}_{AB}({\bf r}) \right]  \nonumber \\
& \sim & \frac{\rho \xi^3}{N} \int {\rm d}^3{\bf x}\; \left[ \langle \phi\rangle g^{\rm inter}_{AA}({\bf x}) +(1-\langle \phi\rangle)g^{\rm inter}_{BB}({\bf x}) - g^{\rm inter}_{AB}({\bf x})\right] \qquad \mbox{with} \qquad {\bf x} = {\bf r}/\xi
\end{eqnarray}
Assuming that the intermolecular correlation functions decay with a characteristic decay length $\xi$, we evaluate the amplitude of non--random mixing effects at
$\langle \phi \rangle=1/2$:
\begin{equation}
g^{\rm inter}_{AA}({\bf r}) - g^{\rm inter}_{AB}({\bf r}) \sim N/\rho R_g^3 \sim 1/\sqrt{N}
\end{equation}
Hence, the difference between the $AA$ and $AB$ correlation functions due to statistical thermal fluctuations decreases with chain length.
These mean field arguments are in agreement with P-RISM calculations by Yethiraj and Schweizer\cite{SCHWEIZER1}.

Monte Carlo simulations suggest\cite{PORE} that one possible mechanism of conformational changes in blends is associated to exchanging energetically unfavorable intermolecular 
contacts for attractive intramolecular contacts upon reducing the molecule's spatial extension. Attributing this shrinking of the minority component to a balance between the entropy loss 
due to deviations from the unperturbed conformations and the energy gain upon shrinking, we can estimate the magnitude of conformational changes\cite{CLUSTER}: Flexible chains are 
describable by the Gaussian chain model. Within this framework
a deviation from the unperturbed chain extension $R_0$ gives rise to an entropic force of the form
\begin{equation}
\frac{{\rm d}S}{{\rm d}R} \sim \frac{(R-R_0)}{R_0^2}
\end{equation}
This is opposed to an enthalpic force ${\rm d}E/{\rm d}R$, where
$E$ denotes the single chain energy. $E$ comprises energetically favorable interactions $N z^{\rm intra}$ among monomers of the same chain and $N z^{\rm inter}$
interactions with monomers of other polymers. The exchange of an intermolecular interaction with an intramolecular one lowers the single chain energy by an amount 
of the order $\chi$. The number of intramolecular interactions per monomer $z^{\rm intra}$ is given by the density of monomers of the same chain inside of its volume
$z^{\rm intra} \sim N/\rho R_0^3$. Under the assumption that the reduction of the chain extension does not affect the total number of interactions, but merely exchanges 
intermolecular interactions into energetically favorable intramolecular ones, we estimate the chain length dependence of the energy change as 
\begin{equation}
\frac{{\rm d}E}{{\rm d}R} \sim - \chi  \frac{N^2}{\rho R_0^4}
\end{equation}
Balancing the entropic force against the enthalpic one, we obtain:
\begin{equation}
\frac{R_0-R}{R_0} \sim \chi N \frac{N}{\rho R_0^3} \sim \chi \sqrt{N}
\end{equation}
These scaling arguments suggest that the perturbation of the chain conformations decreases upon increasing the chain length at $\chi N = {\rm const}$.
The conformations in high molecular weight blends are only very mildly perturbed in the minority phase. If there are no alternative
mechanisms, which give rise to a dependence of the chain conformations on temperature or composition, approximation ({\bf B})
is justified in the high molecular weight limit. Examples for alternative mechanisms shall be discussed in sec IV.

These general considerations show that the mean field theories describe the phase behavior of strictly symmetric blends 
adequately in the limit of large molecular weight, when the $\chi$ parameter is identified via the intermolecular paircorrelation function.
However, they cannot yield information about the explicit form of intermolecular pair correlation functions
and their dependence on the molecular architecture and density. The latter is an important ingredient for predicting structure--property relations reliably.
Moreover, these scaling considerations do not predict the absolute magnitude
of corrections to the leading scaling behavior and do not estimate the lower chain length at which a universal polymeric behavior sets in.

Modern theories of miscibility in polymer blends aim at predicting the influence of the molecular architecture on the intermolecular pair correlation function 
quantitatively and calculate non-random mixing effects as well as entropic contributions to the $\chi$ parameter. Effects of various asymmetries
have been explored.
The Lattice Cluster theory of Freed and coworkers\cite{FREED_REV} investigates the thermodynamics in the framework of a lattice model. Monomers are allowed to occupy 
several sites as to model explicitly various monomer architectures. The partition function is expressed in a systematic double expansion with respect to the 
inverse temperature and inverse coordination number of the underlying lattice. To zero order the approach recovers the original Flory Huggins theory. 
Higher order terms account for geometric packing on the monomer scale and non-random mixing effects The theory has been successful in predicting
various subtle influences of the monomer architecture, including the occurrence of entropic contributions to the Flory Huggins parameter and the
influence of tacticity on the miscibility. An alternative approach for calculating the thermodynamic properties of polymeric systems on a Bethe lattice
has been explored by Gujrati and co-workers\cite{G1}. This computational scheme has also been applied to spatially inhomogeneous systems\cite{G2}.

The P-RISM theory of Schweizer and co-workers\cite{SCHWEIZERR} investigates the structure of polymer blends in the framework of an integral equation theory
which accounts of the fluid-like packing structure and effects of the molecular architecture. In the limit that the chains are modeled as
thin structureless threads, the blend is totally symmetric and incompressible the theory reproduces the predictions of the Flory Huggins theory.
In general, however, there are deviations from the Flory-Huggins theory. The P-RISM approach yields a detailed description of
the intermolecular pair correlation functions taking account of their dependence of the monomeric architecture as well as on the correlation hole
effect on the scale of the chain's extension. Also the effects of the coupling between the chain conformations and temperature and composition
have been explored, and the predictions of the theory have been compared to Monte Carlo simulations. Moreover, the approach has been extended to 
liquid-vapor phase coexistence\cite{SCHWEIZER5}. An alternative integral equation approach has been pursued by Lipson and co-workers\cite{LIPSON}.

Fluctuation corrections to the Flory Huggins theory have been analyzed in the framework of the Gaussian chain model. Field theoretical calculations
by Vilgis and Holyst\cite{VILGIS}, and Garas and Kosmas\cite{KOSMAS} investigate the shift of the critical temperature and composition from the mean field value and the
dependence of the single chain dimensions have been explored. Liu, Fredrickson and Bates\cite{LIU} studied the miscibility behavior of blends with constituents
differing in the bending rigidity in a field theoretical framework. They pointed out that phase separation in blends with components of different flexibility 
or architecture is also promoted by long range composition correlations.

\section{Models and Techniques.}
In the following we shall present Monte Carlo simulations aiming at scrutinizing the various assumptions and refinements of the
theoretical approaches. In particular, we address the role of chain architecture on the qualitative aspects of the miscibility behavior
and the chain length dependence of the critical temperature and chain conformations.

Simulations of the miscibility behavior in polymer blends are considerably more exacting in computational terms than those of small molecules or
magnetic systems. The difficulties stem from the difficulty of dealing with the widely spread time and length scales caused by the extended structure 
of the macromolecules. These start at an atomistic scale of the bond length, encompass the chain extension and increase up to the correlation length of 
composition fluctuations, which grows without bound at the critical point, or the characteristic size of the morphology in phase separated blends. The accurate
determination of the macroscopic behavior while retaining the detailed atomistic chemical structure is not feasible even with state--of--the--art
supercomputers. Since the $\chi$ parameter depends on rather delicate differences in intermolecular energies and packing, predicting $\chi$ on an
atomistic basis is extremely difficult.  Therefore coarse grained models are promising candidates for investigating the universal, qualitative 
characteristic of miscibility. Introducing specific structural modifications or asymmetries, we can highlight their influence on the miscibility 
behavior. A detailed comparison between different computational models is absolutely warranted to investigate the degree of universality of the observed 
effects and to explore the effects of structure on various length scales. Simulational models of various degree of coarse graining have been
employed, ranging from the representation of polymers as self avoiding walks on a simple cubic lattice -- as in the original treatment of Flory
and Huggins\cite{FH} -- to simulations of the effect of branching in hydrocarbon melts in the framework of a united atom model\cite{GREST}. The choice of the
simulation model is a compromise between computational efficiency and a more faithful representation of the details of molecular architecture.

We shall present Monte Carlo simulations in the framework of the bond fluctuation model\cite{BFM}, which incorporates the relevant universal
characteristic of polymer blends: connectivity of the monomers along a chain, excluded volume of the segments, and a thermal interaction between
monomers. We compare our observations
to the results of other computational models. In the framework of this coarse grained lattice model, a monomer occupies the 8 corners of a unit 
cell from further occupancy. Monomers along a polymer are connected by one of 108 bond vectors of length $2,\sqrt{5},\sqrt{6},3$, and $\sqrt{10}$.
The bond vectors are chosen such that the excluded volume interactions prevents a crossing of bond vectors during the motion. Therefore the
algorithm captures the effect of entanglements.  This large number of bond vectors allows for 87 different bond angles -- an indication for the 
rather good approximation of continuous space properties by this complex lattice model. This property also allows for a rather realistic implementation 
of the bending rigidity. 

Here and in the following all length scales are measured in units of the lattice spacing. When atomistically 
detailed simulations are mapped onto the bond fluctuation model a lattice unit corresponds to roughly $2 \AA$ and a monomer in the bond fluctuation model
represents a small number -- say 3 to 5 -- of chemical repeat units\cite{MAP}. If not noted explicitly, we work at a monomer number density of $\rho = 1/16$, i.e.,
due to the extended structure of the monomers half of the lattice sites are occupied. These parameters correspond to a concentrated solution or a melt. 
On the one hand the presence of vacancies allows a reasonably fast equilibration of the chain conformations on the lattice. On the other hand the size 
disparity between (single site) vacancies and extended monomers gives rise to packing effects. Indeed, the monomer--monomer density pair correlation 
function $g({\bf r})$ exhibits oscillations at small distances indicating a fluid--like packing due to the local compressibility. The first neighbor 
shell of the pair correlation function comprises the 54 nearest lattice sites. Moreover, the relation between the osmotic pressure and the density is well 
describable via the Carnahan-Starling equation\cite{CAR} -- an approximation for the equation of state of hard spheres. This exemplifies that this lattice model shares many features 
with off--lattice models. 

The conformations of the polymers on the lattice evolve via local random monomer hopping\cite{BFM} -- a randomly chosen monomer attempts to move one lattice constant in a random
direction -- or slithering snake-like of moves\cite{M0,SS}, -- a segment of the chain is removed at one end of the chain and added at the opposite one. While the former allow for a dynamical
interpretation of the Monte Carlo simulations in terms of a purely diffusional dynamics\cite{BFM}, the latter relax the chain conformations a factor $N$ faster. For all but the simulations
to the single chain dynamics in sec V a combination of both Monte Carlo moves has been applied.

The blends comprise two components -- denoted $A$ and $B$ -- and monomers of the same type attract each other whereas different monomers repel each other via 
a square well potential.  Unless noted otherwise, we employ the most symmetric choice of pairwise interactions
\begin{equation}
\epsilon = - \epsilon_{AA} = - \epsilon_{BB} =  \epsilon_{AB}
\end{equation}
and the potential is extended over the first peak of the pair correlation function, i.e., it incorporates the first 54 neighbors up to a distance $\sqrt{6}$.
The form of the potential is chosen by computational convenience; we expect our results to be qualitatively independent of the specific potentials used.
However, if we were to model the interactions as (strongly) attractive with (slightly) different strengths between unlike species -- a more faithful modeling of
interactions in view of the experiment situation -- the presence of vacancies would allow for a liquid-vapor phase separation between a concentrated polymer melt and a dilute
phase in our ternary system. We believe that this liquid-vapor phase separation is common to both lattice based models and models in continuous space.
The temperature scale of this liquid-vapor coexistence is set by the $\Theta$ temperature, which is chain length independent.
This contrasts with the temperature scale of the liquid-liquid phase separation into an $A$ polymer rich and a $B$ polymer rich phases with similar content of vacancies.
The latter temperature scale increases linearly with the chain length. Therefore, the two phenomena are well separated in blends of high molecular weight.
However, for chain lengths accessible in computer simulations this need not to be the case and our choice of the potentials aims at avoiding an interference between 
the two different types of phase separation. Using Eq.(\ref{eqn:chis}), we obtain a simple estimate for the Flory Huggins parameter
\begin{equation}
\chi = \frac{2 z_c \epsilon}{k_BT} \qquad \mbox{with} \qquad z_c = \rho \int_{r \leq \sqrt{6}} {\rm d}^3{\bf r}\; g^{\rm inter}({\bf r})
\end{equation}
and in analogy to the original treatment of Flory and Huggins we refer to $z_c$ as the effective coordination number.
This equation expresses the $\chi$ parameter in terms of directly accessible quantities in the Monte Carlo simulations.

Being a lattice model, the bond fluctuation model is highly computationally efficient and allows for the investigation of rather large chain lengths and
large system sizes. The latter is necessary to accurately locate the critical temperature via finite size scaling analysis. As we shall illustrate, the
large chain length is crucial for reaching the high molecular weight scaling limit and extrapolating some quantities to experimentally relevant chain 
lengths. For the present investigation chains with up to 512 monomeric units have been employed.

Various simulation techniques have been used for exploring the miscibility properties of polymer blends. The most direct one is the simulation
of both the coexisting phases in the simulation cell. This method has been employed by Madden\cite{MINTER} and Cifra\cite{PINTER} for well segregated blends, and it also yields information 
about interfacial properties. However, it requires rather large simulation cells in order to extract ``bulk'' properties. Especially, the scheme is not very 
well suited to cope with the growing of the correlation length and vanishing of the difference between the phases as the critical point is approached. 
Computationally more efficient seems the direct estimation of the chemical 
potential of each individual species as a function of temperature and composition. At coexistence the chemical potentials of the species in both phases are equal 
and the coexistence curve can be mapped out. This technique has been applied successfully by Kumar\cite{KUMARB} using the incremental chemical potential method\cite{INC}. 
Kumar explored the influence of pressure and compressibility on the miscibility behavior. If one point on the coexistence curve is known a Gibbs Duhem integration 
technique can be employed to obtain the coexistence under constant pressure conditions\cite{dePabloEPL,KOFKE}.
Both methodologies are particularly useful for blends in which the constituents are characterized by very different chain architecture.

Sariban and Binder\cite{SARIBAN} employed simulations in the semi-grandcanonical ensemble for investigating the phase behavior at constant volume. In this ensemble, the total monomer
density is fixed, the composition of the blend fluctuates, and the chemical potential difference $\Delta \mu$ between the species is controlled.
The Monte Carlo scheme comprises two types of moves. Canonical updates relax the conformation of the macromolecules on the lattice, whereas semi-grandcanonical 
ones transfer $A$ polymers into $B$ polymers and vice-versa.  Sariban and Binder investigated strictly symmetric chains, for which the semi-grandcanonical 
moves consists in a mere exchange of labels. The algorithm can be extended to some degree of structural asymmetry (e.g., different chain lengths between the species\cite{M0}).
Overall speaking, it is reasonably efficient for a modest degree of structural asymmetry between the different constituents, but the extension to pronounced structural
asymmetries is a challenging task. Improvement might be achieved via gradually ``mutating''  one species into another\cite{MUTATE}. The major advantages of the methodology stem from 
the fluctuation of the composition of the mixture:
({\bf 1}) The relaxation times are much smaller than in the canonical ensemble, where the composition is conserved and composition fluctuations decay 
via the slow diffusion of polymers in a melt. The semi-grandcanonical ensemble allows the straightforward application of finite size scaling techniques
known from simple mixtures. Therefore we can measure the critical temperature in symmetric and asymmetric mixtures accurately from the Monte Carlo 
simulations of modest system size. ({\bf 2}) Moreover, the SG--EOS which relates the composition of the mixture to the difference of 
the chemical potentials of the species is directly accessible with high accuracy. The latter is particularly important to establish a direct contact to 
analytical approaches outside of the ultimate critical region where 3D Ising critical behavior dominates -- i.e., in the region where the mean field theories
are applicable. ({\bf 3}) Additionally, it is possible to determine excess interfacial properties (e.g., interfacial tension\cite{MBO}, excess energy\cite{COP}, 
enrichment of a third component\cite{COP}). Technical details and results on interfaces in polymer blends have been summarized in ref.\cite{FREV}.

\section{Phase behavior and fluid structure.}
\subsection{Symmetric blends.} 
The most faithful realization of the original lattice model\cite{FH} is a symmetric blend. The chain architecture of the two species is exactly identical and phase 
separation is brought about by an enthalpic symmetric repulsion between monomers of different species. In the following we refer to this most idealized situation 
as a strictly  symmetric blend.  For rather short chain lengths and medium vacancy concentration Sariban and Binder\cite{SARIBAN} studied this model on a simple cubic lattice
in the semi-grandcanonical ensemble. Using finite size scaling techniques, they obtained accurate estimates for the critical temperature and demonstrated that the 
critical point belongs to the 3D Ising universality class. They explored the dependence of the chain conformations and the energy of mixing on temperature and 
composition. Also the miscibility behavior in binary polymer solutions was investigated\cite{BINDER_SOL}. However, the modest range of chain lengths did not allow for a conclusive 
investigation of the scaling behavior with chain length. Moreover, the simple structure of the polymers on the lattice does not permit a fluid-like packing 
on the monomer scale and a realistic modeling of various asymmetries (e.g.\ bending stiffness).

The scaling of the critical temperature with chain length was subsequently studied by Deutsch and Binder\cite{HPD} in the framework of the bond fluctuation model. 
In accord with the predictions of the Flory Huggins theory\cite{FH}, the simulations exhibited a linear scaling of the critical temperature with chain length, which 
was also observed in carefully designed experiments\cite{GEHLEN}. This finding stimulated advances in the development of integral equation theories\cite{SCHWEIZER1}. 
The crossover between the mean field behavior away from the critical point and the ultimate 3D Ising behavior at the critical point was unraveled via a sophisticated analysis of
the Monte Carlo data which coped simultaneously with finite size effects and the crossover from mean field to 3D Ising critical behavior.

\begin{figure}[tbhp]
\begin{minipage}[t]{102mm}%
  \mbox{
       \setlength{\epsfxsize}{9cm}
       \epsffile{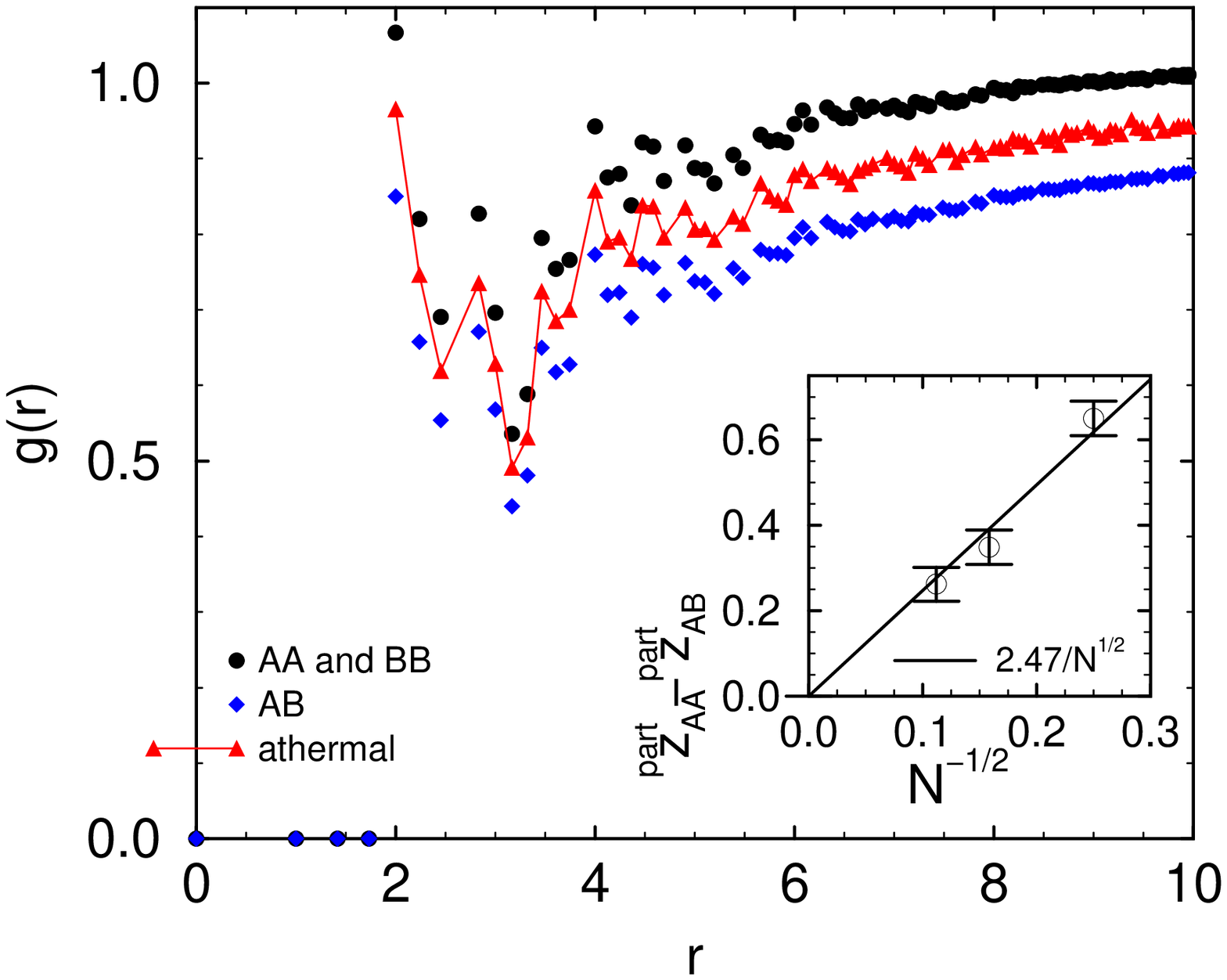}
       }
  \mbox{
       \setlength{\epsfxsize}{9cm}
       \epsffile{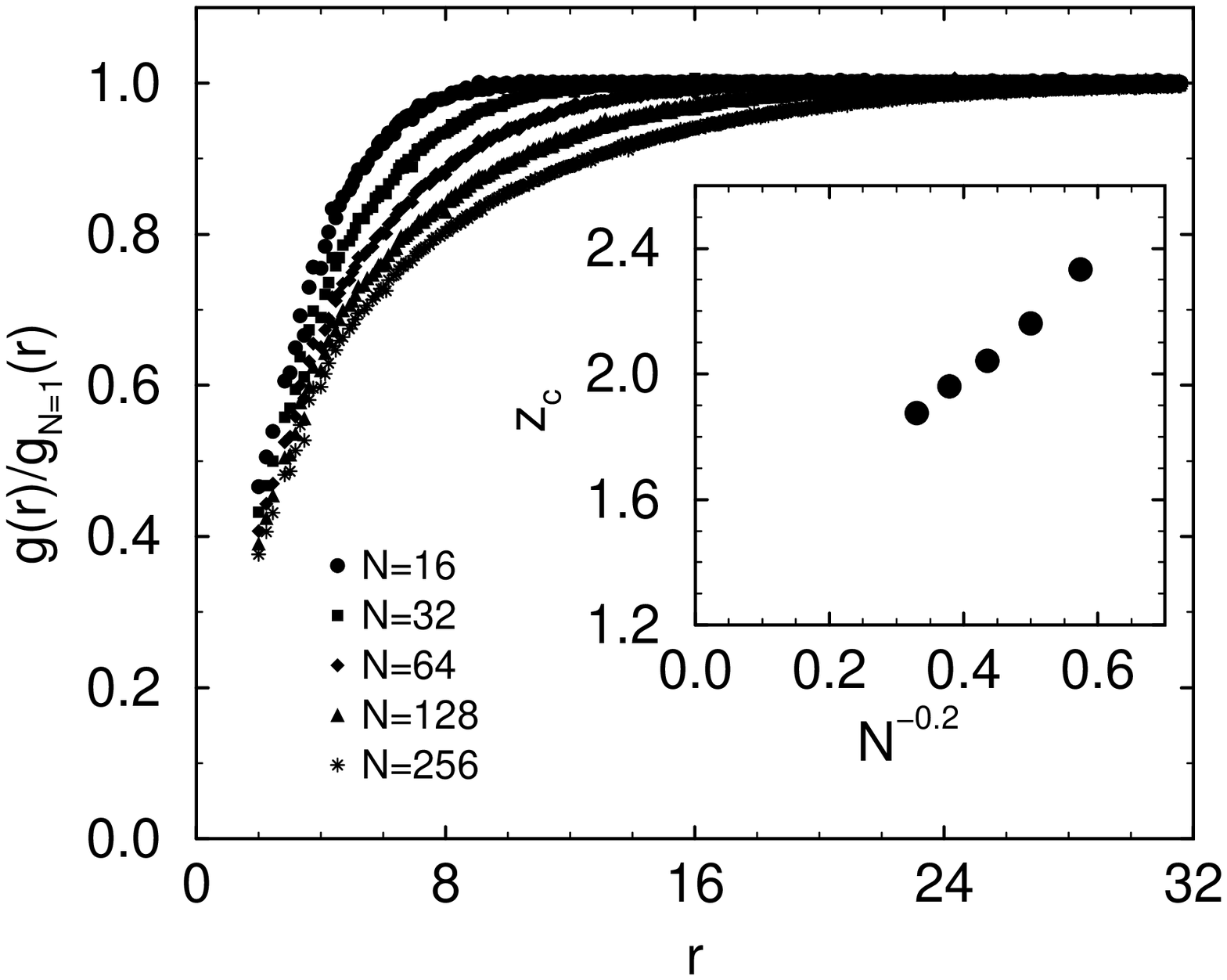}
       }
  \mbox{
       \setlength{\epsfxsize}{9cm}
       \epsffile{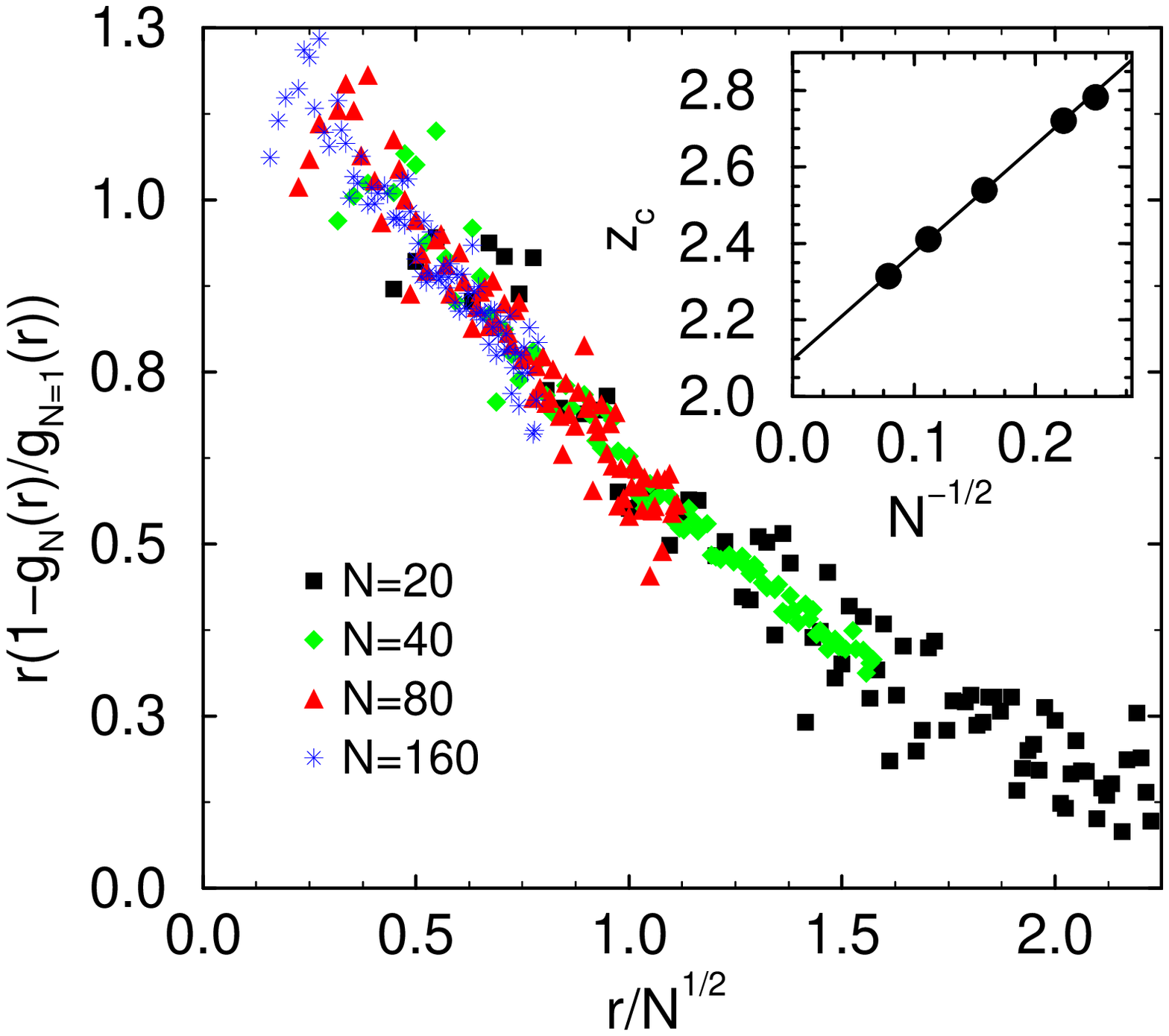}
       }
\end{minipage}%
  \hfill%
\begin{minipage}[c]{54mm}%
\caption{\label{fig:chole}}
({\bf a})Intermolecular paircorrelation function for chain length $N_A=N_B=80$ for the athermal system (triangles)
and at criticality (circles and diamonds). The inset presents the scaling of the non-random mixing with
increasing chain length. $z_{AA}^{\rm part} = \langle \phi \rangle \int {\rm d}^3{\bf r} g^{\rm inter}_{AA}({\bf r})$.
({\bf b}) Intermolecular paircorrelation function of ring polymers.
The inset presents the number of intermolecular contacts as a function of the chain length. The number of contacts is smaller than for
linear chain of the same number of monomers.  From ref.\cite{MWC2}.
({\bf c}) Correlation hole of linear chains: Scaling behavior of the athermal intermolecular pair correlation function with chain length.
The inset shows the chain length dependence of the effective coordination number. The line corresponds to
$z_c = 2.1+2.8/\sqrt{N}$. From ref.\cite{M0}.
\end{minipage}%
\end{figure}

In order to relate the measured critical temperatures to the structure of the polymeric fluid\cite{M0}, we present in Fig.\ref{fig:chole}({\bf a}) the intermolecular pair correlation 
function of strictly symmetric polymer blends in the athermal limit and at the critical point. In the athermal case, the distinction between the two species becomes 
irrelevant. The intermolecular pair correlation function mirrors two effects\cite{M0}: Due to the extended monomer structure the pair correlation function
vanishes for distances $r<2$. The presence of vacancies introduces local packing effects, which give rise to a highly structured function at short distances.
One can identify several neighbor shells, which are characteristic of the monomeric fluid. These packing effects are, of course, absent in simple lattice models where 
a monomeric unit occupies a single lattice site and are less pronounced in the bond fluctuation model than in continuum models. The length scale of these packing effects
is set by the monomeric extension or the statistical segment length; the detailed shape depends strongly on the model and the degree of structure on local length scales.
Furthermore, the extended structure of the macromolecules manifests itself in a reduction of contacts with {\em other} chains on intermediate length scales\cite{DG}.
The length of this polymeric correlation hole is set by the radius of gyration $R_g$ and its shape is characteristic for the large scale conformations of the molecule.

To a first approximation, we assume that the fluid structure is determined by the packing of the hard cubes on the lattice; neither the connectivity of the
monomers along a polymer nor the thermal interactions influence the total pair correlation function. Under this assumption, we can separate the fluid-like packing effects on the
monomer scale and the polymeric correlation hole effects and approximate the intermolecular pair correlation function by\cite{M0}:
\begin{equation}
g^{\rm inter}(r) = g_1(r) \left( 1- \frac{1}{r^\alpha}f(r/R_g) \right)
\label{eqn:ch}
\end{equation}
where $g_1$ denotes the pair correlation function of the monomer fluid and the function $f$ parameterizes the structure of the molecules on the scale of the 
radius of gyration. The exponent $\alpha= 3-1/\nu$ is determined by the requirement that the correlation hole contains $N$ monomers. $\nu$ denotes the scaling exponent 
of the chain extension with the chain length $R \sim N^\nu$. Indeed, this factorization works excellently for flexible molecules in the bond fluctuation model. For 
linear chains this is demonstrated
in ref\cite{M0}, while Fig.\ref{fig:chole}({\bf b}) presents the data for an athermal melt of non--concatenated and not self--knotted polymer rings. The ratio 
$g^{\rm inter}(r)/ g_1(r)$ is largely independent of packing effects and permits a distinction between monomeric packing effects and polymeric correlation hole effects 
in the simulations -- though the length scales are not clearly separated for short rings or chains. Not surprisingly, the correlation hole becomes deeper and wider as we 
increase the molecular weight. The scaling behavior of the correlation hole is shown in Fig.\ref{fig:chole}({\bf c}) for linear athermal chains.  It
imparts a chain length dependence on the Flory Huggins parameter or the effective coordination number, respectively.
\begin{equation}
z_c = \rho \int {\rm d}^3{\bf r}\; g^{\rm inter}(r) = z_c^{\infty} \left(1 + \frac{{\rm const}}{N^{3\nu-1}}\right)
\end{equation}
The scaling of the effective coordination numbers for flexible linear chains and flexible ring polymers are presented in the inset of the figures.
The effective coordination number approaches its limiting scaling behavior with a $1/N^{3\nu-1}$ correction. Rings in a melt are (in our model)
neither concatenated nor self-knotted\cite{MWC1}. These topological constraints squeeze the ring and for the investigated range of chain lengths and density, the
ring extension is characterized by an effective Flory exponent close to $2/5$, in agreement with a Flory-like estimate by Cates and Deutsch\cite{CATES}. Since the structure of the
rings is more compact than that of linear chains, the number of intermolecular contacts is about $30\%$ smaller than for linear chains. Even though the {\em local} molecular 
structures of flexible ring polymers and linear chains are identical, we anticipate a blend of ring polymers to be more miscible than a blend of chemically identical
pairs of homopolymers.

In Fig.\ref{fig:chole}({\bf a}) we also present the intermolecular pair correlation functions $g_{AA}$ and $g_{AB}$ for chain length $N=80$ close to criticality.
In accord with intuition, $AA$ contacts are more likely than $AB$ ones and, hence, $g^{\rm inter}_{AA} > g^{\rm inter}_{BB}$. However, note that
the sum of $AA$ and $AB$ correlations can be well approximated by the intermolecular paircorrelation function $ g^{\rm inter}_{\rm atherm}({\bf r})$
in the athermal limit:
\begin{equation}
\frac{g^{\rm inter}_{AA}({\bf r}) + g^{\rm inter}_{AB}({\bf r})}{2} \approx g^{\rm inter}_{\rm atherm}({\bf r})
\end{equation}
This relation shows that the weak difference in interactions between the monomers $\chi \sim 1/N$ does not alter the structure of the monomer fluid.
The system's energy is mainly determined by composition fluctuations. This approximative decoupling between density fluctuations/packing effects and 
composition fluctuations in our model makes it possible to use the athermal value of the intermolecular pair correlation functions in Eq.(\ref{eqn:chis}). 
This identification corresponds to the high temperature approximation in the framework of the P-RISM theory\cite{SCHWEIZER1}. In this sense, the miscibility behavior in this 
model can be described to a good approximation by a purely enthalpic $\chi$ parameter. At much lower temperatures the blend is strongly segregated ($\langle \phi \rangle \approx 0$ 
or $\langle \phi \rangle \approx 1$), and the system cannot lower its energy by composition fluctuations. In this temperature regime $\chi N \gg 1$ the structure 
of the fluid depends on temperature and the system lowers its energy by density fluctuations upon decreasing the temperature. For the present choice of interactions 
these two mechanisms are rather well separated: concentration fluctuations prevail for $\chi N \sim {\cal O}(1)$ whereas density fluctuations are important for 
$\chi \sim 1/T_\Theta$ or $\chi N \gg 1$. This suggests that this decoupling is characteristic for the behavior of strictly symmetric blends of high
molecular weight macromolecules. If we were to model the monomeric interactions by attractive potentials with a slight difference between the species, however,
the fluid structure would depend on the temperature even on the temperature scale where the blend demixes, and the two phenomena would exhibit an interesting interplay. 

The inset of Fig.\ref{fig:chole}({\bf a}) presents the integral of the correlation functions over the range of the square well potential, i.e., 
$z^{\rm part}_{AA} = \langle \phi \rangle \int_{r \leq \sqrt{6}} {\rm d}^3{\bf r}\; g^{\rm inter}_{AA}({\bf r})$ and similarly for $z_{AB}^{\rm part}$.
In agreement with the considerations in Sec.II and P-RISM calculations, the Monte Carlo simulations show that the difference between the $AA$ and $AB$ 
contacts decreases like $1/\sqrt{N}$, when $\chi N$ is held constant. This exemplifies that the random mixing approximation ({\bf B}) is justified 
in the limit $N \to \infty$.

\begin{figure}[htbp]
\begin{minipage}[t]{102mm}%
\setlength{\epsfxsize}{10cm}
\mbox{\epsffile{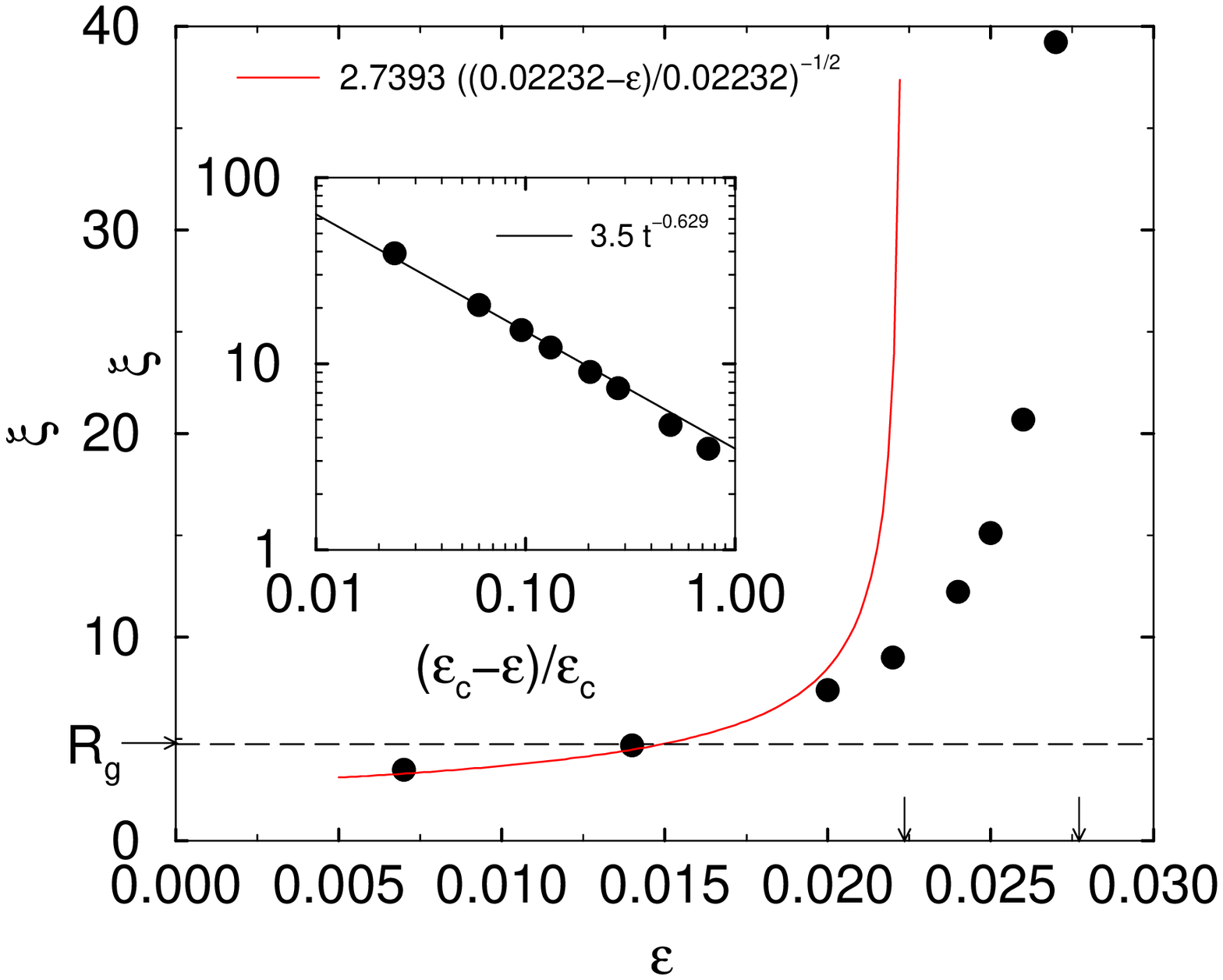}}
\end{minipage}%
\hfill%
\begin{minipage}[t]{54mm}%
\caption{\label{fig:xi}}
         Temperature dependence of the correlation length, and comparison to the
         Mean Field prediction (solid line). The dashed line marks the gyration radius.
         Arrows indicate the actual critical
         temperature and the mean field prediction. \newline
         inset: divergence of the correlation length at criticality. From ref.\cite{DYN}.
\end{minipage}%
\end{figure}

Upon approaching the critical point, the correlation length of composition fluctuations grows. This can be assessed directly in the simulations via
the correlation function of composition fluctuations\cite{DYN}. From its exponential decay at long distances, we extract the correlation length $\xi$ of composition fluctuations.
According to the Ginzburg criterion\cite{GINZBURG}, the mean field theory fails when $|\chi - \chi_c|/\chi_c > 1/N$ or $\xi > N$. In Fig.\ref{fig:xi} we compare
the measured correlation length at $\langle \phi \rangle = 1/2$ for chain length $N=16$ and system size $L=128$ with the mean field prediction (cf.\ 
Eq.(\ref{eqn:xi})). Far above the critical temperature, the mean field yields a rather good description. However, upon approaching the critical temperature 
from above, the mean field theory overestimates the correlation lengths and yields a mean field critical temperature which is about $20\%$ to high. The inset 
shows the divergence of the correlation length at the critical point; it is compatible with a form $\xi \sim [(\chi_c - \chi)/\chi_c]^{-\nu}$ where the 
exponent $\nu=0.629$ is characteristic of the 3D Ising universality class. In agreement with a detailed analysis of Deutsch and Binder\cite{HPD}, this result shows that
the non--mean field regime around the critical point is quite extended for chain length $N=16$.

\begin{figure}[tbhp]
\begin{minipage}[c]{54mm}%
  \mbox{
       \setlength{\epsfxsize}{10cm}
       \epsffile{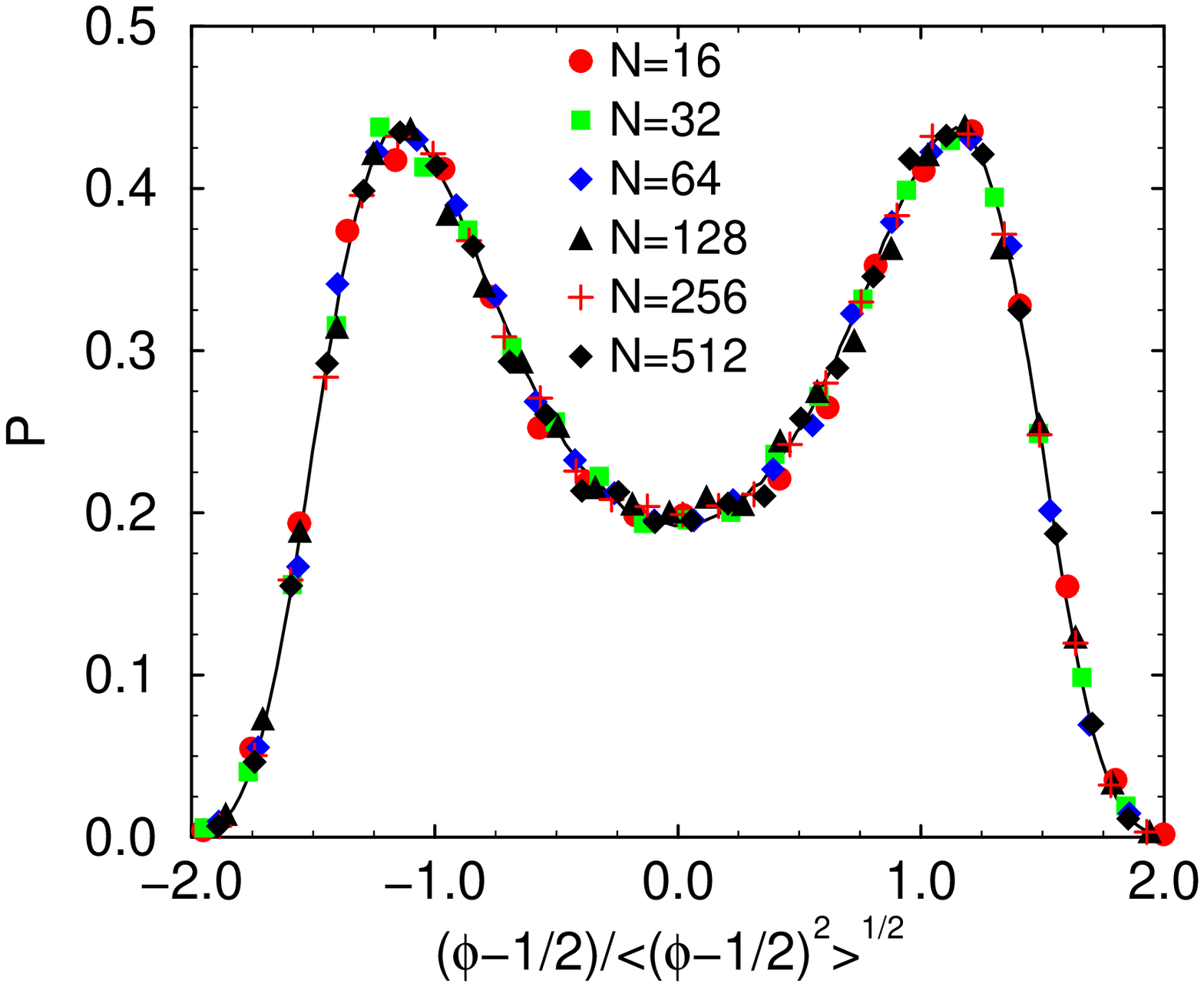}
       }
\end{minipage}%
  \hfill%
\begin{minipage}[c]{54mm}%
\caption{\label{fig:prob}}
Accurate location of the critical temperature via mapping the probability distribution function of the order parameter
fluctuations onto the universal distibution of the 3D Ising universality class. The latter is shown as a solid line,
while the symbols denote the Monte Carlo results of a symmetric blend of ring polymers with chain length $N=16$ to $512$.
\end{minipage}%
\end{figure}

If the correlation length $\xi$ becomes comparable to the linear dimension $L$ of the simulation cell, the Monte Carlo data exhibit a pronounced dependence on the
system size. A finite size analysis\cite{FSS,NIGEL} is necessary to extract reliable estimates for the critical temperature. For symmetric as well as for asymmetric  blends
there exist sophisticated techniques to cope with the divergence of the correlation length at the critical point. If the universality class of the critical point
is known from symmetry considerations and the dimensionality of the order parameter, a very efficient and accurate method consists of mapping the distribution function 
of the order parameter (or ordering operator in asymmetric blends) onto the universal functions which describe the universality class. For a symmetric binary polymer blend, the order parameter is the
deviation of the composition from its critical value $\phi_c$. The results of such a mapping are presented in Fig.\ref{fig:prob} for a symmetric blend of ring polymers.
The success of the mapping of the measured probability distribution onto the universal 3D Ising curve not only accurately determines the value of the critical temperature but
furthermore corroborates that the system size $L=128$ is large enough to reveal the true 3D Ising character of the critical point.
A detailed analysis for asymmetric polymer blends has been presented in ref.\cite{MW}, and the method is generally applicable\cite{NIGEL} to a wide variety of systems.
A finite size scaling analysis also yields the binodals in the vicinity of the critical point.

\begin{figure}[tbhp]
\begin{minipage}[t]{102mm}%
  \mbox{
       \setlength{\epsfxsize}{10cm}
       \epsffile{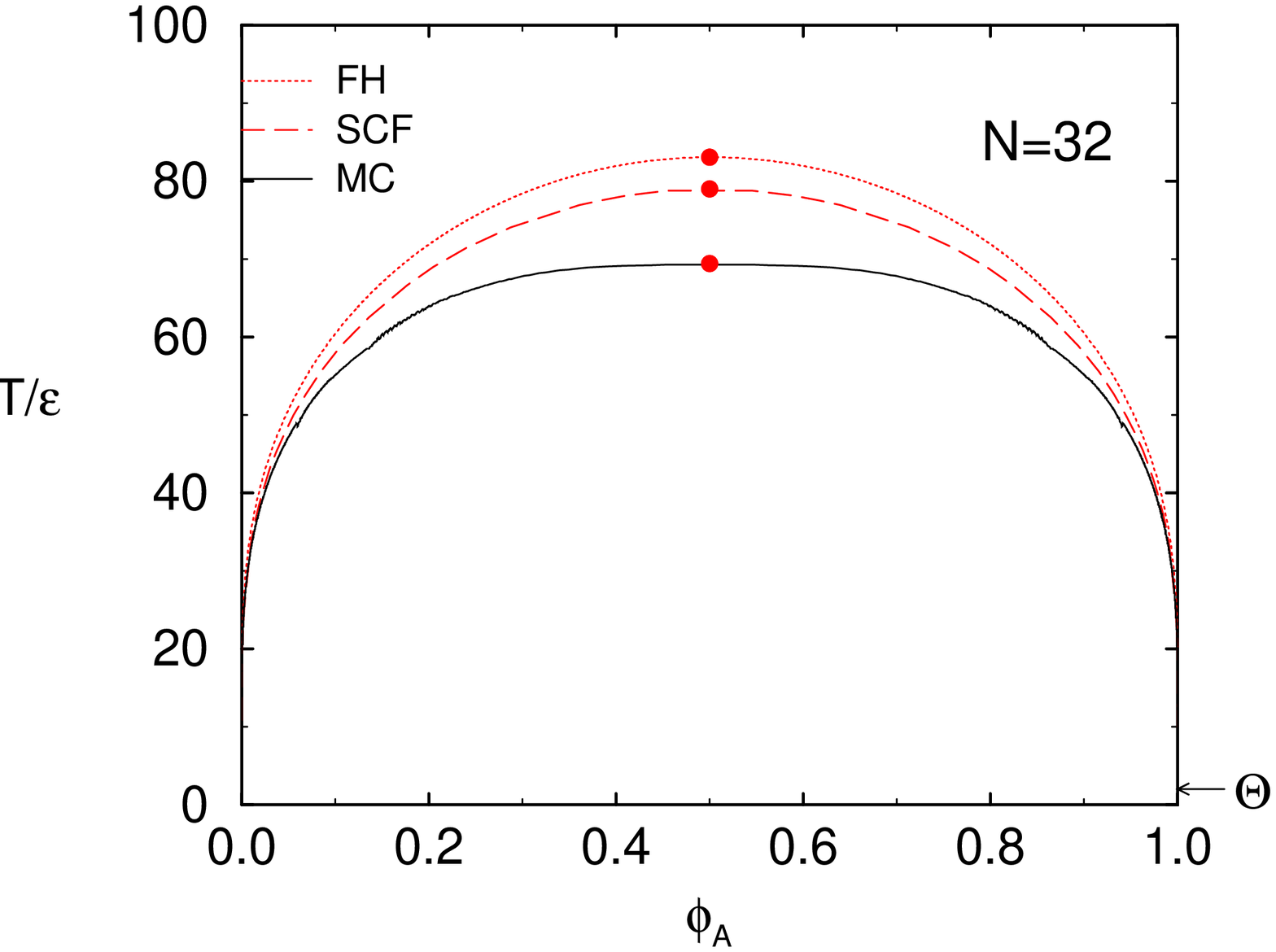}
       }
\end{minipage}%
  \hfill%
\begin{minipage}[c]{54mm}%
\caption{\label{fig:pd}}
Phase diagram for a symmetric polymer blend at chain length $N=32$. The solid line is extracted from
the Monte Carlo simulations using finite size scaling techniques. The dotted line corresponds to the
Flory Huggins theory, whereas the dashed line shows the result of a more sophisticated mean field theory.
The arrow at the left hand side marks the $\Theta$ temperature.
From ref.\cite{CLUSTER}.
\end{minipage}%
\end{figure}

A comparison between the phase diagram of a binary polymer blends and the mean field prediction, where we have identified the $\chi$ parameter according to 
Eq.(\ref{eqn:chis}) via the athermal intermolecular pair correlation function is presented in Fig.\ref{fig:pd} for a symmetric blend of linear chains of length
$N=32$. The mean field theory overestimates the critical temperature and the shape of the binodals in the vicinity of the critical point differ. In the mean field 
theory they are parabolic ($\beta^{\rm MF}=1/2$), while they are flatter ($\beta^{3DI}=0.324$) in the simulations. However, the overestimation of the critical
temperature is smaller than for chain length $N=16$. The dashed line shows a mean field estimate\cite{CLUSTER}, which includes composition fluctuations on
the length scale of a small cluster which consists of $n_c=14$ polymers, but treats long range composition fluctuations in a mean field fashion. Of course, this
theory also yields a parabolic shape of the binodals, but results in an improved estimate of the critical temperature. The arrow on the left hand side marks the $\Theta$
temperature ($\epsilon/k_BT_\Theta=0.495(5)$\cite{LG}) , below which a liquid-vapor phase separation sets in. For our choice of monomeric interactions the two
phenomena are well separated.

\begin{figure}[tbhp]
\begin{minipage}[c]{54mm}%
  \mbox{
       \setlength{\epsfxsize}{10cm}
       \epsffile{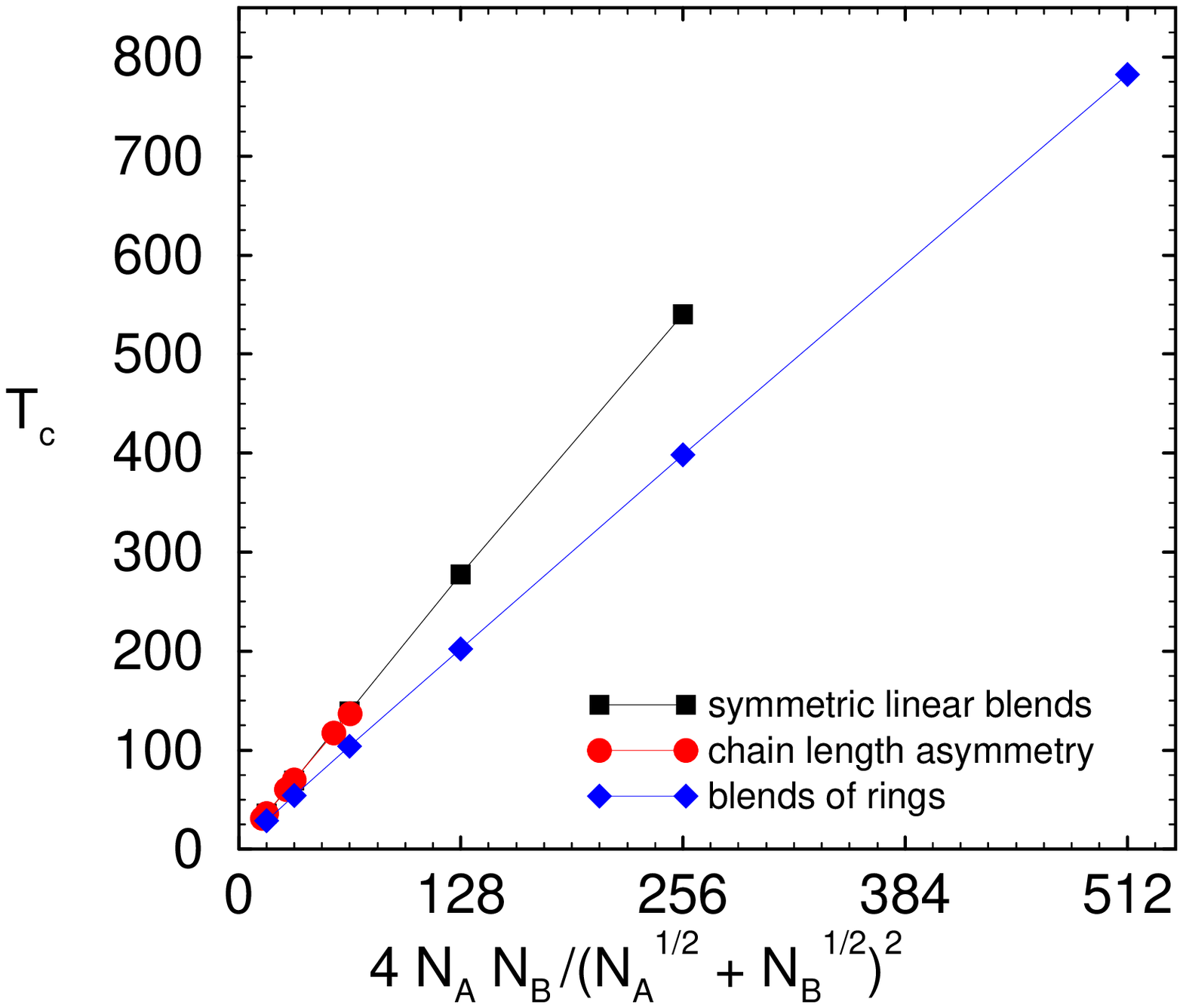}
       }
\end{minipage}%
  \hfill%
\begin{minipage}[c]{54mm}%
\caption{\label{fig:tcscal}}
Scaling of the critical temperature with chain length. The figure presents data for symmetric linear chain\cite{HPD},
results for blends with different chain lengths\cite{M0}, and for blends of ring polymers. 
\end{minipage}%
\end{figure}

\begin{figure}[tbhp]
\begin{minipage}[c]{54mm}%
  \mbox{
       \setlength{\epsfxsize}{10cm}
       \epsffile{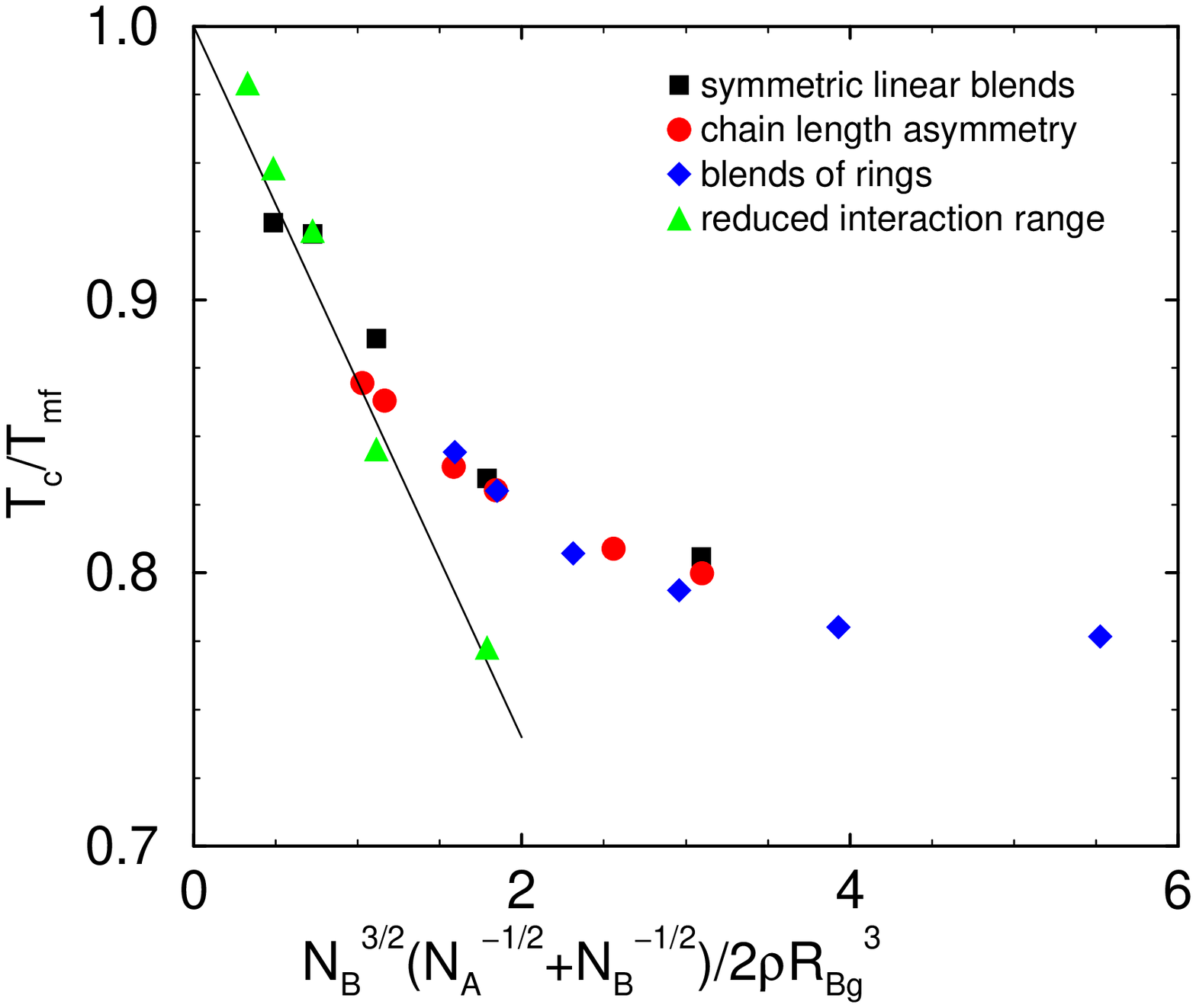}
       }
\end{minipage}%
  \hfill%
\begin{minipage}[c]{54mm}%
\caption{\label{fig:tc1}}
Ratio of the critical temperature (as determined in Monte Carlo simulations) and the Flory Huggins estimate
for binary blends. Using the scaling variable $\sqrt{\chi}/\rho\bar{b}^3$ the Monte Carlo results for
blends of linear chains and ring polymers collapse onto a common curve.
\end{minipage}%
\end{figure}

The linear dependence of the critical temperature on the chain length is presented in Fig.\ref{fig:tcscal}. The figure displays also the critical temperatures
of mixtures of different chain lengths and blends of ring polymers. This linear scaling of the critical temperature at constant density is also reproduced in
off-lattice models\cite{KREMER,SMILE}. The overestimation of the critical temperature is investigated in more detail in Fig.\ref{fig:tc1}. 
Here, we plot the ratio between the Monte Carlo results and our simple mean field estimate of the critical temperature.
Upon increasing the chain length the difference between the Monte Carlo result and the mean field estimate decreases.
The figure includes the ratios between the critical temperature and the mean field estimate for mixtures of different chain lengths and blends of ring polymers. Using the 
scaling variable $\sqrt{\chi_c}/\rho \bar{b}^3$, we achieve a collapse of all data onto a common curve within the accuracy of the Monte Carlo data which is of the order $1-5\%$.
The figure also displays results for two choices of interactions ranges (for symmetric linear chains)\cite{HPD}. Circles represent the results of a model, where
the square well potential is extended over the first 54 lattice sites (as in the remainder of this paper), while triangles denote the results of a model, where the
interaction comprises only the 6 nearest lattice sites. For large chain lengths the Monte Carlo results are consistent with an linear dependence on the
scaling variable. The collapse of the ratio  $T_c/T_c^{\rm MF}$ with $\sqrt{\chi_c}/\rho \bar{b}^3$ for all blends marks the regime of chain lengths where 
the universal polymeric behavior dominates. This universal behavior is indicated as a straight line.

The bending towards a constant value of $T_c/T_c^{\rm MF}$ for small chains might be rationalized as follows: The deviations from the mean field behavior depend on 
the correlation length $\xi$. The length scale is set by the amplitude of the square gradient term in the expansion of the free energy per chain with
respect to long range composition fluctuations. In a symmetric polymer mixture the prefactor is of the form $b^2N/36\phi(1-\phi)$. Hence, the free energy costs
of an inhomogeneous composition are due to the configurational entropy. However, for finite--ranged interactions there is an enthalpic contribution
to the square gradient term with a prefactor of the order $\chi N \sigma_e^2$, where $\sigma_e$ denotes the range of the monomeric interaction potentials.
For long chain lengths the entropic contribution dominates, while the enthalpic term becomes important when the range of the thermal interactions becomes
comparable to the chains radius of gyration. Consequently, the range of the interactions does not enter the Ginzburg criterion\cite{GINZBURG} to leading order.
For very small polymers or rather long--ranged monomeric interactions, the interaction range $\sigma_e$ might increase the correlation length and the
shift between mean field and true critical temperature is smaller than estimated by $\sqrt{\chi_c}/\rho \bar{b}^3$. When the range of interactions $\sigma_e$ 
is decreased this effect should set in at smaller chain lengths. This is consistent with the simulation data: the data with the reduced interaction range
show larger deviations at small chain lengths.

In the long chain length limit, strictly symmetric blends in the framework of the bond fluctuation model are very well describable by mean field theory,
if the $\chi$ parameter is identified via the intermolecular pair correlation function of the athermal blend. The decoupling of composition 
and density fluctuations and the temperature independence of the structure of the underlying monomer fluid in the temperature range,
where the phase separation in the binary blends occurs, give rise to a purely enthalpic $\chi$ parameter. For finite chain length, 
deviation between the Monte Carlo results can be traced back to composition fluctuations and a mild perturbation of the chain conformations (see sec.\ V).
At the critical point, however, the mean field theory fails, and we observe 3D Ising critical behavior in accord with general arguments and the
observed deviations ({\bf iii}) from the mean field theory.

The simple form of the $\chi$ parameter is based on the decoupling of the density fluctuations or packing effects from the molecular architecture, the 
composition of the blend, and the temperature. To illustrate the consequences of chain architecture on the miscibility behavior, let us discuss two examples: 
a blend of ring polymers and a blend of polymers with indented monomer shapes. The first one shall highlight the consequences of the chain architecture on long length scales, 
while the second example serves to demonstrate the effect of architecture on the local scale of the monomers.

In spite of having identical molecular architectures, the topological constraint that the rings are neither concatenated with each other nor self-knotted results 
in a shrinking of the ring extension compared to linear chains\cite{CATES}. This partial collapse leads, in turn, to a
smaller number of intermolecular contacts as discussed above. The critical temperatures as well as their ratio to the mean field estimate are compared to
the miscibility behavior of linear chains in Figs.\ref{fig:tcscal} and \ref{fig:tc1}. As anticipated from the analysis of the intermolecular paircorrelation function, topological 
constraints decrease the critical temperature by $30\%$. This effect is quite different from the lowering of the ordering temperature in melts of cyclic
diblock copolymers. In the latter case, the spatial structure influences the ordering temperature directly via the single chain structure factor at
finite wavevector. Since the molecules are more compact, they are less polarized into $A$ rich and $B$ rich zones. Hence, a stronger incompatibility is
necessary to bring about a spatial ordering into mesophases. This has been analyzed in the framework of the random phase approximation by Benmouna\cite{COPRINGS}.
There are Monte Carlo simulations on this system by Weyersberg and Vilgis\cite{VILGISCOP}. 

\begin{figure}[htbp]
\begin{minipage}[t]{102mm}%
\setlength{\epsfxsize}{10cm}
\mbox{\epsffile{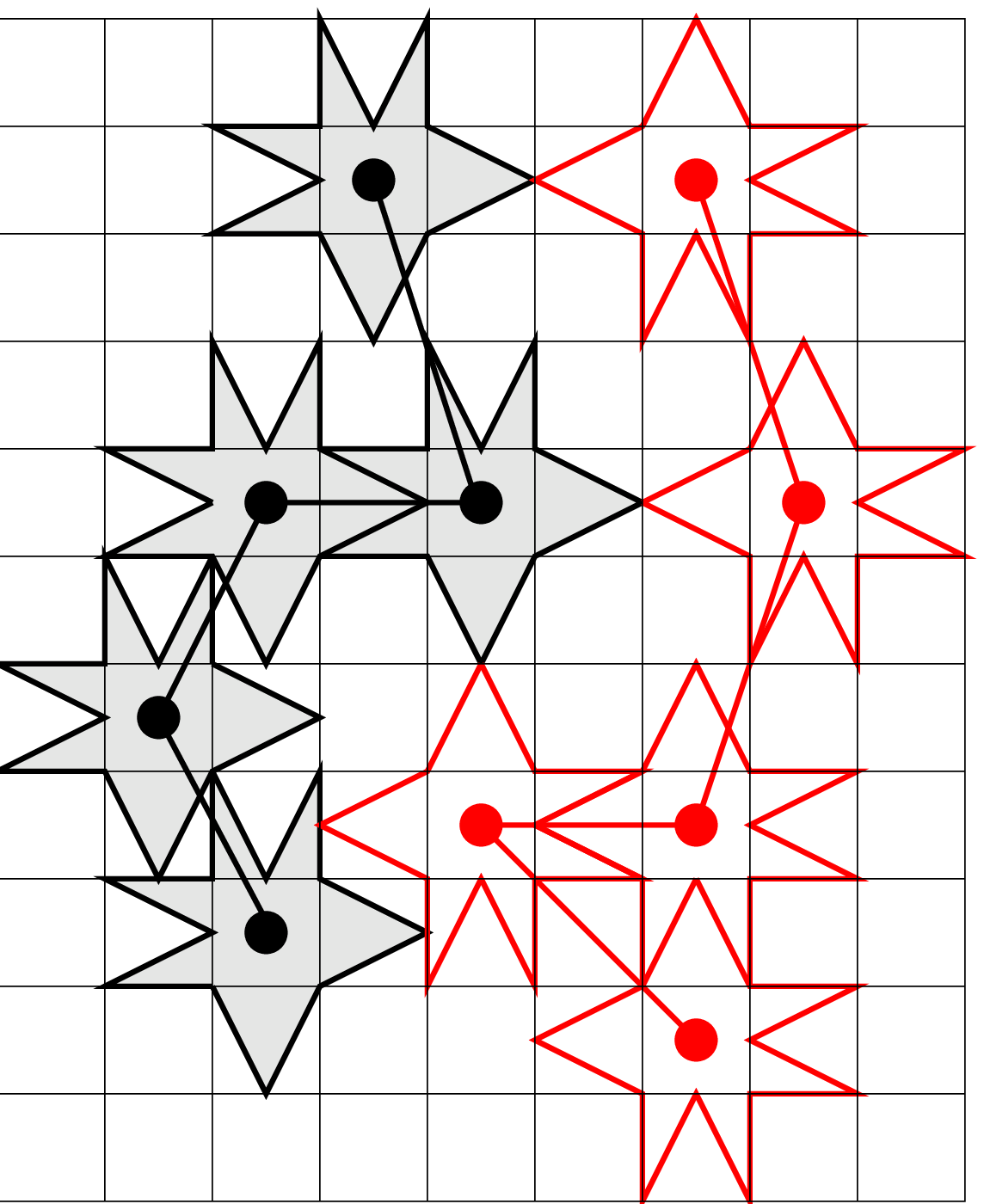}}
\end{minipage}%
\hfill%
\begin{minipage}[c]{54mm}%
\caption{\label{fig:zacken}}
         Illustration of the intended monomer structure. A monomers are shaded, B monomers open. The effect of the indentation
         can be described by a non-additive hard core repulsion between monomers. As alluded, there is a strong entropic packing
         advantage which leads to phase separation. From ref.\cite{STIFF1}.
\end{minipage}%
\end{figure}

The consequences of the molecular architecture on a local scale can be even more pronounced. In Fig.\ref{fig:zacken} we sketch a symmetric mixture with indented
monomer shapes\cite{STIFF1}. As one can observe, monomers of different types are separated by a spatial distance of at least $\sqrt{3}$, whereas monomers of the same species
can approach each other up to a distance $2$. The properties of the pure phases are not altered, because the packing constraints only restrict the minimal distance 
between unlike species.  Of course, this extremely simple monomer shape is no faithful representation of realistic monomer packing effects on a microscopic length scale. 
However, in the spirit that a monomer in the bond fluctuation model corresponds to a small number of chemical repeat units, we expect the model to capture some universal, 
long wavelength properties on the coarse grained length scale of a Kuhnian segment. The shape of the monomers leads to a non-additive packing between monomers of different 
species. In an $A$-rich phase an $A$ polymer possesses more conformational freedom as in a $B$--rich environment.

We cannot resort to the simple estimate
of the $\chi$ parameter (\ref{eqn:chis}), but have to go back a step and  approximate the single chain partition function (\ref{eqn:single}). 
The energy $\delta E$ of inserting an $A$ chain into a mixture of composition $\phi$ can still be expressed via the integral of the intermolecular correlation function
over the range of the interactions. However, the number of chain conformations is now composition dependent. Without the non--additive packing constraint, the
number of chain conformations can be approximated by $V z_b^{N_A-1}$, where $z_b$ denotes the effective number of possibilities to add a segment without violating
the excluded volume constraint. The non-additivity
reduces the number of conformations, because none of the $z_{\rm na}$ monomers at distance $2$ away from each monomer of the inserted $A$ polymer must be occupied by
species $B$. $z_{\rm na}$ can be approximated via the intermolecular pair correlation function in an additive mixture: $z_{\rm na} = \rho \int_{r \leq 2} {\rm d}^3{\bf r}\;
g^{\rm inter}({\bf r})$. This constraint reduces the number of conformations on average by a factor $[1-(1- \langle \phi \rangle)z_{\rm na}]^{N_A}$. 
We approximate the conformational entropy $s_A$ per monomer by\cite{STIFF1}:
\begin{equation}
s_A = \frac{1}{N_A} \ln \left(z_b^{N_A-1} [1-(1-\langle \phi \rangle)z_{\rm na}]^{N_A} \right) \approx \ln z_b  - (1-\langle \phi \rangle)z_{\rm na}
\end{equation}
Using a similar expression for $s_B$  the $\chi_{\rm SANS}$ parameter (\ref{eqn:sans}) takes the form:
\begin{equation}
\chi_{\rm SANS} = \frac{2\epsilon z_c}{k_BT}+z_{\rm na}
\label{eqn:lcsp}
\end{equation}
The enthalpic term is identical to the additive blends, where the monomers are not indented but simple cubes. The second contribution is a temperature
independent, positive contribution to the $\chi$ parameter. 

\begin{figure}[htbp]
\begin{minipage}[t]{102mm}%
\setlength{\epsfxsize}{10cm}
\mbox{\epsffile{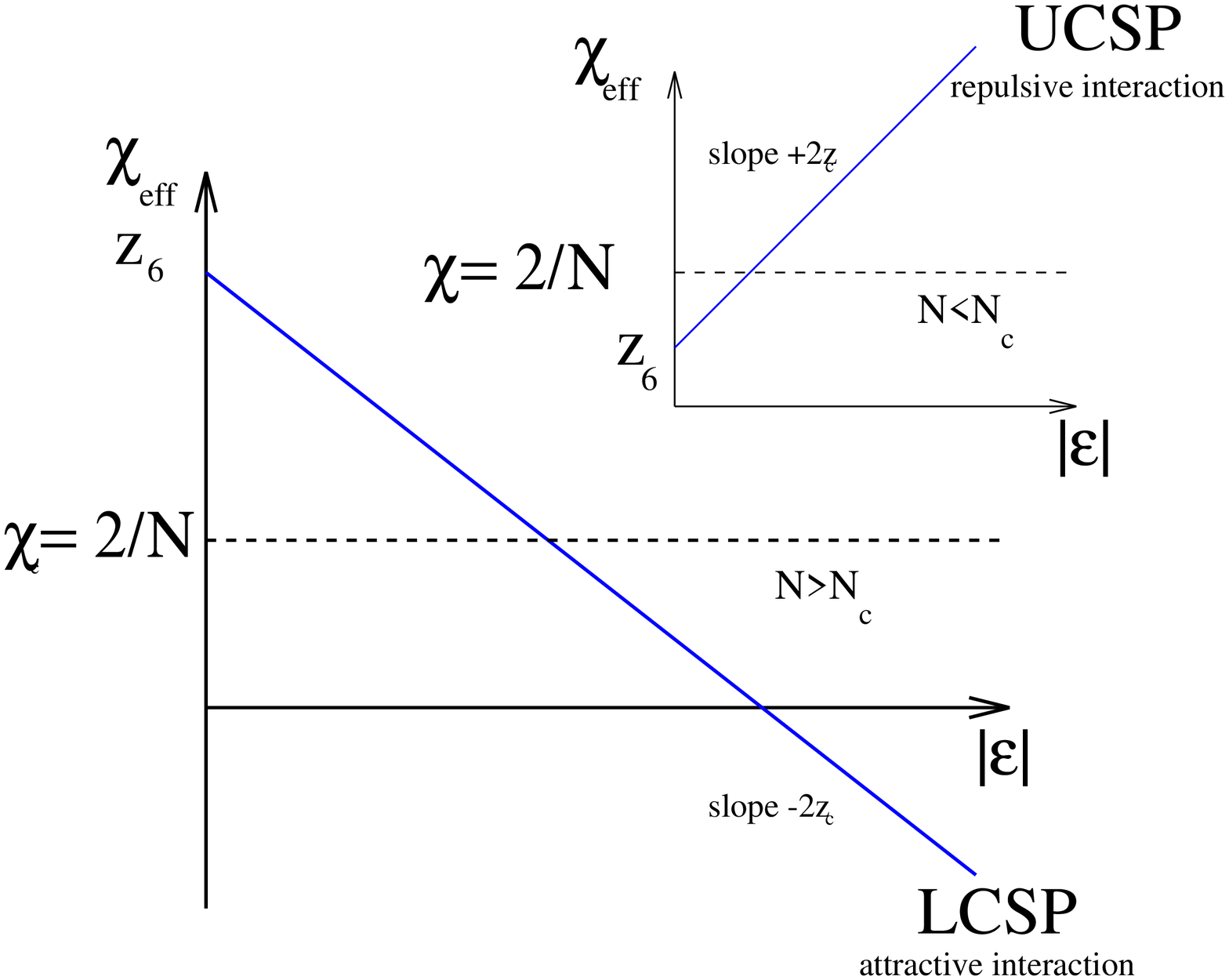}}
\end{minipage}%
\hfill%
\begin{minipage}[c]{54mm}%
\caption{\label{fig:chiz}}
Qualitative behavior of the $\chi$-parameter for short chains ({\em UCSP}) and long chains ({\em LCSP}).
From ref.\cite{STIFF1}.
\end{minipage}%
\end{figure}

The consequences for the miscibility behavior are discussed in Fig.\ref{fig:chiz}: If the chain length is small enough, the translational entropy, which favors mixing
dominates over the positive entropic contribution to the $\chi$ parameter. For $N <N_c\equiv 2/z_{\rm na}$ the athermal blend is completely miscible. Upon lowering
the temperature the blend phase separates at an upper critical solution temperature (UCST). For longer chains ($N>N_c$) the athermal blend is only partially miscible and to bring 
about a phase transition, we have to assume an attractive interaction between unlike species. Keeping with the notation of additive blends the attraction corresponds to
negative values of $\epsilon$. Upon increasing the absolute magnitude of the interactions $|\epsilon|$, the blend becomes miscible at a lower critical solution temperature (LCST):
\begin{equation}
\frac{1}{T_c} \sim -\epsilon_c =  \frac{z_{\rm na}}{2z_c} \left( 1-\frac{2}{z_{\rm na}N} \right)
\end{equation}
The scaling of the lower critical solution temperature is in marked contrast to the scaling at the upper critical solution point with temperature.
In the latter case the transition temperature is determined by a competition between the translational entropy of a polymer versus the monomeric repulsion;
the critical temperature (UCST) increases linearly with chain length. In the former the conformational entropy per segment is balanced against the monomeric 
interactions and a chain length independent lower critical solution temperature (LCST) is approached from above. When expressed in terms of the $\chi$ parameter the Ginzburg
criterion\cite{GINZBURG} and the critical amplitudes of the magnetization take the same form than for the upper critical solution points of the symmetric, additive mixture.
However, when these quantities are written in terms of temperature the combination $(\chi_c - \chi)/\chi_c$ takes the form $
N z_c |T_c - T|/TT_c \approx 0.5 N z_{\rm na} |T - T_c|/T$. This gives rise to an additional chain length dependence of the critical amplitudes and
the Ginzburg number, when the temperature instead of the $\chi$ parameter is used in the vicinity of a lower critical solution point. For instance,
the binodals in the temperature--composition plane open the wider the larger the chain length and the Ginzburg number which measures the relative distance between
the critical temperature and the temperature where mean field behavior sets in becomes proportional to $N^{-2}$. The dependence of the width of the non-mean field
like region has been investigated experimentally by Schwahn and co-workers\cite{SCHWAHN}. The experiments reveal a strong dependence on the molecular architecture
and the pressure. These effects have been explored in the framework of the Lattice Cluster Theory\cite{FREED_GINZ}.

An important alternative mechanism for the occurance of lower critical solution points or close loop phase behavior are specific interactions (e.g., hydrogen bonds) 
between monomers. The effects of hydrogen bonds on the phase behavior have attracted longstanding interest\cite{HB1,HB2,HB3,HB4}.

\begin{figure}[htbp]
\begin{minipage}[t]{102mm}%
\setlength{\epsfxsize}{10cm}
\mbox{\epsffile{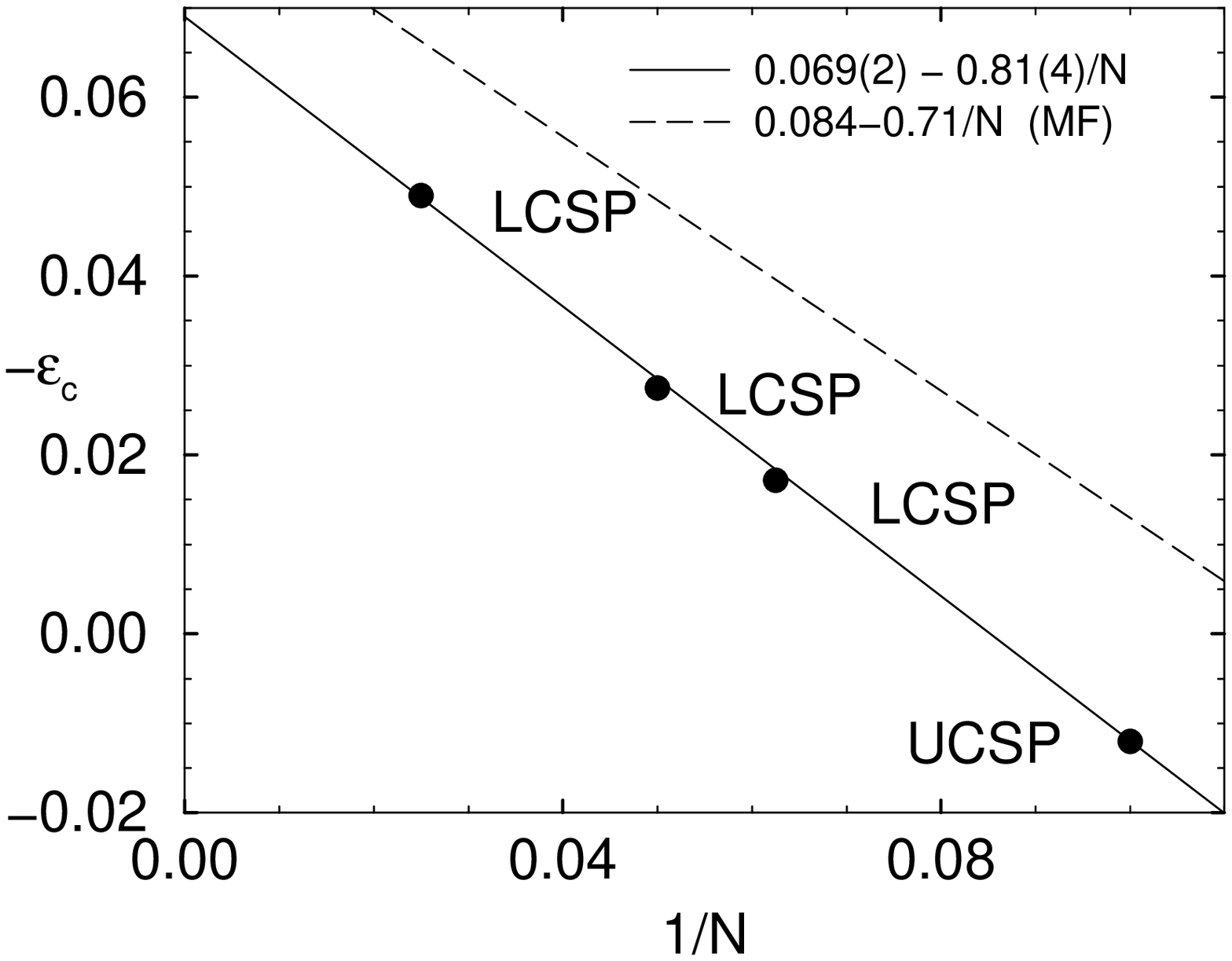}}
\end{minipage}%
\hfill%
\begin{minipage}[c]{54mm}%
\caption{\label{fig:tcz}}
         Chain length dependence of the critical temperature for mixtures of non-additive monomers.
         The straight line is given by $\epsilon_c=0.069(2)-0.81(4)\frac{1}{N}$.
	 From ref.\cite{STIFF1}.
\end{minipage}%
\end{figure}

This mean field picture has been investigated via Monte Carlo simulations\cite{STIFF1} in the bond fluctuation model with the indented monomers sketched in Fig.\ref{fig:zacken}.
Due to the additional excluded volume between unlike species the simulations have been performed at a reduced monomer density $\rho=0.35/8$ in order to
allow for a reasonable acceptance ratio of the semi-grandcanonical identity switches. Using a finite size scaling analysis we have determined the critical temperatures
accurately. The dependence of the critical temperature on the chain length is summarized in Fig.\ref{fig:tcz}. For chain length $N=10$ we
find an upper critical solution temperature while for chain length $N=16$ and larger we observe  lower critical solution points. We find Ising critical behavior
at the lower critical solution points and the temperatures approach a limiting value upon increasing the chain length. The critical temperatures are describable
by a dependence of the form: $-\epsilon_c=0.069(2)-0.81(4)/N$. For chain length $N=20$ we have measured $z_c=1.41$ and $z_{\rm na}=0.238$ in the Monte Carlo simulations.
This yields $\epsilon_c=0.084-0.71/N$ for the critical temperatures. The corresponding mean field estimate is also displayed in the Fig.\ref{fig:tcz}. It describes
the simulation data only qualitatively. Deviations are partially due to the chain length dependence of the parameters $z_c$ and $z_{\rm na}$ via the correlation hole effect 
and composition fluctuations. Moreover, pronounced deviations stem from the crude approximation of the composition dependence of the the configurational entropy which lead to the
simple expression(\ref{eqn:lcsp}). Our approximation treats the (rather strong) non additivity only perturbatively and neglects, e.g., the composition dependence of the
chain conformations (cf.\ sec.\ V).

If monomer of the same species pack less efficient than monomers of the same species, a negative entropic contribution to the $\chi$ parameter results.
An entropic contribution to the $\chi$ parameter occurs not only for non--additive monomer shapes but also for large enough disparities in the segment size.
Mixtures of small and large spheres demix\cite{FRENKEL1} when the size difference between the species is large enough. We expect the consequences to be even more pronounced
for polymers due to the small entropy on mixing. These effects have been explored in the framework of the Lattice Cluster Theory\cite{FREED_REV}.

Some of the deviations of the original Flory Huggins theory can be qualitatively modeled in the framework of the bond fluctuation model, because the structure of
the complex lattice model allows for non trivial chain topologies and packing effects. This allows us to investigate the non-Flory-Huggins effects ({\bf i}) and ({\bf ii})
discussed in the introduction.

\subsection{Asymmetric blends.} 
Generally, the constituents of a blend are not symmetric, and asymmetry might have pronounced effects on the miscibility behavior.
Since the properties of a blend deviate from the linear superposition of the individual properties of its components, the blend has
new, possibly favorable characteristics. In this section we shall present Monte Carlo simulations of various asymmetries in polymer blends.
We shall begin with the weakest asymmetry, namely an asymmetry in adhesive forces between the constituents\cite{HPD2}, proceed via chain length asymmetry\cite{M0}
and finally examine blends of polymers with different stiffness\cite{STIFF1}.

A difference in adhesive forces has been investigated by Deutsch and Binder\cite{HPD2}. They explored the miscibility behavior of a blend which constituents had
identical chain lengths, but monomers interacted via square well potentials of different strength:
\begin{equation}
\epsilon_{AB} = -\epsilon_{BB} \equiv \epsilon \qquad \mbox{and} \qquad \epsilon_{AA} = -\lambda \epsilon
\end{equation}
The parameter $\lambda$ controls the asymmetry of the mixture, and $\lambda=1$ corresponds to the symmetric mixture considered above. In the previous section
we have demonstrated that the weak thermal interactions leave the fluid structure almost unaltered and we expect this approximation to hold in this asymmetric situation
as well. Since the energy of mixing remains still quadratic in the composition, the asymmetry merely changes the $\chi$ parameter and adds a linear term to the
free energy of mixing. The latter can be absorbed into the chemical potential and is immaterial for the miscibility behavior. Hence, within the mean field 
theory the model can be mapped onto a symmetric model. In particularly, the critical composition is independent from $\lambda$, i.e., $\phi_c = 1/2$.
Estimating the $\chi$ parameter via Eq.(\ref{eqn:chis}), we obtain: $\chi = z_c \epsilon (3+\lambda)/2k_BT$. Indeed, the linear dependence of the critical temperature 
on the combination $2/(3+\lambda)$ has been observed in the simulations\cite{HPD2} for chain length $N=32$ and $\lambda=1.5,2$, and $5$. Moreover, within the accuracy 
of the simulations the phase diagram is symmetric around the composition $\phi=1/2$

The semi-grandcanonical algorithm has been generalized to chain length asymmetry for the case that the chain lengths of the constituents are a small multiple of 
each other, i.e., $N_B = k N_A$ with $k=1,2$ and $3$. Within a semi-grandcanonical Monte Carlo move, $k$ $A$ chains are connected to form a long $B$ polymer or 
a $B$ polymer is cut into $k$ pieces which subsequently are relabeled $A$. The acceptance criterion not only involves the energy change associated with the relabeling 
of the monomers but also accounts for the number of possibilities to connect the short $A$ chains to a long $B$ chain. Moreover, the finite size scaling analysis
of the Monte Carlo data becomes more involved, because due account of the asymmetry of the system has to be taken to locate reliably the critical point\cite{MW,NIGEL}.

\begin{figure}[tbhp]
\begin{minipage}[t]{80mm}%
  \mbox{
       \setlength{\epsfxsize}{9cm}
       \epsffile{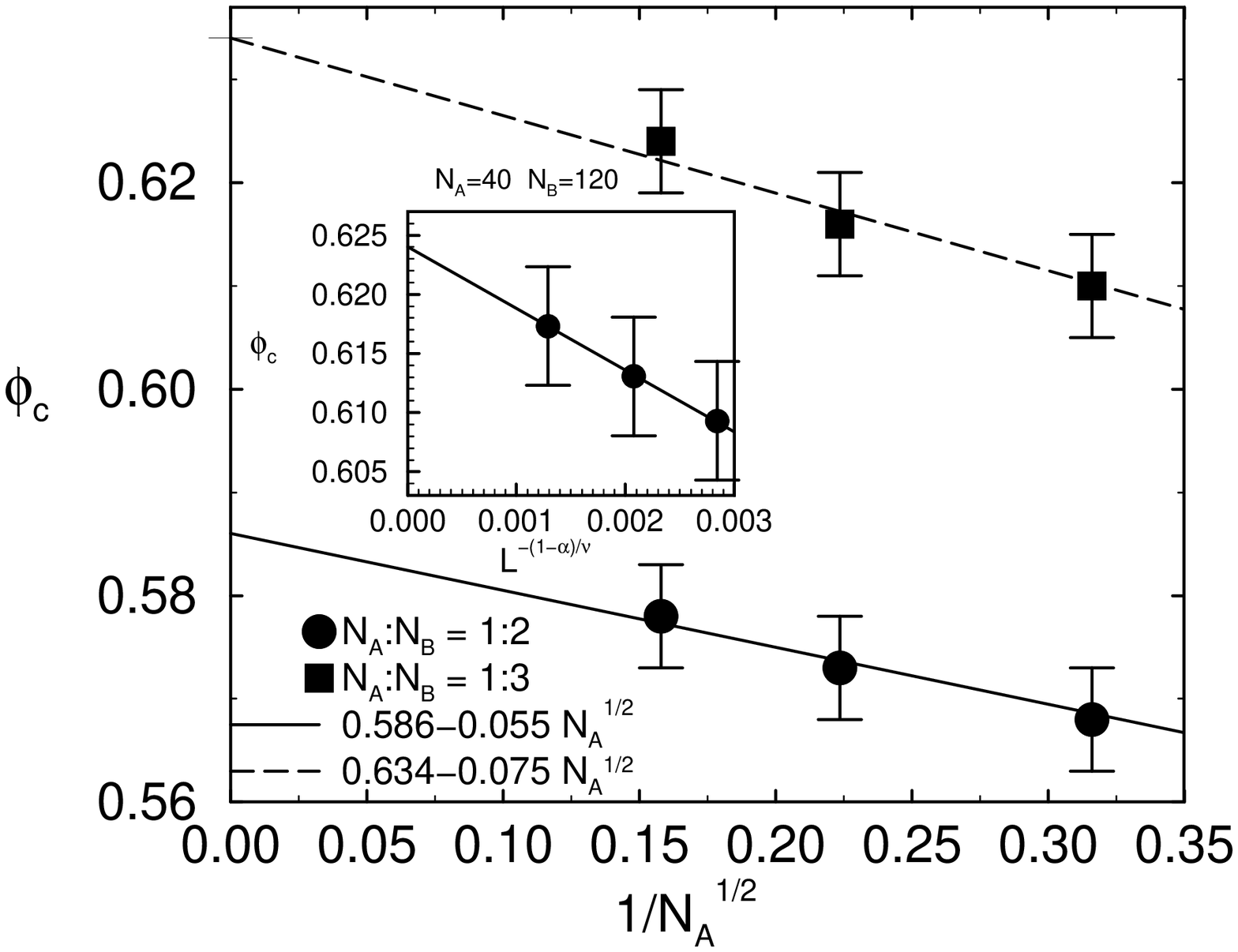}
       }
  \mbox{
       \setlength{\epsfxsize}{9cm}
       \epsffile{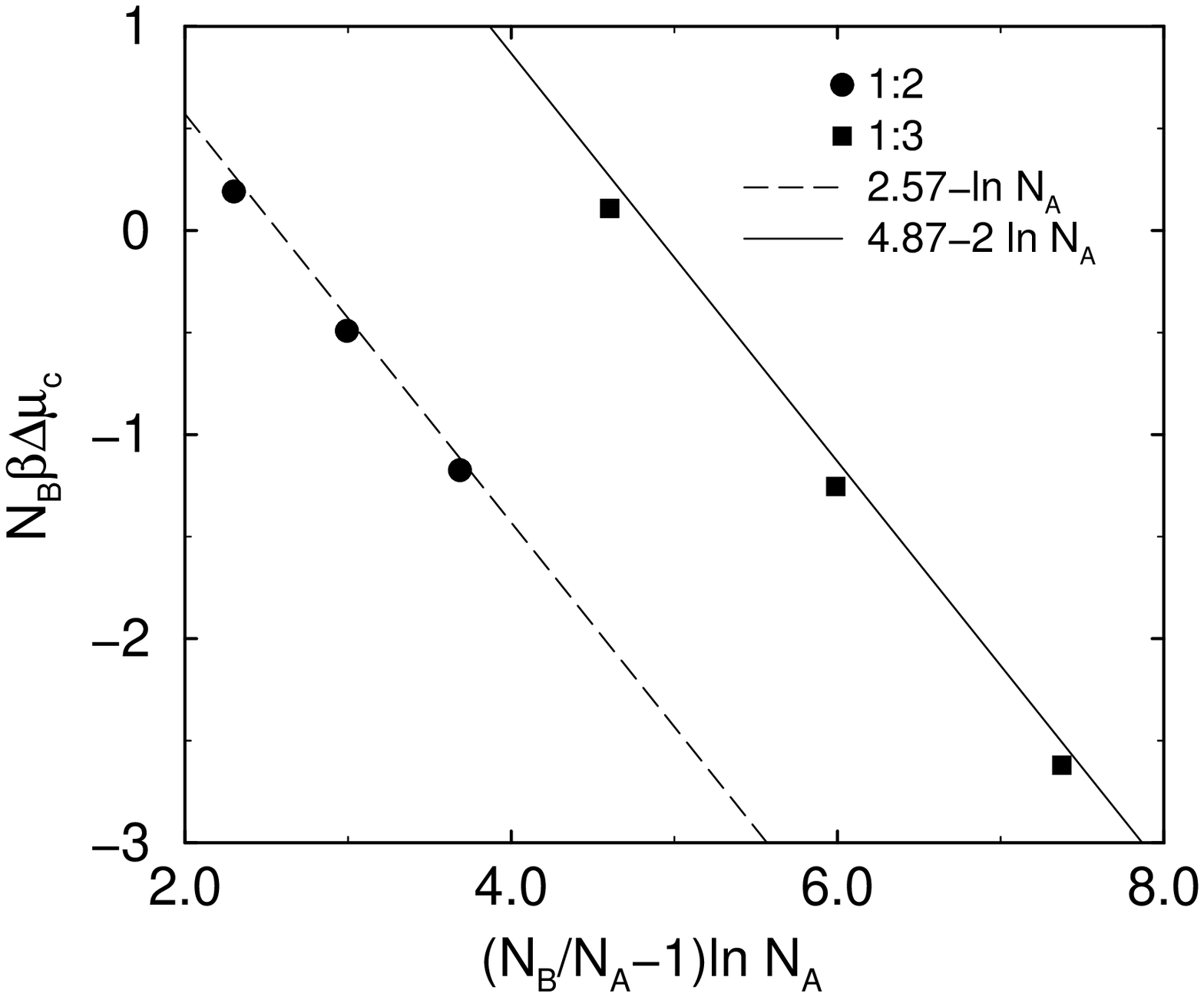}
       }
\end{minipage}%
  \hfill%
\begin{minipage}[c]{54mm}%
\caption{\label{fig:asym}}
Critical density ({\bf a}) and critical chemical potential difference $\Delta \mu_c$ ({\bf b}) for
mixtures of different chain lengths. The critical densities approach the Flory Huggins estimate
from below upon increasing the chain length. The inset presents finite size effects of the critical
density due to field mixing effects. From ref.\cite{M0}.
\end{minipage}%
\end{figure}

The chain length asymmetry modifies the translational entropy term and alters all terms in the expansion of the free energy with respect to the composition.
This results in a shift in the critical composition and a non trivial coexistence curve $\Delta \mu_{\rm coex}(\epsilon/k_BT)$. 
Invoking the assumptions ({\bf A}) and ({\bf B}) the segmental energy differences are given by $e_A - e_B = 2 \chi \phi$. Again, we approximate
the intermolecular pair correlation functions by their athermal values, which depend on the chain length. Moreover, we use $z_{AB} \approx (z_{AA}+z_{BB})/2$\cite{STIFF2}
and obtain $\chi=(z_{AA}^{\rm atherm}+z_{BB}^{\rm atherm})\epsilon/k_BT$.
The segmental entropies $s_A - s_B = (N_A-1) \ln(z_b)/N_A - (N_B-1)\ln(z_b)/N_B$ are independent of composition. In the crudest approximation, the number of
possibilities $z_b$ to add a monomer is the number of allowed bond vectors in the model, i.e. 108. However, the number is slightly reduced because the chain cannot
fold back and we obtain $z_b=104(1)$ for the bond fluctuation model.
The segmental entropies do not contribute to the $\chi$ parameter but shift the chemical potential. The SG--EOS takes the form:
\begin{equation}
\frac{\Delta \mu}{k_BT} = \frac{1}{N_A} \ln \frac{\phi}{N_A} - \frac{1}{N_B} \ln \frac{(1-\phi)}{N_B} - \chi(2\phi-1) 
+ \frac{1}{N_B} \ln \left( \frac{N_B}{2z_b\rho} \right)
- \frac{1}{N_A} \ln \left( \frac{N_A}{2z_b\rho} \right)
\label{eqn:mu}
\end{equation} 
The critical temperature and composition are given by Eq.(\ref{eqn:crit}) whereas the shift in the chemical potential is given by
\begin{equation}
\frac{\Delta \mu_c}{k_BT} = - \frac{1}{N_A} \ln \left( \frac{\sqrt{e}N_A(1+\sqrt{N_A/N_B})}{2z_b\rho}\right)
			    +  \frac{1}{N_B} \ln \left( \frac{\sqrt{e}N_B(1+\sqrt{N_B/N_A})}{2z_b\rho}\right)
\label{eqn:muc}
\end{equation}
Indeed, the scaling of the critical parameters $T_c$ (cf.\ Fig.\ref{fig:tcscal}), $\phi_c$ , and $\Delta \mu_c$ (cf.\ Fig.\ref{fig:asym})
are well describable by their mean field predictions in the long chain length limit. The deviations between the mean field estimate and the Monte Carlo result for the
critical temperature scale in the same way than for mixtures of equal chain length. Moreover, in accord with predictions of Holyst and Vilgis\cite{VILGIS}, the critical composition
is approached from below upon increasing the chain length with a $1/\sqrt{N}$ correction (cf.\ Fig.\ref{fig:asym}({\bf a})). The segmental entropies do not affect $\chi$, they are, however, 
detectable in the shift of the exchange chemical potential.  The prediction (\ref{eqn:muc}) is compared to the Monte Carlo results in Fig.\ref{fig:asym}({\bf b}) without any adjustable 
parameter. The good agreement indicates that the dominant contributions to the segmental entropy are captured by the considerations above. 

\begin{figure}[tbhp]
\begin{minipage}[t]{80mm}%
  \mbox{
       \setlength{\epsfxsize}{10cm}
       \epsffile{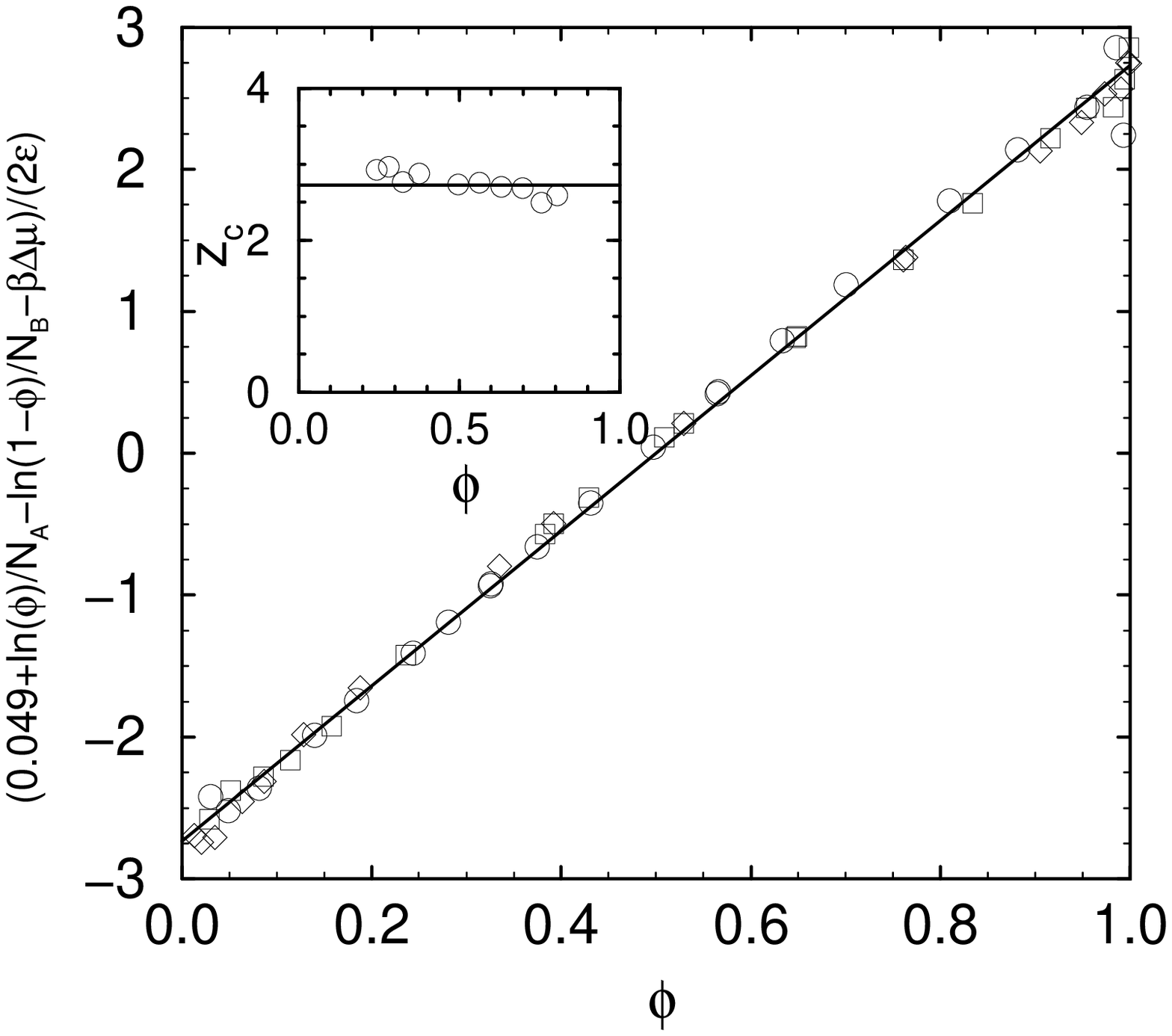}
       }
\end{minipage}%
  \hfill%
\begin{minipage}[c]{54mm}%
\caption{\label{fig:eos}}
$\chi$-parameter and thermal equation of state: difference of the chemical potential ($N_A=10$ $N_B=20$) from its athermal value
for different temperatures. $\epsilon=0.01$ (circles),$\epsilon=0.02$ (squares),$\epsilon=0.025$ (diamonds). \newline
The straight line is given by $(\beta \Delta \mu_{\rm athermal} - \beta \Delta \mu)/2\epsilon= 2.735 (2\rho-1)$ 
The inset shows the determination via the collective structure factor for $\epsilon=0.01$. The solid line marks the value
$z_c=2.735$. From ref.\cite{M0}.
\end{minipage}%
\end{figure}

A direct relation between the simulations in the semi-grandcanonical ensemble and the thermodynamics can be established via the SG--EOS
(\ref{eqn:mu}) which relates the composition $\phi$ of the mixture to the exchange chemical potential $\Delta \mu$. This is particularly useful for comparing
Monte Carlo simulations and mean field theory outside the critical region where composition fluctuations are dominant. A direct test of Eq.(\ref{eqn:mu}) at high temperatures
is presented in Fig.\ref{fig:eos}. The Monte Carlo data are well describable by the SG--EOS.  The inset shows the measurement of the $\chi$ parameter via 
the inverse collective structure factor (cf.\ Eq.(\ref{eqn:sans})). It also yields an estimate for the effective coordination number $z_c$ which is consistent with the value extracted 
from the intermolecular pair correlation function (at this temperature).

The influence of composition fluctuations on the $\chi$ parameter becomes even more pronounced when we approach the critical 
temperature. For the chain length accessible in the simulations the region where 3D Ising critical behavior dominates is rather extended. In this region
the mean field theory underestimates the strength of fluctuations dramatically. Using the mean field expression (\ref{eqn:sans}) to extract an effective
Flory Huggins parameter in the 3D Ising critical regime leads to a strong composition dependence of the $\chi_{\rm SANS}$ parameter. To an excellent approximation
the left hand side of Eq.(\ref{eqn:sans}) is much smaller in the 3D Ising critical region than expected by mean field theory and we estimate
$\chi_{\rm SANS} \approx 1/2N_A\phi+1/2N_B(1-\phi)=1/2N\phi(1-\phi)$ for symmetric mixtures\cite{M0}. This corresponds to a pronounced upward parabolic 
composition dependence of the $\chi$ parameter which is in the vicinity of the critical temperature much stronger than the effect of the composition
dependence of the segmental entropies. This observation is consistent with simulations\cite{HPD}.

For much larger asymmetries ($N_B \gg N_A$) more pronounced effects might occur. This limit corresponds to the phase separation between a polymer and a solvent. 
In this case the Ginzburg criterion\cite{GINZBURG} is dominated by the short chain length $N_A$ and the conformations of the long chain depend on composition (cf.\ sec.\ V).

Another common asymmetry in polymer blends are differences in the statistical segment length\cite{BATES}. This effect has attracted much attention recently
because of novel synthesis techniques for saturated hydrocarbon with a controlled degree of branching and their practical applications\cite{GRAESSLEY}. Bates\cite{BATES} suggested
that the degree of branching can be represented on a coarse grained scale by a difference in statistical segment lengths. Graessley and co-workers\cite{GRAESSLEY} have
studied systematically the miscibility behavior of this class of polymers. Many - though not all blends -- were describable in terms of Hildebrands solubility\cite{HILDEBRAND}
parameters. This suggests that the incompatibility is chiefly determined by enthalpic effects.

\begin{figure}[tbhp]
\begin{minipage}[t]{80mm}%
  \mbox{
       \setlength{\epsfxsize}{10cm}
       \epsffile{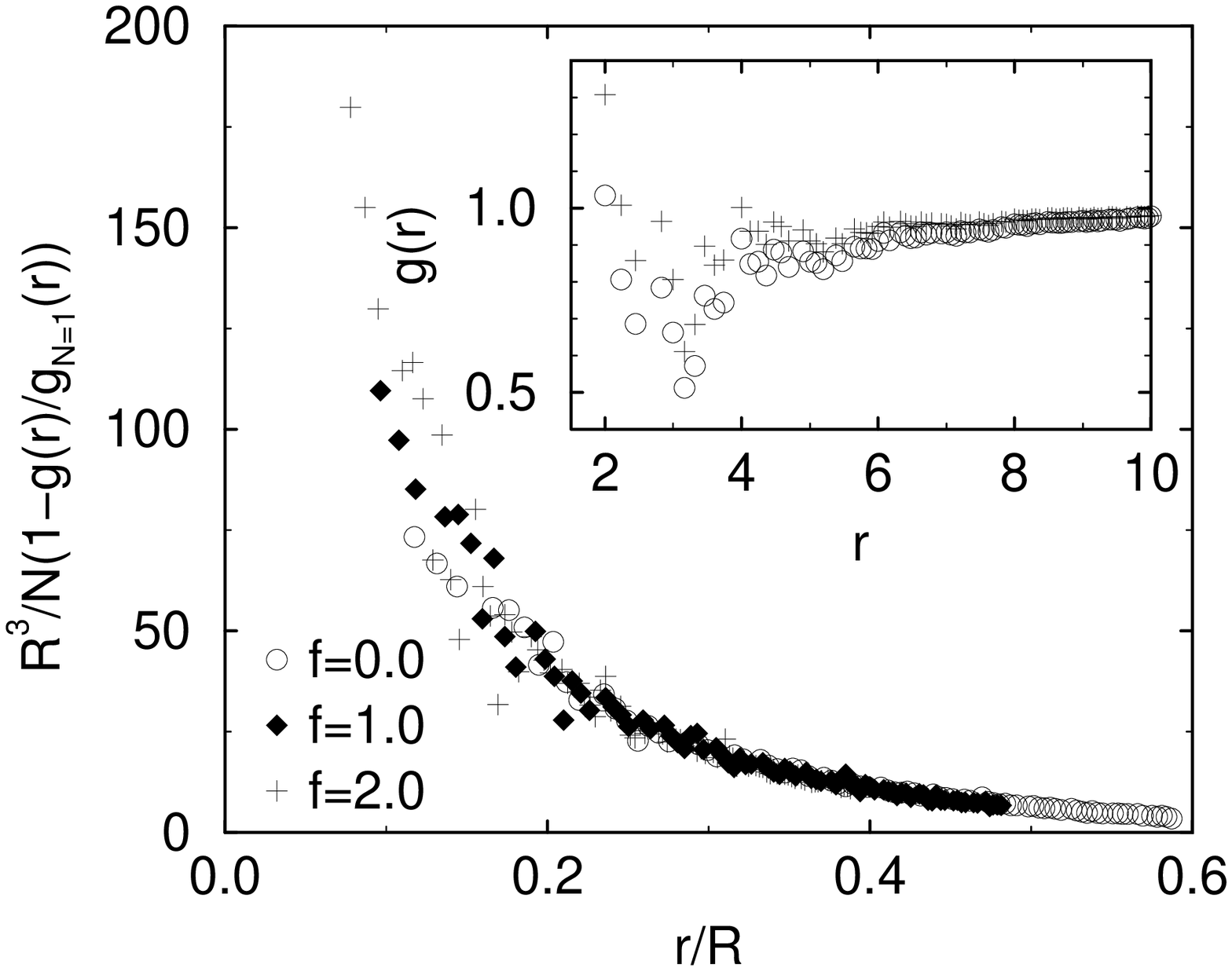}
       }
\end{minipage}%
  \hfill%
\begin{minipage}[c]{54mm}%
\caption{\label{fig:cstiff}}
Scaling of the intermolecular pair correlation function for semi-flexible linear chains at chain length $N=32$ and stiffnesses $f=0.0$ 
(i.e., flexible),$1$, and $2$. The inset compares the unscaled data for flexible chains and semi-flexible ($f=2$) ones.
From ref.\cite{STIFF2}.
\end{minipage}%
\end{figure}

In the framework of the bond fluctuation model the effect of stiffness can be incorporated via an intramolecular bond angle potential of the form\cite{JOJO,STIFF1,STIFF2}:
\begin{equation}
E(\theta) = f k_BT \cos (\theta)
\end{equation}
where $\theta$ denotes the complementary angle to two successive bonds. Increasing the stiffness parameter $f$ we energetically favor straight bond angles
and increase the spatial extension of the molecule. The more open the molecule, the larger the number of intermolecular contacts. Upon increasing the
stiffness parameter $f$ from $0$ (flexible chains) to $2$ (semi-flexible), the chain extension for a $32$ segment polymer increases about a factor 1.5 and the number
of intermolecular contacts $z_c$ increases from 2.65 to 3.29 at $\epsilon=0.05$ because the bond stiffness makes a folding back of the chain less probable\cite{STIFF1,STIFF2}. 
This is investigated in detail in Fig.\ref{fig:cstiff}. The inset presents the intermolecular 
pair correlation function for flexible chains $f=0$ and semi-flexible ones $f=2$. The increase of the intermolecular pair correlation function at small distances
results in an increase of the enthalpic part of the $\chi$ parameter according to Eq.(\ref{eqn:chis}). 

Unlike the situation for flexible chains, however, the decoupling of the local packing structure of the monomers and the chain connectivity breaks down for 
semi-flexible chains. A similar analysis as in Fig.\ref{fig:chole}({\bf b}), which aims at separating the local packing effects from the correlation hole, is presented in 
Fig.\ref{fig:cstiff}. We plot the ratio of the intermolecular paircorrelation function and the correlation function of the monomer fluid for flexible ($f=0$) and
semi-flexible ($f=1,2$) chains of lengths $N=32$.  We observe a data collapse only at large distances,
whereas there are systematic deviations at short distances. This breaking down of the simple scaling (\ref{eqn:ch}) can be interpreted in two ways: On the 
one hand the chain structure is characterized by two length scales -- the radius of gyration and the statistical segment length. Hence, the function $f(r)$ 
not only depends on the ratio $r/R_g$ but also on $r/b(f)$, where $b(f)$ denotes the stiffness dependent statistical segment length. On the other hand,
the bond angle potential between neighboring monomers influences the packing structure of the polymeric liquid, which for semi-flexible chains differs 
from the packing of the monomer fluid. The interplay between the packing arrangement of the monomers and the local conformations favored by the bond angle 
potential gives rise to an entropic contribution to the Flory Huggins parameter.

\begin{figure}[tbhp]
\begin{minipage}[t]{80mm}%
  \mbox{
       \setlength{\epsfxsize}{10cm}
       \epsffile{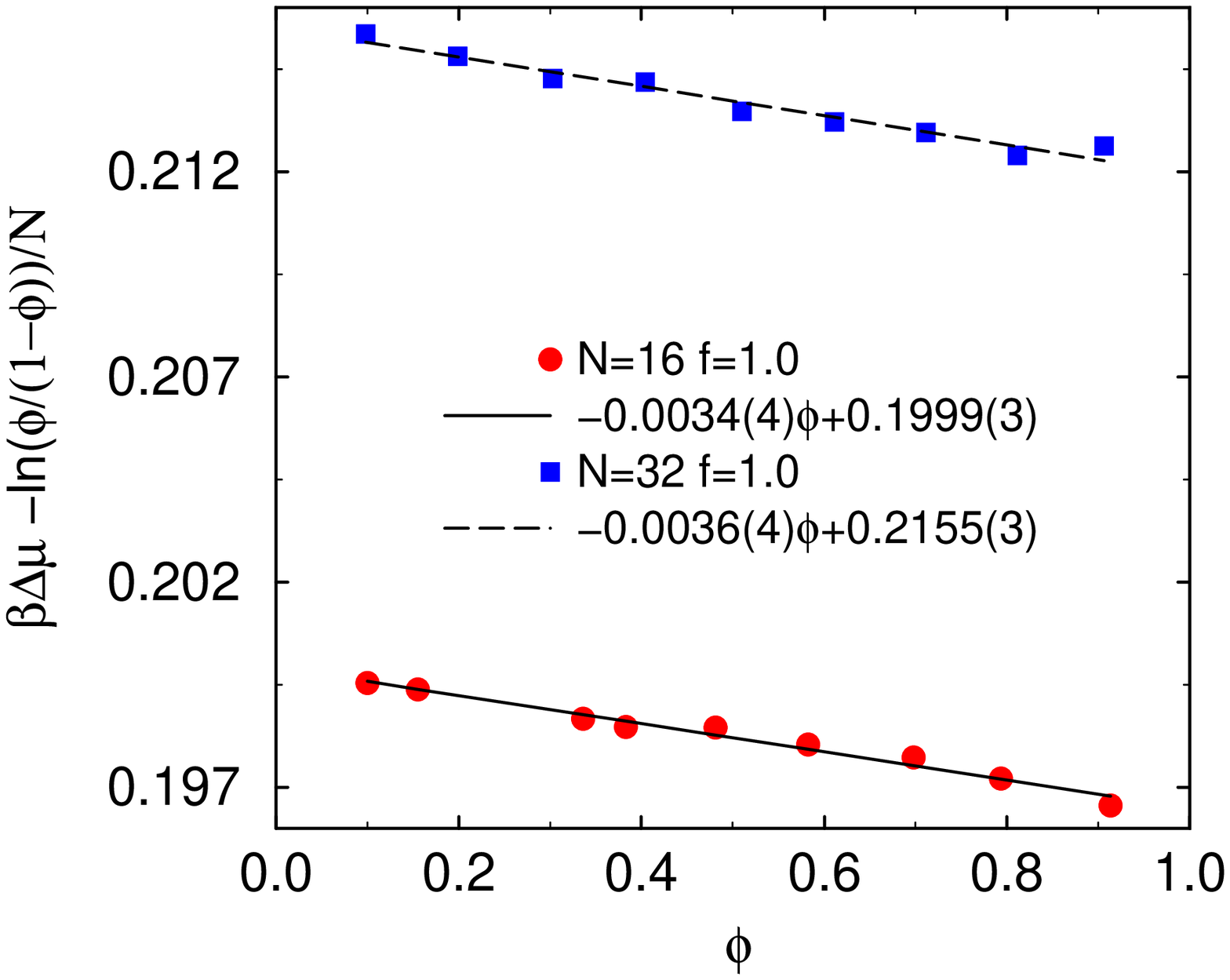}
       }
\end{minipage}%
  \hfill%
\begin{minipage}[c]{54mm}%
\caption{\label{fig:chis}}
Deviations from the semi-grandcanonical equation of state (SG-EOS) for blends of polymers with different stiffness.
The (negative) slope is proportional to the entropic contribution to the $\chi$ parameter.
From ref.\cite{STIFF1}.
\end{minipage}%
\end{figure}

To explore the possibility of entropic contributions to the $\chi$ parameter\cite{STIFF1}, we accurately measure the dependence of the chemical exchange potential $\Delta \mu$
on the composition of the mixture. If there were no entropic contributions to the $\chi$ parameter, only the (exactly known) translational entropy would determine the
SG--EOS. In Fig.\ref{fig:chis} we present the deviations $\beta \Delta \mu - 1/N \ln[\phi/(1-\phi)]$ from the 
ideal mixing behavior. An entropic contribution to $\chi$ is related to a composition dependence in the form $-\chi(2\phi-1)$ according to Eq.(\ref{eqn:mu}). Indeed, the
Monte Carlo results do reveal a composition dependence of this form and we accurately extract a small positive entropic contribution $\Delta \chi$ to the Flory Huggins parameter\cite{STIFF1}.
$\Delta \chi = 0.0017(2)$ for $N=16$ and $f=1$, $\Delta \chi = 0.0018(2)$ for $N=32$ and $f=1$, and $\Delta \chi = 0.0031(3)$ for $N=16$ and $f=1.5$.
For the parameters investigated, the entropic contribution $\Delta \chi$ increases with stiffness disparity in the blend and is independent of chain length.
For the chain lengths considered, it is only a few per cent of the critical value $\chi_c = 2/N$ and we anticipate only a small increase in the critical 
temperature. If the chain length independence of the entropic contributions remains true in the long chain length limit, the data suggest that for long macromolecules 
($N \approx {\cal O}(1000)$) the stiffness disparity alone will be sufficient to cause phase separation or LCST behavior\cite{STIFF1}.
The positive entropic contribution to the $\chi$ parameter is consistent with field theoretical studies by Liu, Fredrickson and Bates\cite{LIU}, recent P-RISM calculations by 
Singh and Schweizer\cite{SCHWEIZER3} and Lattice Cluster Theory by Foreman and Freed\cite{FREEDS}.

Unfortunately, the semi-grandcanonical Monte Carlo moves rapidly become less efficient as the chain length or the stiffness disparity is increased, because the
typical conformations of the two species differ strongly. For the chain lengths accessible in the simulations the shift in the critical temperature is small and not chiefly
determined by the entropic contribution to the $\chi$ parameter. Most notably, the stiffness increases the size of the molecules and the number of intermolecular
contacts. Hence, the enthalpic contribution to the $\chi$ parameter increases as well. The increase of the enthalpic and entropic contributions are both of the
order of a few percent. We expect both effects to persist in the long chain length limit. 

For the short chain length considered in the simulation the mean
field theory overestimates the critical temperature by about $20\%$. According to the Ginzburg criterion the deviation between the critical temperature and the mean 
field estimate decreases with the statistical segment lengths (cf.\ Eq.(\ref{eqn:ginz})). We expect composition fluctuations to shift the critical temperature down
the less the higher stiffness. Even if the $\chi$ parameter remained unaltered, the critical temperature in the Monte Carlo simulations (for short chains)
would increase.

\begin{figure}[tbhp]
\begin{minipage}[t]{80mm}%
  \mbox{
       \setlength{\epsfxsize}{10cm}
       \epsffile{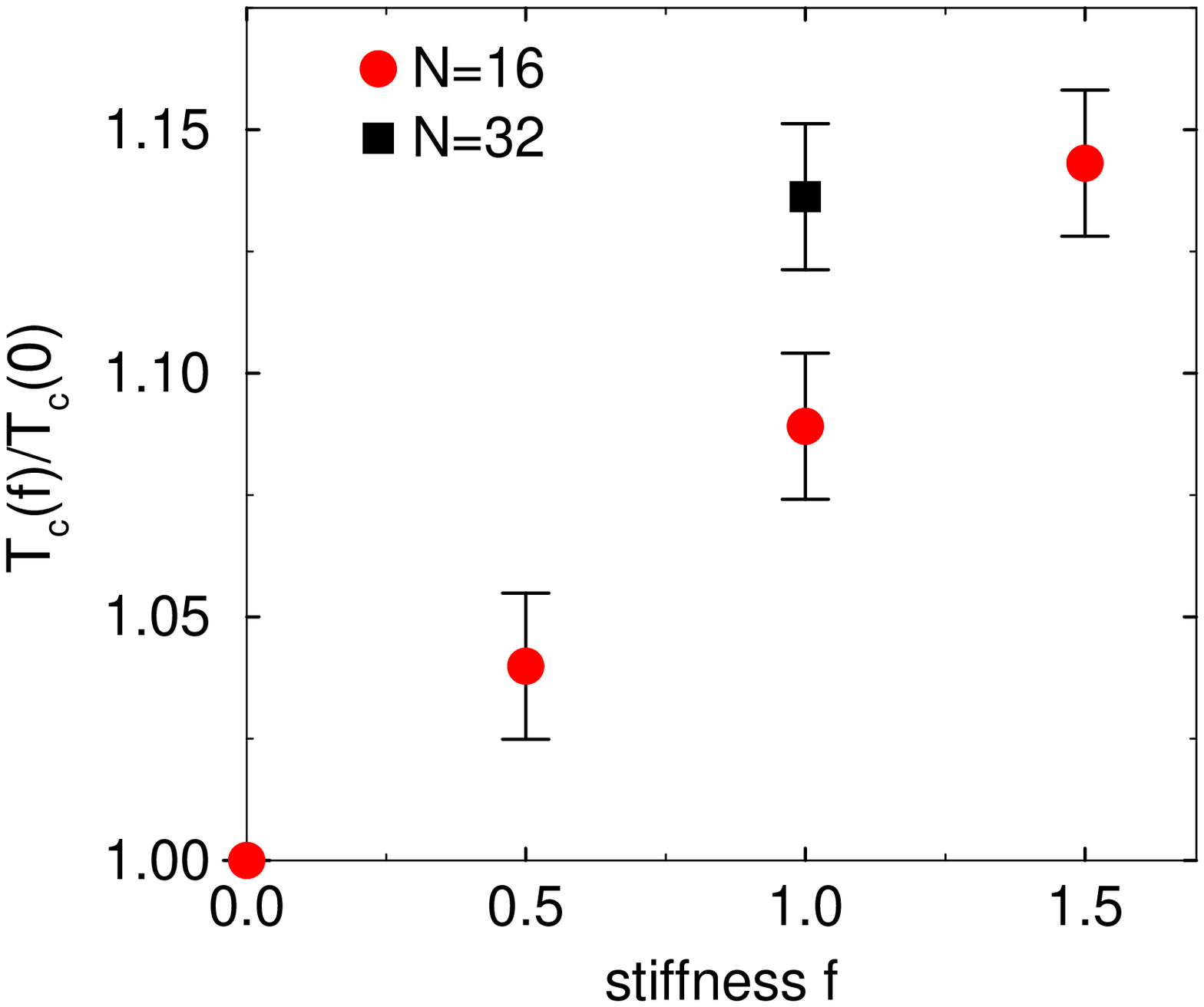}
       }
\end{minipage}%
  \hfill%
\begin{minipage}[c]{54mm}%
\caption{\label{fig:tcs}}
Dependence of the critical temperature on the stiffness $f$ for chain length $N=16$ and $32$.
From ref.\cite{STIFF1}.
\end{minipage}%
\end{figure}

The measured shift of the critical temperature is presented in Fig.\ref{fig:tcs}. Upon increasing the stiffness or the chain length the ratio of critical temperatures
between the strictly symmetric blend and the blend with stiffness disparity increases. For $N=16$ and $f=1$ we obtain a relative shift of $9\%$ for $T_c$.
The entropic contribution in the athermal system amounts to $2 \Delta \chi/N = 0.014$ while the relative increase of the effective coordination number is $6\%$
as measured in the Monte Carlo simulations. The remaining deviation is consistent with the dependence of the ratio $T_c^{\rm MF}/T_c$  on the persistence length.

\begin{figure}[htbp]
\begin{minipage}[t]{102mm}%
\setlength{\epsfxsize}{10cm}
\mbox{\epsffile{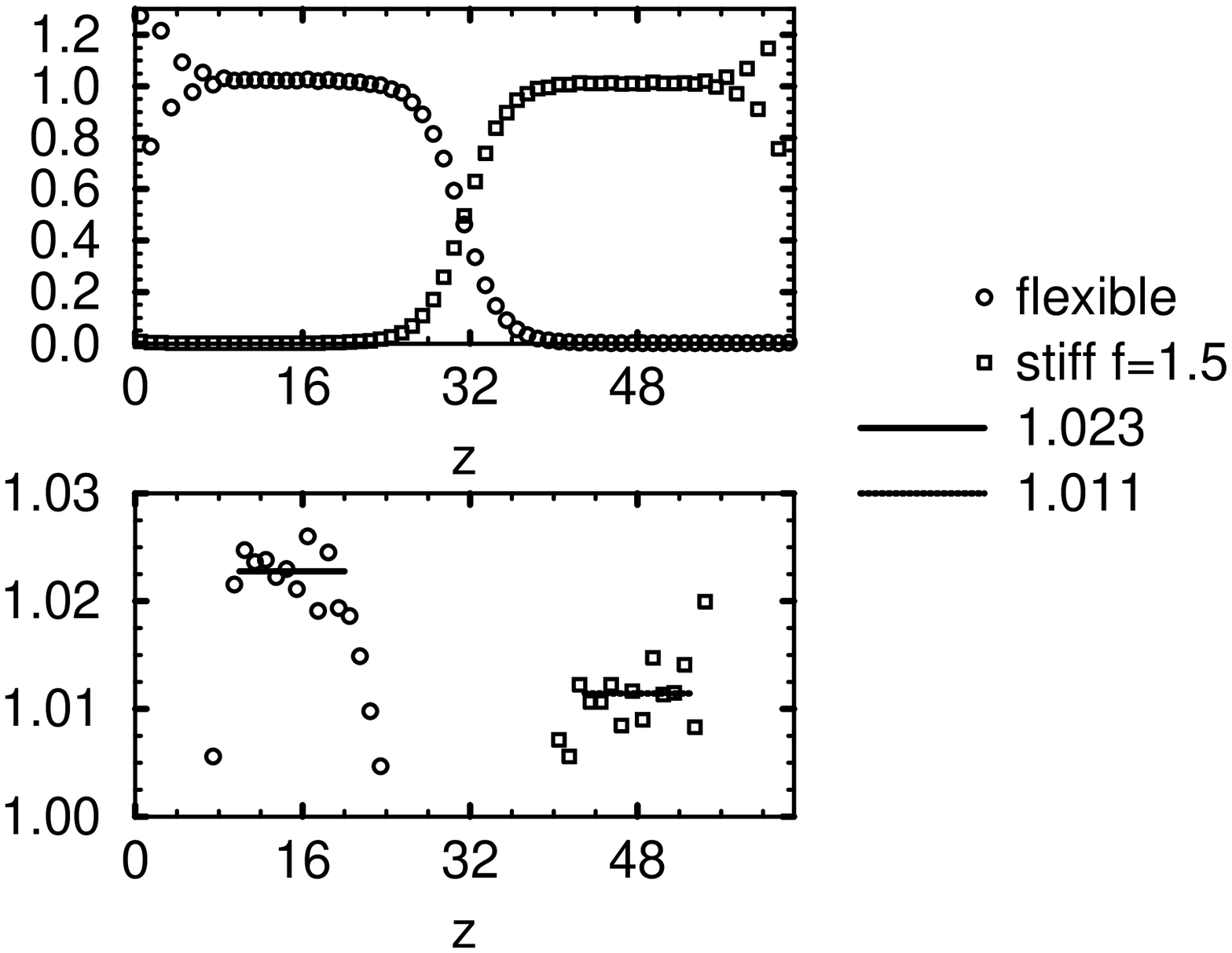}}
\end{minipage}%
\hfill%
\begin{minipage}[c]{54mm}%
\caption{\label{fig:pressure}}
         Interfacial profile of the normalized monomer density in a blend of stiff ($f=1.5$ and flexible polymers
         well below criticality $T=0.346 T_c$. The simulation box $32 \times 32 \times 64$ has hard walls
         in z direction. The enlargement shows, that vacancies are enriched in the stiffer phase (right hand side).
	 From ref.\cite{STIFF1}.
\end{minipage}%
\end{figure}

Below the critical point the blend phase separates into regions in which the flexible or the stiff component are enriched. The two phases have different
osmotic pressures at equal monomer density -- the stiffer component has the higher osmotic pressure. If the two phases coexist in the simulation box the vacancies 
(or the solvent) redistribute as to restore equal osmotic pressure. At equal pressure, the density of the phase rich in the flexible component is higher than
the density of the phase in which the stiffer polymers are concentrated. The Fig.\ref{fig:pressure} displays density profiles across an interface between the coexisting 
phases well below the critical temperature. One observes that the density difference is about $1\%$.

A similar study on the consequences of stiffness disparity was pursued in an off-lattice model by Weinhold et al.\cite{WEINHOLD}. Using the increment chemical potential method\cite{INC} 
the authors explored the miscibility behavior.  Upon blending the chemical potential of the flexible chains increases, while the stiffer component lowers its free energy. This effect was 
rationalized via equation of state effects: At constant density the blend has a higher osmotic pressure than the pure flexible component and a lower osmotic pressure than the stiff component.
The authors stated that the behavior is in almost quantitative agreement\cite{WEINHOLD} to the simulations in the bond fluctuation model. Again this indicates that the lattice structure in the 
bond fluctuation model is a good approximation for the continuum space properties. The net excess free energy per monomer obtained from the simulations is essentially zero to within the 
$\pm 0.005 k_BT$ statistical error\cite{WEINHOLD}. This is consistent with the small values $\Delta \chi$ found in our simulations.

They also illustrated the significance of the intermolecular pair correlation function for the $\chi$ parameter. Therefore they choose the interaction parameters in
Eq.(\ref{eqn:org}) such that the original Flory Huggins expression for the $\chi$ parameter, which does not incorporate the fluid structure, vanishes. However, since 
the intermolecular pair correlation functions of stiff and flexible blends differ, there is an enthalpic $\chi$ parameter according to Eq.(\ref{eqn:chis}) and phase 
separation is observed\cite{KW}. 

The effect of pressure on the miscibility behavior has attracted much interest. There is an interesting interplay between equation of state effects and 
the phase behavior. This has also a practical importance when a blend is mixed in an extruder or during injection molding of plastics\cite{SUPER}.
In the framework of the bond fluctuation model the local fluid
structure is mainly determined by athermal packing effects. In the athermal melt, the osmotic pressure is independent of the chain length\cite{MCP}.
This chain length independence of the pressure at high densities is a universal property of polymer melts\cite{DG}.
The excess free energy change upon mixing per monomer is of the order $\chi \sim 1/N$. The free energy cost for a density fluctuation (on the length scale of a monomer)
is proportional to the compressibility and, hence, chain length independent. Therefore, in the one phase region, the interactions lead only to a small
excess volume change upon mixing for high molecular weight (compatible) polymer blends. These kind of compressibility effects have been observed, e.g., at polymer--polymer interfaces, 
where the energetic unfavorable interactions at the interface result in a slight decrease of the density. However, even for rather strongly segregated blends\cite{MBO} 
($\chi N < 20$) in the bond fluctuation model the effect yields only a density reduction by a few percent. This is also in agreement with the decoupling
of composition and density fluctuations. The insignificance of compressibility effects in weakly interacting blends and the consequences for the
analysis of neutron scattering data has been explored in ref.\cite{KP,B}.Gromov and de Pablo\cite{SMILE2} found in Monte Carlo simulations of a model of 
Lennard-Jones chains of chain length $N=16$ a volume change of approximately $10\%$ at constant pressure.

For our specific choice of interactions the total energy density per monomer is also of the order $\chi$. Hence, the fluid structure
corresponds to that of an athermal melt in the limit of long chain lengths and $\chi N =$const. This observation is in accord with the
temperature independence of the packing and the effective coordination number in the temperature range where the phase separation occurs.
If we were to simulate at constant pressure, the density around the phase transition would be chiefly determined by the value of the athermal
system in the limit of long chains. Therefore, we expect not much change in the chain length dependence of the miscibility at constant
volume or at constant density.

However, we would like to emphasize that the approximate decoupling between  the fluid structure/density and the temperature (at constant pressure) is not 
a universal property and does depend on the specific choice of the interactions.
Unlike the excess free energy change upon mixing, the total energy density per monomer needs not to be small for long polymers at $\chi N=$const.
In many experimental instances, concentrated polymer solutions and melts exhibit a temperature dependent equation of state. Therefore
the density $\rho$ is a function of temperature $T$ at a given fixed pressure. If the $\chi$ parameter is mainly enthalpic,
we still can use Eq.(\ref{eqn:chis}) to calculate the $\chi$ parameter. In a very crude approximation the intermolecular
pair correlation function is independent of the density, and the incompatibility $\chi$ is proportional to the density $\rho(T,p)$.
The critical temperature of the blend at constant pressure scales like $T_c \sim N/\rho(T_c,p)$. This leads to a weaker dependence of the
critical temperature with chain length. Such a dependence has been observed in various experiments. In a recent manuscript  Escobedo and
de Pablo\cite{SMILE} have investigated the scaling of the critical temperature in a symmetric blend under constant pressure. They found that the critical
temperature increases approximately like $\sqrt{N}$ for the range of chain length investigated. Moreover, recent experiments on polyolefin blends
by Lohse and co-workers find evidence for a temperature-pressure superposition\cite{SUPER}: far from the UCST the pressure and density dependence of interaction energies
is only a function of the density. However, deviations from this scaling are found for blends which demix upon heating. This might indicate an
additional dependence of the local packing arrangements with the density. 

If the polymer concentration becomes small -- e.g., a binary polymer blend in a common solvent -- very interesting deviations from the Flory-Huggins theory can be expected.
This has been worked out in detail by Leibler\cite{LEIBLER} and Sch{\"a}fer\cite{SCHAEFER}. Due to the rather small interpenetration of coils in semi-dilute solutions
the Flory Huggins parameter depends strongly on the density $\chi \sim \rho^{(1+x)/(3\nu-1)}$, where $x=0.22$ denotes the correction to scaling exponent.

\section{Single chain conformations.} 
A crucial assumption ({\bf A}) in the derivation of the Flory Huggins theory is the ideality of chain conformations: the conformations of the chains are independent 
from the composition of the mixture and the conformational entropy does not contribute to the excess entropy of mixing. The latter is only determined by the
translational term which is of the order $1/N$ on the segmental scale. The above examples reveal that there might be significant conformational entropy contributions.
Hence, it is important to explore the dependence of the chain conformations on the composition of the mixture. In the following section we shall illustrate the
dependence of the chain conformations on the composition for symmetric and asymmetric blends.

\begin{figure}[tbhp]
\begin{minipage}[t]{80mm}%
  \mbox{
       \setlength{\epsfxsize}{10cm}
       \epsffile{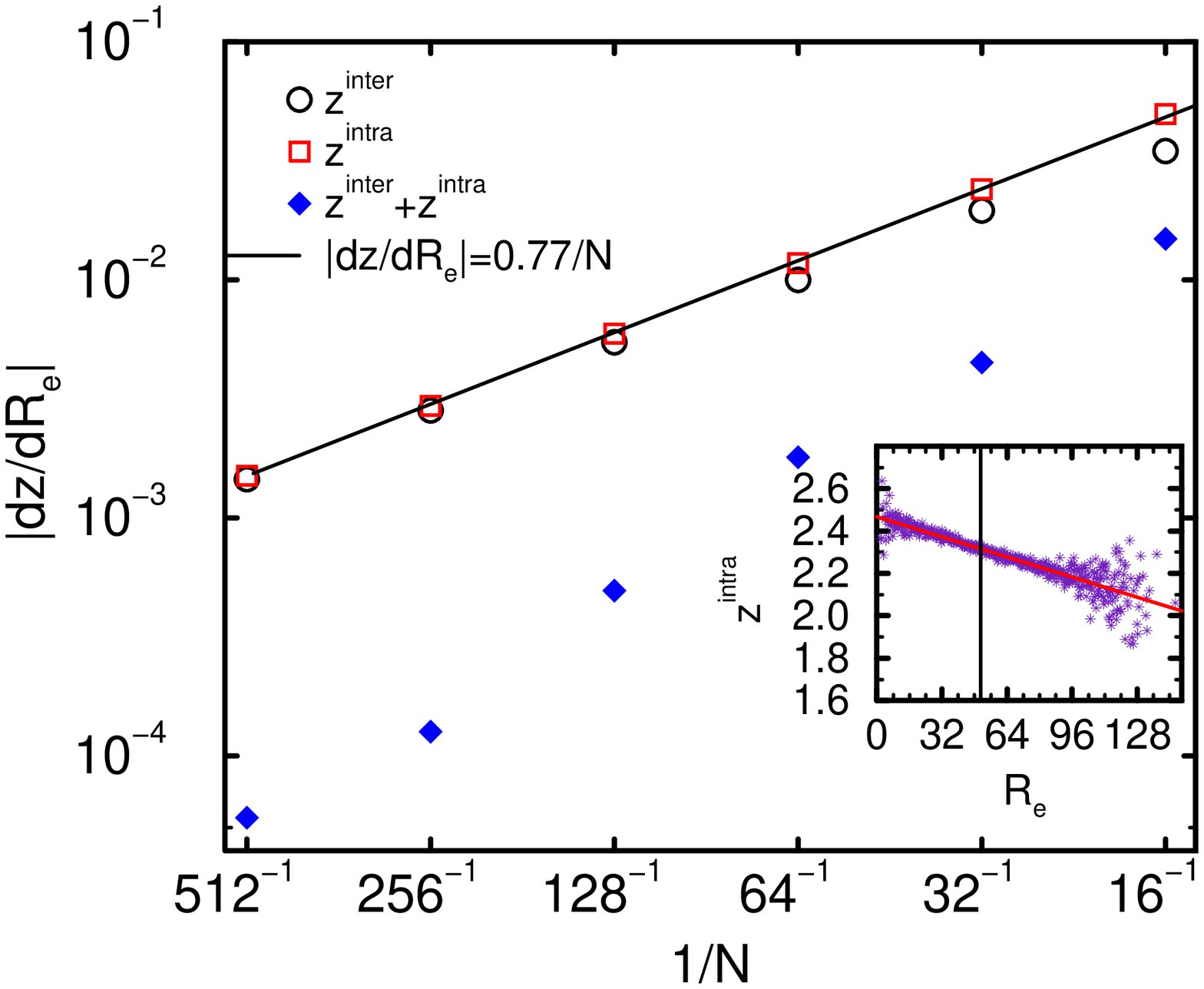}
       }
\end{minipage}%
  \hfill%
\begin{minipage}[c]{54mm}%
\caption{\label{fig:zvsr}}
Correlation between the chain extension and the number of intermolecular and intramolecular contacts.
The straight line marks the prediction of the scaling considerations ${\rm d}z/{\rm d}R \sim 1/N$.
From ref.\cite{CLUSTER}
\end{minipage}%
\end{figure}

For strictly symmetric additive blends Monte Carlo simulations by Sariban and Binder\cite{SARIBAN} reveal a shrinking of the chain extension in the minority
phase. This has been attributed to an exchange of energetically unfavorable intermolecular contacts and favorable intramolecular contacts upon shrinking\cite{PORE}.
The scaling considerations in the introduction show that the relative decrease in the chain extension at fixed $\chi N$, i.e., at fixed concentration of
the minority component at coexistence, decreases like $\chi N/\sqrt{N}$. The derivation of the scaling arguments relies on the number of
intramolecular contacts and its dependence of the chain extension. Using simple scaling arguments similar to those for the collapse of a chain at the
$\Theta$ point we find that $z^{\rm intra}$ scales like $N/\rho R_0^3$, where $R_0$ denotes the unperturbed chain extension. Obviously, this estimate 
excludes contributions from the neighbors along the polymer, which give rise to $z^{\rm intra} \sim N$. Though the neighbor contribution is important for 
the scaling of the number of intramolecular contacts with chain length, we assume them to be independent from the instantaneous shape/extension of the polymer\cite{COMMENT}.
Using the scaling estimate above, the dependence of the $z^{\rm intra}$ on the instantaneous extension $R$ at fixed chain length is given by:
${\rm d}z^{\rm intra}/{\rm d}R \sim N/\rho R_0^4 \sim 1/N$. Clearly, a detailed verification of this scaling behavior by Monte Carlo simulations is warranted. 
Such a test is presented in
Fig.\ref{fig:zvsr}. The inset presents the average number of intramolecular contacts at fixed end-to-end distance for an athermal melt of chain length $N=256$. 
At the mean end-to-end distance $\sqrt{\langle R^2 \rangle_0}$ we determine the slope ${\rm d}z/{\rm d}R$ as indicated by linear regression.
The chain length dependence of the derivatives of the number of inter- and intramolecular contacts with respect to the chain extension decreases like
$1/N$ for large chain lengths. This confirms the scaling predictions. Moreover, the sum of intermolecular and intramolecular contacts depends much weaker on the
spatial extension $R_e$ and the dependence decreases faster than $1/N$. This indicates that the fluid structure of the monomers is mainly determined by packing
and approximately decouples from the chain conformations. For long chains, the conformational changes merely result in an exchange of inter- and intramolecular contacts.
From the Monte Carlo data we estimate ${\rm d z}/{\rm d R} = 0.77(7)/N$ for the bond fluctuation model at density $\rho=1/16$.

The scaling predictions can be made more quantitatively in the framework of the Gaussian chain model. Let $P({\bf R})$ denote the probability distribution 
of the end-to-end vector ${\bf R}$ which incorporates the dependence of the single chain energy $E$ on the chain extension.
\begin{equation}
E({\bf R}) \approx E(\sqrt{\langle R_e^2 \rangle_0}) + \frac{{\rm d}E}{{\rm d}R} \left[|{\bf R}|-\sqrt{\langle R_e^2 \rangle_0}\right]
\end{equation}
where $\langle R_e^2 \rangle_0 = b^2 N$ denotes the end-to-end distance in the athermal limit.
Assuming Gaussian statistics for the unperturbed chain, we can write the probability distribution for a $B$ polymer in the form
\begin{equation}
P({\bf R}) \sim \exp\left( -\frac{3{\bf R}^2}{2\langle R_e^2\rangle_0} - \frac{1}{k_BT}\frac{{\rm d}E}{{\rm d}R} \left[|{\bf R}|-\sqrt{\langle R_e^2 \rangle_0}\right] \right)
\end{equation}
The total energy change associated with the transfer of two intermolecular contacts of a $B$ polymer $[\epsilon(2\langle\phi\rangle-1)]$ into an intermolecular contact
$[-\epsilon]$ and a contact between monomers not belonging to this $B$ polymer $[-\epsilon(2\langle\phi\rangle-1)^2]$ amounts to 
$\Delta E = 4\epsilon \langle \phi \rangle^2$. The number of intramolecular contacts per monomer increases by 2, and the number of 
intermolecular contacts decreases by the same amount. Therefore ${\rm d}E/{\rm d}R$ equals $2\epsilon\langle \phi \rangle^2{\rm d}z/{\rm d}R$.
For symmetric blends we identify the $\chi$ parameter according to $\chi = 2 z_c \epsilon/K_BT$.
Using this estimate and assuming that the conformational changes are small, we calculate the mean square end-to-end distance:
\begin{equation}
\langle R_e^2  \rangle \approx \langle R_e^2  \rangle_0 \left( 1 - \sqrt{\frac{8}{27\pi}} \frac{b}{z_c\sqrt{N}} \left[ \frac{{\rm d}z}{{\rm d}R}N\right]
\chi N \langle \phi \rangle^2 \right)
\label{eqn:shrink}
\end{equation}
We expect this asymptotic expression to hold only for very small values of $\chi \sqrt{N}$, where the conformational changes can be treated perturbatively.
This expression predicts a quadratic dependence of the chain extension on the composition of the mixture. The effect increases linearly with the
$\chi$ parameter, but decreases at fixed $\chi N$ like $1/\sqrt{N}$. Hence, the conformations of long macromolecules are only very weakly dependent on
composition. This justifies assumption ({\bf A}) in the derivation of the Flory Huggins theory. This is also in qualitative agreement with field theoretical calculations
of Vilgis and Holyst\cite{VILGIS}, and Garas and Kosmas\cite{KOSMAS}.

A more detailed self-consistent field (SCF) theory for single chain conformations in dense blends has been developed in ref.\cite{CLUSTER}. In the framework of the SCF calculations 
the statistical mechanics of a cluster of $n_c$ neighboring polymers is solved. Interactions inside the cluster are treated exactly, whereas interactions 
between the $n_c$ members of the cluster and polymers outside the cluster are treated in a mean field approximation. The technique
takes due account of the coupling between single chain conformations and the energy. Composition fluctuations on the length scale of the cluster
are partially incorporated. If the cluster consists of a single thread-like chain, the theory recovers the Flory Huggins predictions. If the cluster
becomes very large, the interactions with polymers outside the cluster which are treated approximatively become mere surface effects.
In the calculations, however, the cluster size $n_c \leq 14$ remains small. The results of this approach have been compared to Monte Carlo data
in the framework of the bond fluctuation model. In the long chain length limit, these SCF calculations agree with the simple estimates presented above.

\begin{figure}[tbhp]
\begin{minipage}[t]{102mm}%
\mbox{
\setlength{\epsfxsize}{9cm}
\epsffile{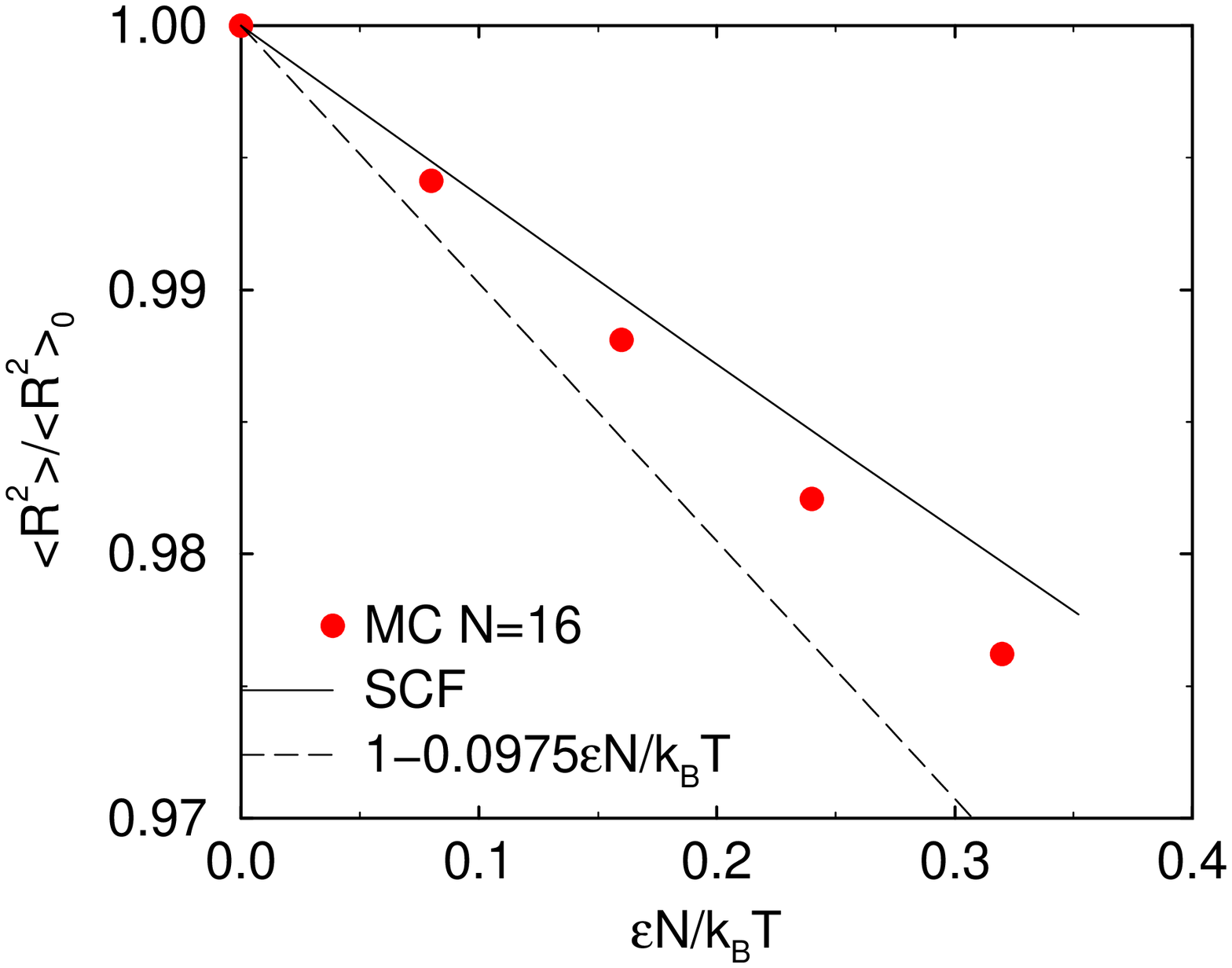}
}
\mbox{
\setlength{\epsfxsize}{9cm}
\epsffile{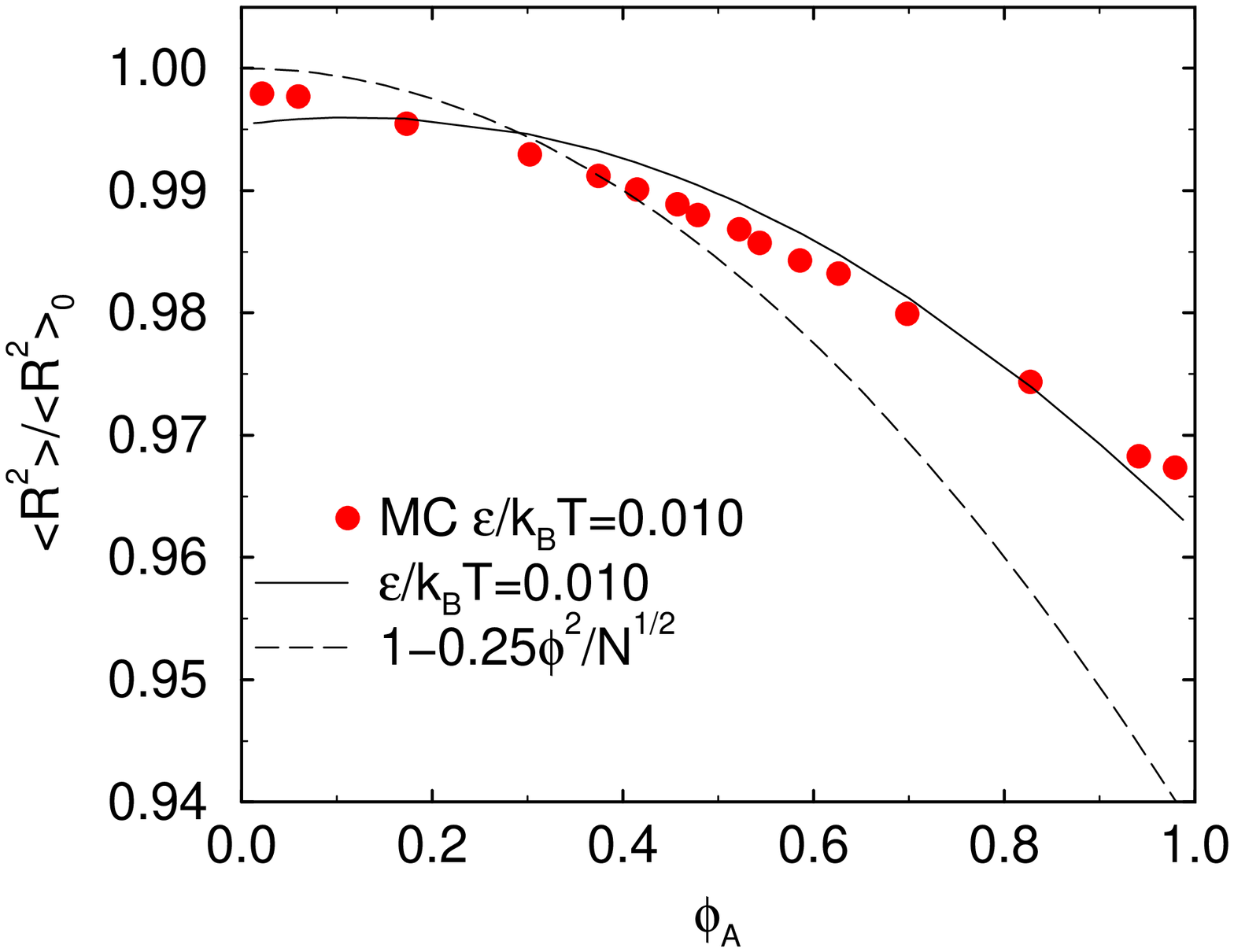}
}
\end{minipage}%
\hfill%
\begin{minipage}[c]{54mm}%
\caption{\label{fig:ret}}
         ({\bf a}) Temperature and chain length dependence of the end-to-end distance at $\bar{\phi}_A=0.5$. The lines present the results of the
         SCF calculations ($n_c=14$) for chain length $N=16,32,64,128,256$ and $512$ (from bottom to top), while the symbols display
         the results of a Monte Carlo simulation for chain length $N=16$.
	 ({\bf b}) Composition dependence of the chain extension for chain length $N=16$ above the critical point.  From ref.\cite{CLUSTER}.
\end{minipage}%
\end{figure}

Fig.\ref{fig:ret} ({\bf a}) presents the temperature dependence of the chain extension in a strictly symmetric blend for chain length $N=16$ at $\langle \phi \rangle =1/2$. Upon decreasing the temperature
close to the critical temperature, the chains shrink by about $3\%$. The asymptotic expression overestimates the effect slightly, while the full SCF calculations
yield an improved estimate. The composition dependence of the chain conformations in the one phase region is investigated in Fig.\ref{fig:ret}({\bf b}). The Monte Carlo
simulations are well describable by an parabolic dependence and the SCF calculations agree almost quantitative with the results of the Monte Carlo simulations
without any adjustable parameter. The simple estimates gives a qualitatively reasonable description, but quantitatively overestimates the degree of shrinking.

\begin{figure}[tbhp]
\begin{minipage}[t]{80mm}%
\mbox{
\setlength{\epsfxsize}{9cm}
\epsffile{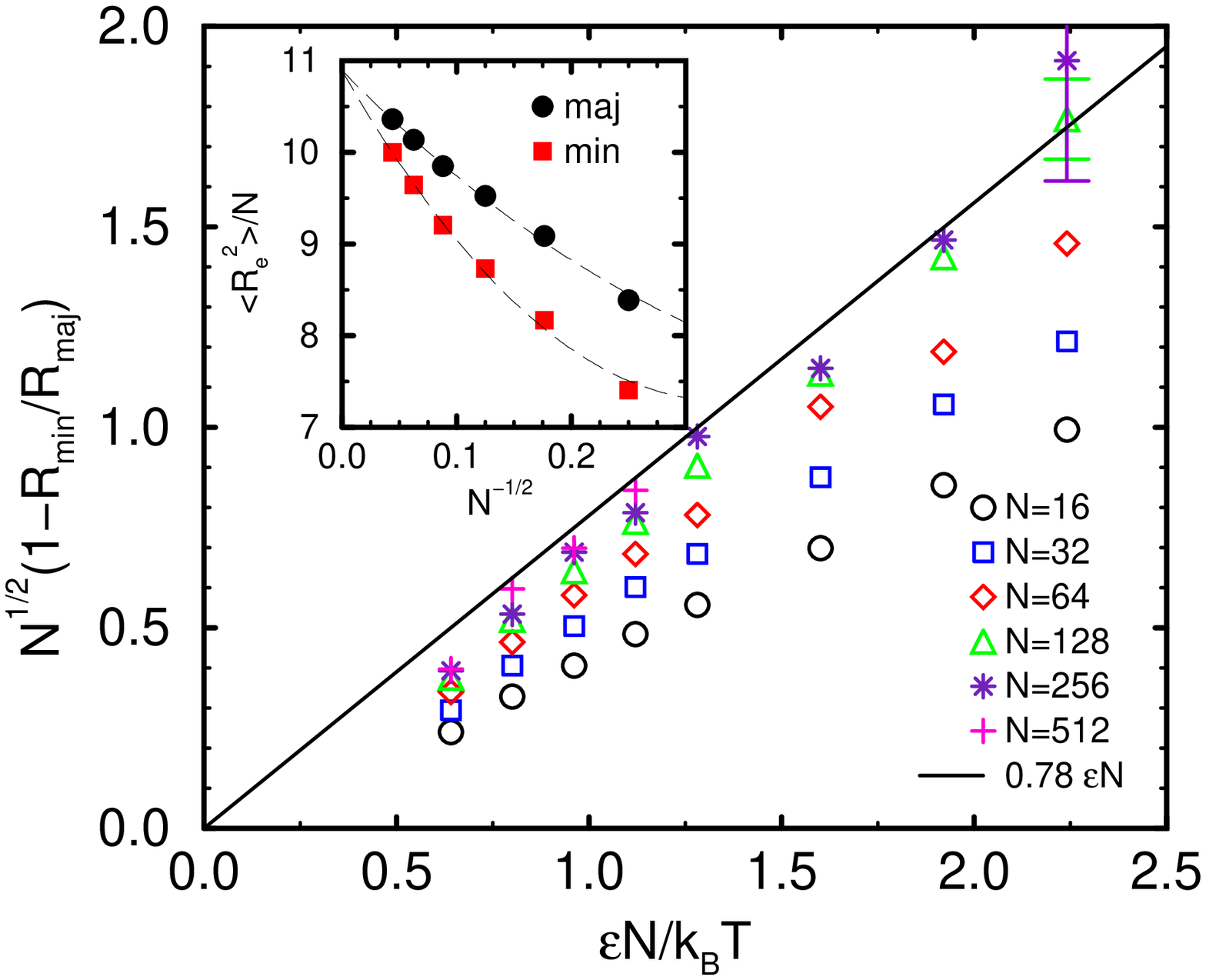}
}
\mbox{
\setlength{\epsfxsize}{9cm}
\epsffile{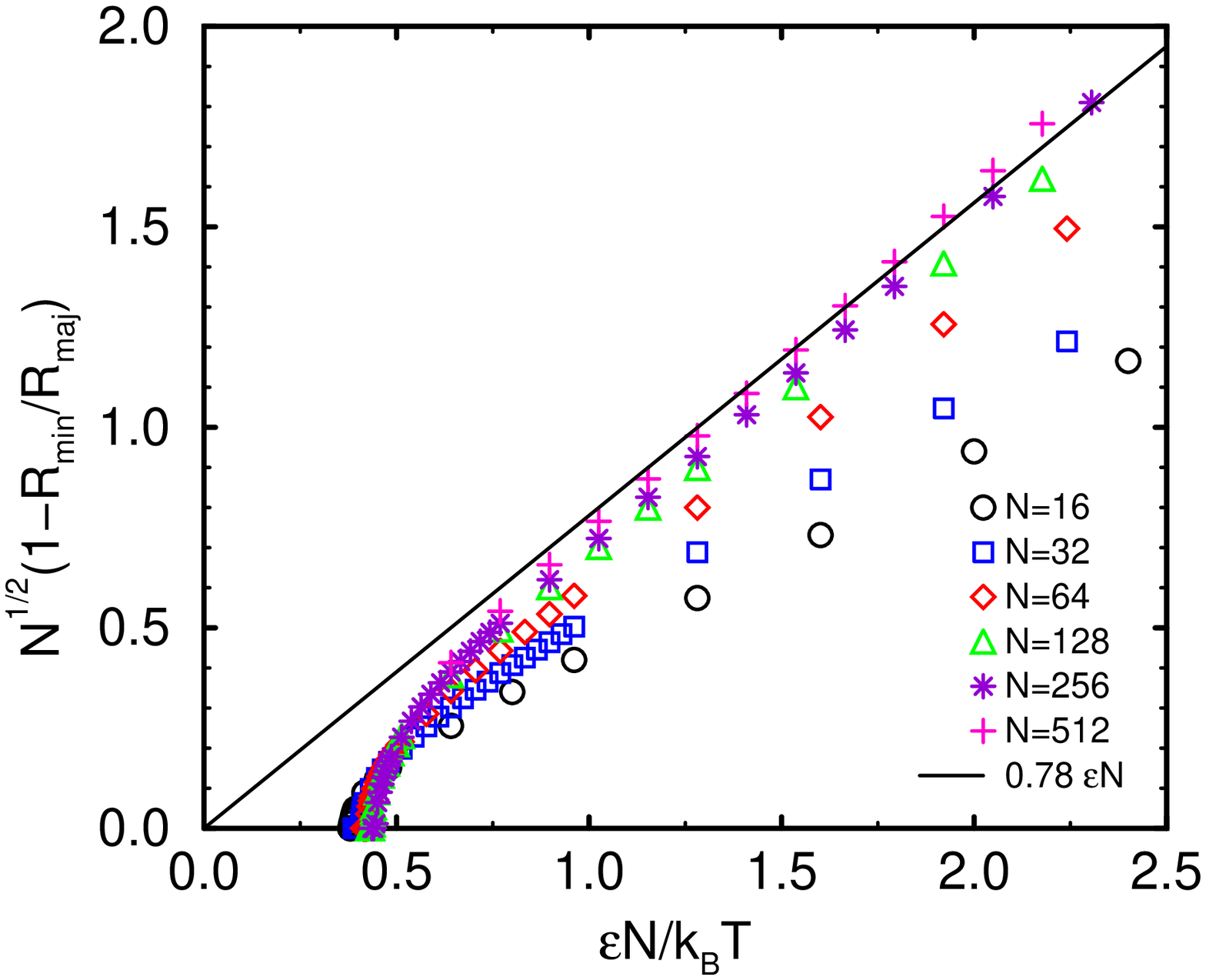}
}
\mbox{
\setlength{\epsfxsize}{9cm}
\epsffile{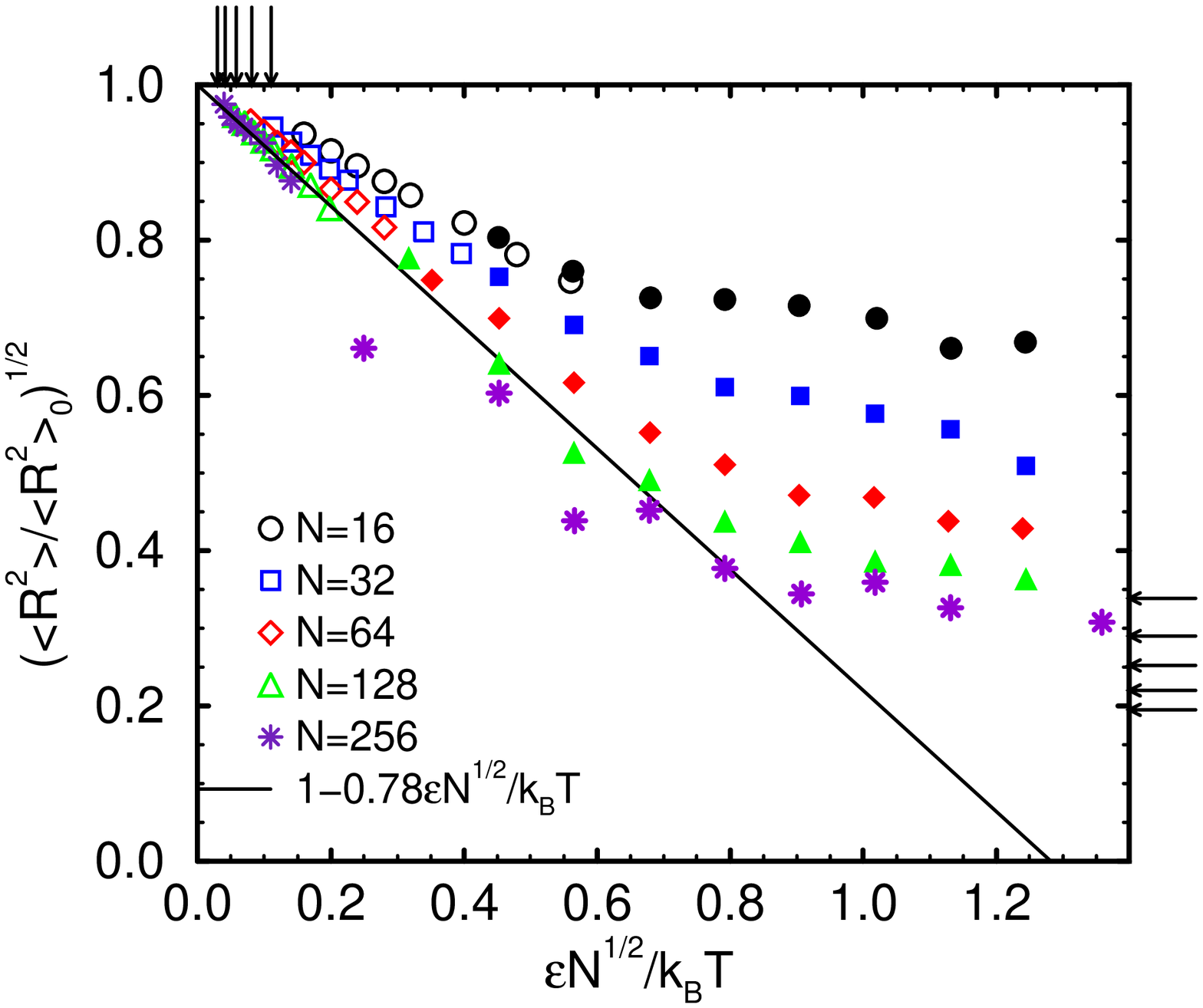}
}
\end{minipage}%
\hfill%
\begin{minipage}[c]{54mm}%
\caption{\label{fig:rescal}}
         Shrinking of the end-to-end extension in the Monte Carlo simulations ({\bf a}) and the SCF calculations ($n_c=12$) ({\bf b})
         below the critical temperature. The solid line marks the expected behavior for long chain lengths. The inset in ({\bf a}) shows the
         chain length dependence of end-to-end distance of the majority and minority component at constant $\epsilon N = 0.64$. Lines in
	 the inset are only guides to the eye.
         Panal ({\bf c}) presents the extension of the Monte Carlo simulations to stronger segregation. From ref.\cite{CLUSTER}.
\end{minipage}%
\end{figure}

In Fig.\ref{fig:rescal} we explore the scaling of the shrinking of the chains in the minority phase at two phase coexistence. The simple estimate (\ref{eqn:shrink})
predicts that the chains in the majority phase are unperturbed, while the minority component reduces its size in a well segregated blend. The relative shrinking 
$1-R_{\rm min}/R_{\rm maj}$ increases linearly with $\epsilon N$ and is at fixed $\epsilon N$ of the order $1/\sqrt{N}$. This is in qualitative agreement with
the simulation data for the end-to-end distance presented in Fig.\ref{fig:rescal}({\bf a}). The straight line represents the prediction of Eq.(\ref{eqn:shrink}).
Surprisingly, the simulation data approach the asymptotic behavior  very slowly. Only for chain length $N\geq 128$ the simulations reach the scaling limit;
for smaller chain length the estimate(\ref{eqn:shrink}) overpredicts the shrinking. In Fig.\ref{fig:rescal}({\bf b}) the predictions of the SCF calculations for the same parameters are
presented. The calculations approach for long chain lengths the estimate(\ref{eqn:shrink}), but they also quantitatively capture the corrections to the
asymptotic law for short chain lengths. There are at least two possible reasons for the pronounced small chain length corrections to the scaling behavior:
First, the considerations hold only in the regime $2 \ll  \chi N \ll \sqrt{N}$. The first limit is set by the condition that the blend is well segregated, i.e.,
$\langle \phi \rangle_{\rm min} \ll 1$. The second requirement corresponds to small conformational changes. This temperature regime is experimentally relevant
because the concentration
of the minority component is small but does not vanish for long chain lengths. However, for short chain lengths these conditions are rather restrictive.
For very strong segregation the linear decrease of the chain dimensions with $\chi\sqrt{N}$ will certainly break down and the changes cannot be treated perturbatively.
A more general scaling behavior 
might emerge for the collapse of the chains. If the chains were in a collapsed state for high segregation, one would expect the free energy per chain to scale as\cite{COLLAPSE}
\begin{equation}
F(R) \sim \frac{N}{R^2} - \chi N \frac{N}{R^3} + wN\left(\frac{N}{R^3}\right)^2
\end{equation}
The first term describes the conformational entropy loss of a Gaussian chain upon confinement into a cavity of linear dimension $R$.
The second contribution stems from the exchange of inter- and intramolecular contacts upon shrinking, while the last term comprises the
third viral coefficient in the expansion in terms of the intrachain monomer density. Temperature and chain length independent prefactor have been ignored.
Minimizing this free energy with respect to the chain extension $R$, we obtain $R \sim (wN\chi)^{1/3} \sim (w/\chi\sqrt{N})^{1/3}\sqrt{N}$
similar to the collapse of a chain in a bad solvent. Again the scaling variable $\chi\sqrt{N}$ emerges from this consideration.

A second source of corrections to asymptotic scaling behavior might be deviations from the Gaussian chain statistics upon shrinking. The ratio between the
end-to-end distance and the radius of a completely collapsed coil $(3/4\pi b^3\rho\sqrt{N})^{1/3}$ decreases only very weakly with chain length. Even for the 
chain length $N=256$ the end-to-end distance exceeds the radius of the densely packed coil only by a factor of 5. If the extension of the shrunken chain becomes 
comparable to the size of the completely collapsed coil, it cannot reduce its size much further. In this case the data would not scale as a function of $\chi\sqrt{N}$, but the data
would crossover to a temperature independent end-to-end distance the earlier the smaller the chain length.

The behavior for stronger segregation in investigated in Fig.\ref{fig:rescal}({\bf c}) in the Monte Carlo simulations. The asymptotic estimate corresponds to the
solid line; it describes the Monte Carlo data reasonably well for small $\chi\sqrt{N}$. The arrows on the top of the figure mark the critical point, while the arrows on the
right side mark the radius of a collapsed coil. The regime where the linear scaling is expected to apply is very small for short chains. For stronger segregation 
the scaling of the chain extension with $\chi\sqrt{N}$ breaks down, and the
end-to-end distance gradually approaches a temperature independent value which is about 3 times the radius in the densely collapsed state. This indicates that 
the deviations from the scaling behavior are chiefly due to deviations from the Gaussian chain statistics in the strongly shrunken state for large $\chi\sqrt{N}$.
An estimate of the asymptotic scaling behavior at larger $\chi\sqrt{N}$ would require even longer chains.

\begin{figure}[tbhp]
\begin{minipage}[t]{80mm}%
  \mbox{
       \setlength{\epsfxsize}{10cm}
       \epsffile{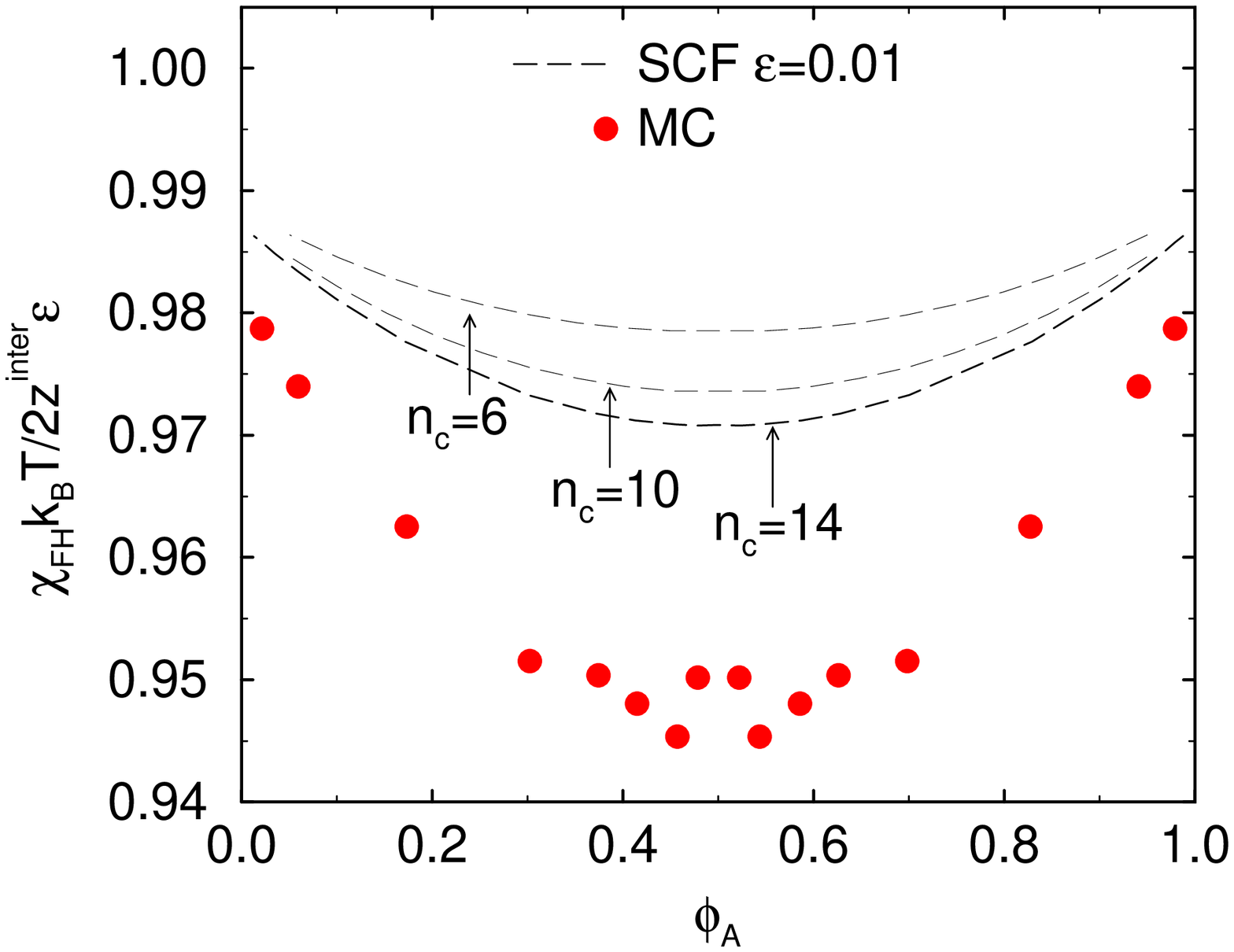}
       }
\end{minipage}%
  \hfill%
\begin{minipage}[c]{54mm}%
\caption{\label{fig:chiphi}}
Composition dependence of the effective Flory Huggins parameter (as extracted from the Flory Huggins equation of state).
From ref.\cite{CLUSTER}.
\end{minipage}%
\end{figure}

The conformational changes result in a composition dependence of the segmental entropies and, hence, give rise to a composition dependence of the $\chi$ parameter.
The scaling arguments suggest that the conformational changes alone might only produce an effect of the order $1/N$ per chain. 
Using the SG--EOS we can measure a $\chi$ parameter and the results are displayed in Fig.\ref{fig:chiphi}. In the one phase region far above
the critical temperature we observe an upward parabolic dependence of the $\chi$ parameter on the composition. The composition dependence is about $4\%$. 
The results of the SCF calculations for various cluster sizes $n_c$ are shown as lines in the figure. Upon increasing the size of the cluster $n_c$ the composition dependence becomes larger.
This is indicative of composition fluctuations. Upon increasing the size of the cluster we increase the length scale of local composition fluctuations
incorporated into the calculations and the agreement between the calculations and the Monte Carlo results improves. However, for the cluster sizes available, we 
achieve only qualitative agreement. For the rather short chain length $N=16$ composition fluctuations become increasingly important as we approach the critical 
temperature. These composition fluctuations will lead to a very pronounced upward parabolic dependence of the $\chi$ parameter (cf.\ sec IV) which 
becomes much larger than the effect of the conformational entropy near $T_c$.

\begin{figure}[tbhp]
\begin{minipage}[t]{80mm}%
  \mbox{
       \setlength{\epsfxsize}{9cm}
       \epsffile{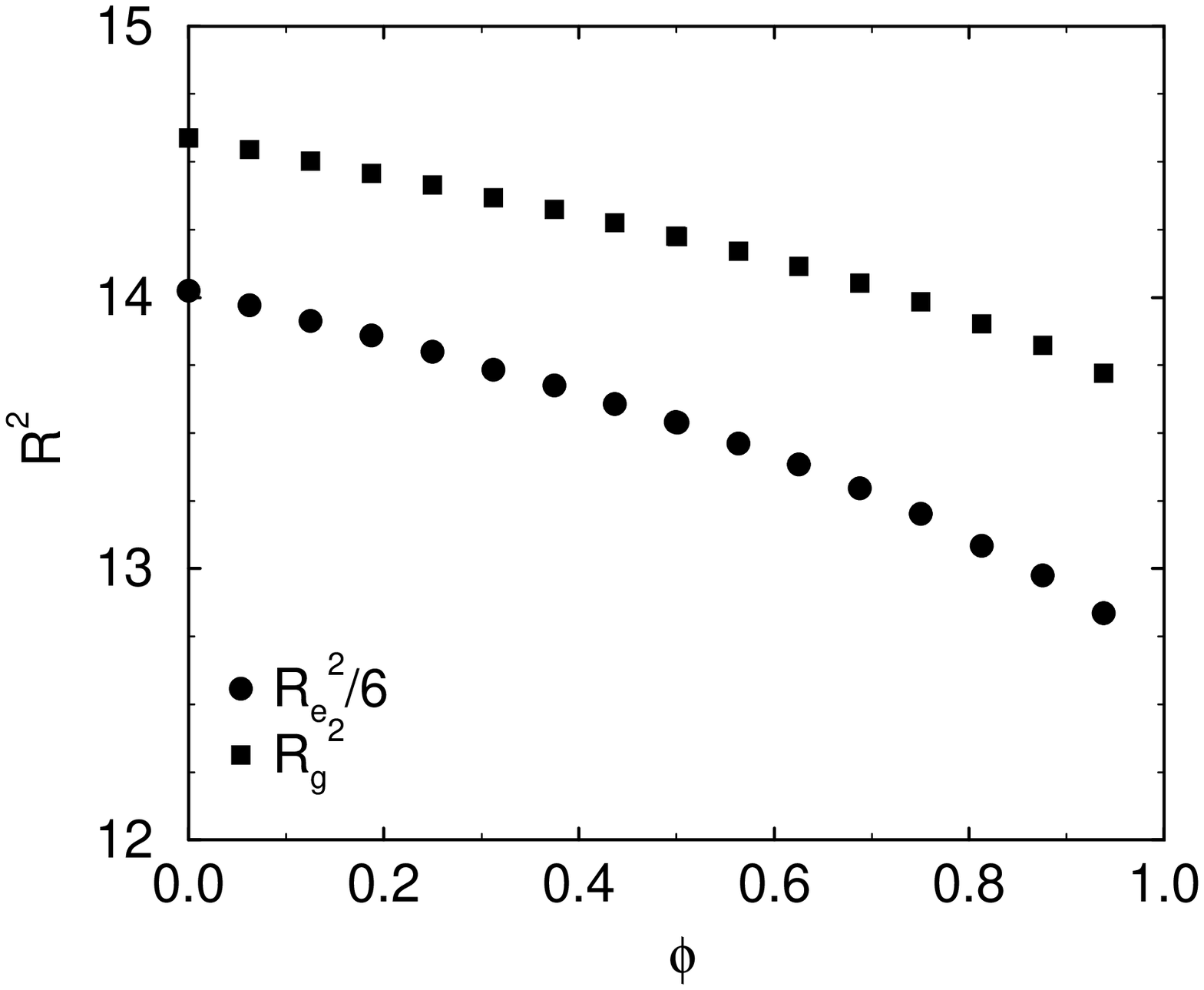}
       }
  \mbox{
       \setlength{\epsfxsize}{9cm}
       \epsffile{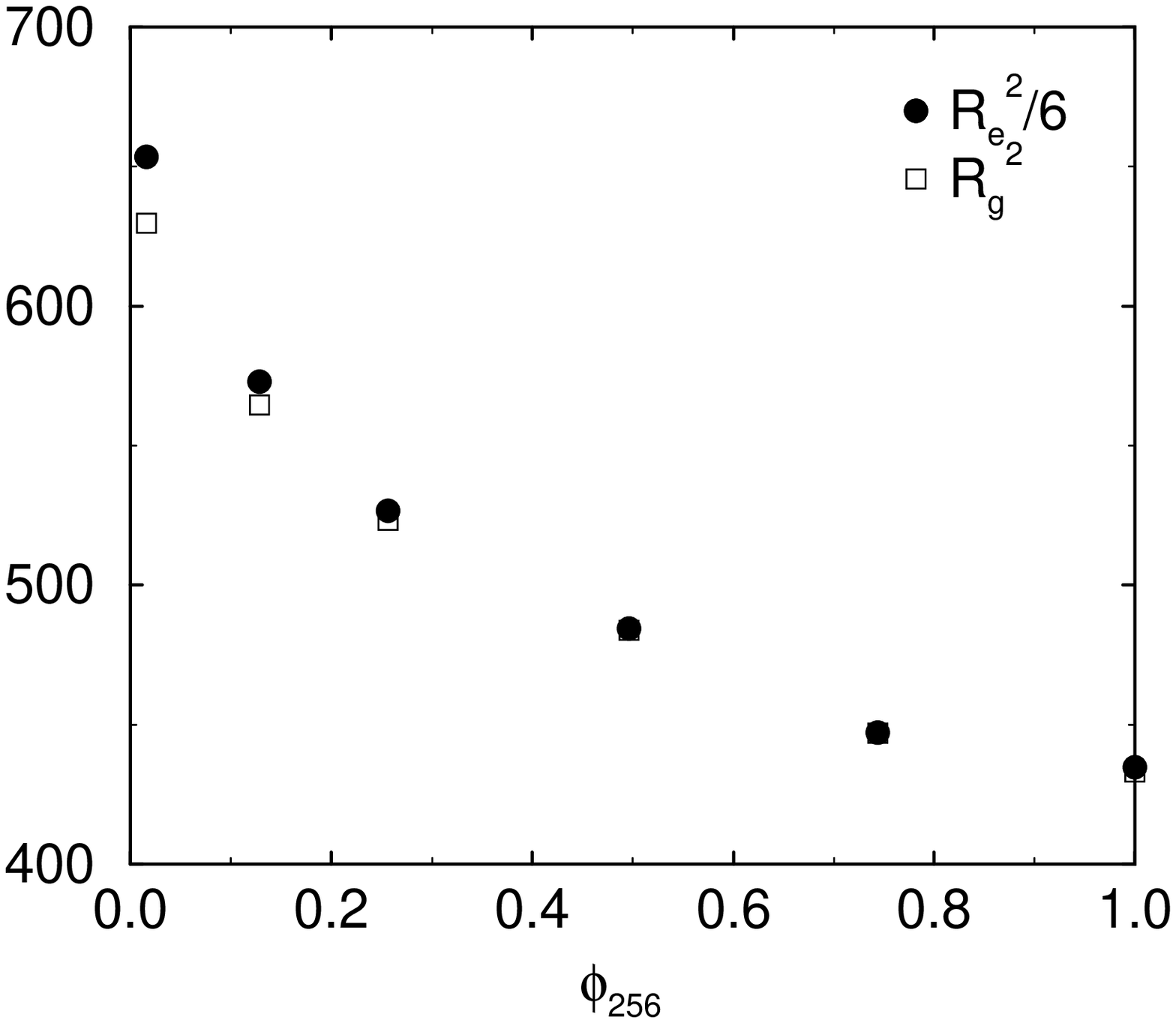}
       }
\end{minipage}%
  \hfill%
\begin{minipage}[c]{54mm}%
\caption{\label{fig:long}}
({\bf a}) Composition dependence of the $B$ chain extension in a non additive mixture of chain length $N=10$ at density $\rho=0.35/8$.
	  In the minority phase the ``effective density'' is higher and the chains are less extended.
({\bf b}) Composition dependence of the chain extension of a mixture of long chains $N=256$ and dimers at density $\rho=0.5/8$.
          $\phi_{512}$ denotes the concentration of the long chains. The dimer do not screen the excluded volume along the long 
	  polymers, which are swollen in an environment of dimers but Gaussian in a melt of identical polymers.
\end{minipage}%
\end{figure}

In the symmetric additive blend the composition dependence of the chain conformations can be quantitatively described via an exchange of inter- and intramolecular 
contacts. This indicates that this exchange is the dominant mechanism for the chain shrinking in symmetric additive mixtures. However, in more complex models
other mechanism might become important. In the case of a symmetric blend with a non-additive packing of the monomers, the conformational entropy depends strongly
on composition and gives rise to a large entropic contribution to the $\chi$ parameter. In the mixed state, the effective density is higher and we expect the chains to 
be less extended. The shrinking of the chains in the minority phase due to this density effect is presented in Fig.\ref{fig:long} ({\bf a}) for an  athermal blend of 
chain length $N=10$.
For this small chain length the athermal blend is completely miscible and the whole composition range is accessible. The simulations reveal a reduction of the
chain extension by about $10\%$ in the minority phase. 

In asymmetric blends we expect these conformational entropy effects to become more important.
Even a very weak asymmetry like the chain length asymmetry might give rise to a composition dependence of the chain extensions in an athermal melt provided the
ratio of chain lengths is extremely high.   In Fig.\ref{fig:long}({\bf b}) we display the chain extension of long polymers of chain length $N=256$ blended with
dimers as a function of the composition. If the concentration $\phi_{256}$ of the long polymer is small the dimers do not screen the excluded volume interactions along
the long polymers. Hence, the long polymers are swollen in a solution of dimers of identical monomer architecture. If the concentration of the longer polymers is
increased, the long polymers overlap and the excluded volume interaction is screened. The polymers adopt Gaussian chain conformations and are less extended 
than in a solution of dimers. Note that this is not a small effect. Our Monte Carlo data show an increase of the squared end-to-end distance of the large chains 
by a factor 1.5 in a solution of dimers compared to a melt of polymers at the same density.
Moreover the ratio $R_e^2/R_g^2$ differs from the value 6, indicating that the chains in a dimer solvent are swollen.
de Gennes\cite{DG} argued that this effect occurs only for rather large differences in chain length $N_A \ll \sqrt{N_B}$. This explains why the simulations 
for rather weak chain length asymmetries $N_B=kN_A$ ($k=2$ and $3$) have not detected entropic contributions to the $\chi$ parameter.

If the pure components exhibit a strong dependence of the density upon pressure and temperatures, this will also affect the conformations of the molecules and give rise
to a composition dependence in a blend. The density dependence of the chain conformations on the density has been investigated by Weinhold et al.\cite{SZLEIFER} and Wang et al.\cite{CIFRA2}.
They derived the density dependence of the chain conformations due to the correlation between the intermolecular contacts and the extension of the chains.  The comparison of their approach with 
Monte Carlo simulations exhibited reasonable agreement for short and intermediate chain lengths. 
The dependence of the chain extension with the density is, 
however, a difficult problem. The second virial coefficient $v$ of the free energy as a function of monomer density stems from the excluded volume or the strong repulsive interactions between 
monomers at a close distance. Unlike the exchange mechanism discussed at the beginning of this section, this coefficient is of the order $v \sim {\cal O}(1)$ and not of the order $\chi \sim 1/N$. 
Consequentially, the conformational changes
are of the order $v\sqrt{N}$ and cannot be captured perturbatively\cite{PERT}. 
The importance of a self-consistent treatment in order to analyze the
dependence of the density on pressure has been demonstrate by a comparison between non-self-consistent P-RISM calculations and Monte Carlo simulations\cite{JANS}.
Recently, the self-consistent solution of the screening of excluded volume interactions by 
Edwards\cite{EDWARDS} has been combined with P-RISM theory to incorporate the coupling between packing and chain conformations self-consistently. The numerical technique
involves the determination of the solvation potential, which mimics the effective intrachain interaction mediated via the surrounding polymers. This solvation potential,
in turn, depends on the intramolecular correlations of a chain in the solvation potential. The effect of the solvation potential on the chain conformations has been treated variationally
or by Monte Carlo simulations\cite{MALENK,SCHWEIZER}. Also diblock copolymers have been considered.\cite{DBLOCK} A similar procedure was employed by Khalatur and coworkers\cite{KHALATUR} to investigate the collapse of a single chain
in a mixture of hard spheres (colloid). They estimated a gain of 2-3 orders of magnitude in computational efficiency compared to direct Monte Carlo simulations of the system.
An entropy driven collapse of the chain conformations in a solution of large spheres has been investigated by Frenkel and coworkers\cite{FRENKEL1}
The dependence of the chain conformations on the solvent becomes even more pronounced when the solvent is close to its liquid-vapor critical point. Aspects of
single chain conformations in supercritical solvents have been explored by Luna-Barcenas et al.\cite{LUNA}. The conformations in the vicinity of the critical 
point of the solvent are reduced\cite{SOLV1}, and in the presence of strong attraction between solvent and polymer pronounced effects have been observed\cite{K2}.
These effects have been investigated experimentally\cite{TO}, analytically\cite{V1} and by self-consistent P-RISM/MC calculations\cite{K2}.

\section{Single chain dynamics.} 
Dynamical properties in polymeric multicomponent systems are pertinent to many practical applications: 
The dynamics of single chains in the blend influences rheological and transport properties of the material.
There might be, e.g., an entropic barrier associated with the transport of a single chain molecule across a sharp 
interface\cite{MUTH}. The single chain dynamics in blends is also pertinent to the understanding of reactive blending techniques, 
which are routinely applied for processing polymeric composites. Last not least, the dynamics of individual chains
is an important prerequisite for describing the dynamics of collective composition fluctuations and the morphology 
of the blend. This understanding is important for controlling mechanical properties associated with the arrangement of 
internal interfaces in the composite material.

However, even the single chain dynamics in one-component systems shows a rich and complex behavior and has attracted
abiding interest for the last decades. The dynamics of an isolated chain in the absence 
of hydrodynamic interactions is describable in the framework of the Rouse model\cite{ROUSE}, i.e., on each segment of the chain acts an independent 
stochastic force and a monomeric friction. Both the strength of the stochastic force and the friction are assumed to be chain length 
independent. Due to the connectivity along the chain the center of mass diffusion constant $D_s$ is proportional to $1/N$.
In concentrated solutions or melts the effect of entanglements among chains become important. In the simplest ansatz, the motion of the chain can be 
pictured as a curvilinear diffusion along its contour. This leads to a diffusion constant $D_s \sim N^{-2}$ for long chain lengths\cite{DOI}.
The variable bond length in the bond fluctuation model allows the implementation of a local monomer hopping dynamics. The dynamical properties
of athermal melts and solutions have been carefully investigated by Paul et al.\cite{PAUL} in the framework of the athermal bond fluctuation model. 
The simulations show a crossover to a reptation-like
dynamics for long chain lengths $N=200$ at density $\rho=1/16$. For short and intermediate chain lengths the self-diffusion constant at high densities is describable
by an apparent power-law dependence of the form $D_s \sim N^{-3/2}$\cite{MWC1} which is between the predictions of the Rouse model\cite{ROUSE} and the reptation scenario\cite{DOI}.
Phenomenologically, one might rationalize this scaling of the diffusion constant with chain length by assuming that a chain drags along its neighbors
in the correlation hole during their motion. This number of neighbors in the correlation hole scales like $\rho R^3 \sim N^{3/2}$. This scaling form is 
also compatible with a dynamical crossover scaling from dilute solutions to semi-dilute solutions $D_s(N,\rho)=D_s(N,\rho=0)f(\rho/\rho^*)$\cite{MWC2},
where the diffusion constant of an isolated chain $D_s(N,\rho=0)$ scales like $1/N$, and $\rho^*=N/R^3$ denotes the overlap density. The ratio
$\rho/\rho^*$ controls the crossover of the static properties between dilute and semi-dilute solutions\cite{DG}.

The dynamics in multi-component systems is much less understood than their equilibrium behavior. Monte Carlo simulations on well characterized model
systems might contribute to the understanding. It is well known that the thermodynamic state has pronounced effects on the dynamics of collective
composition fluctuations. In the vicinity of the critical point the ``thermodynamic force'' which drives the system to its equilibrium is small and
composition fluctuations decay very slowly. This critical slowing down is shared by all binary mixtures\cite{HOHENBERG}. 

\begin{figure}[htbp]
\begin{minipage}[t]{102mm}%
\setlength{\epsfxsize}{10cm}
\mbox{\epsffile{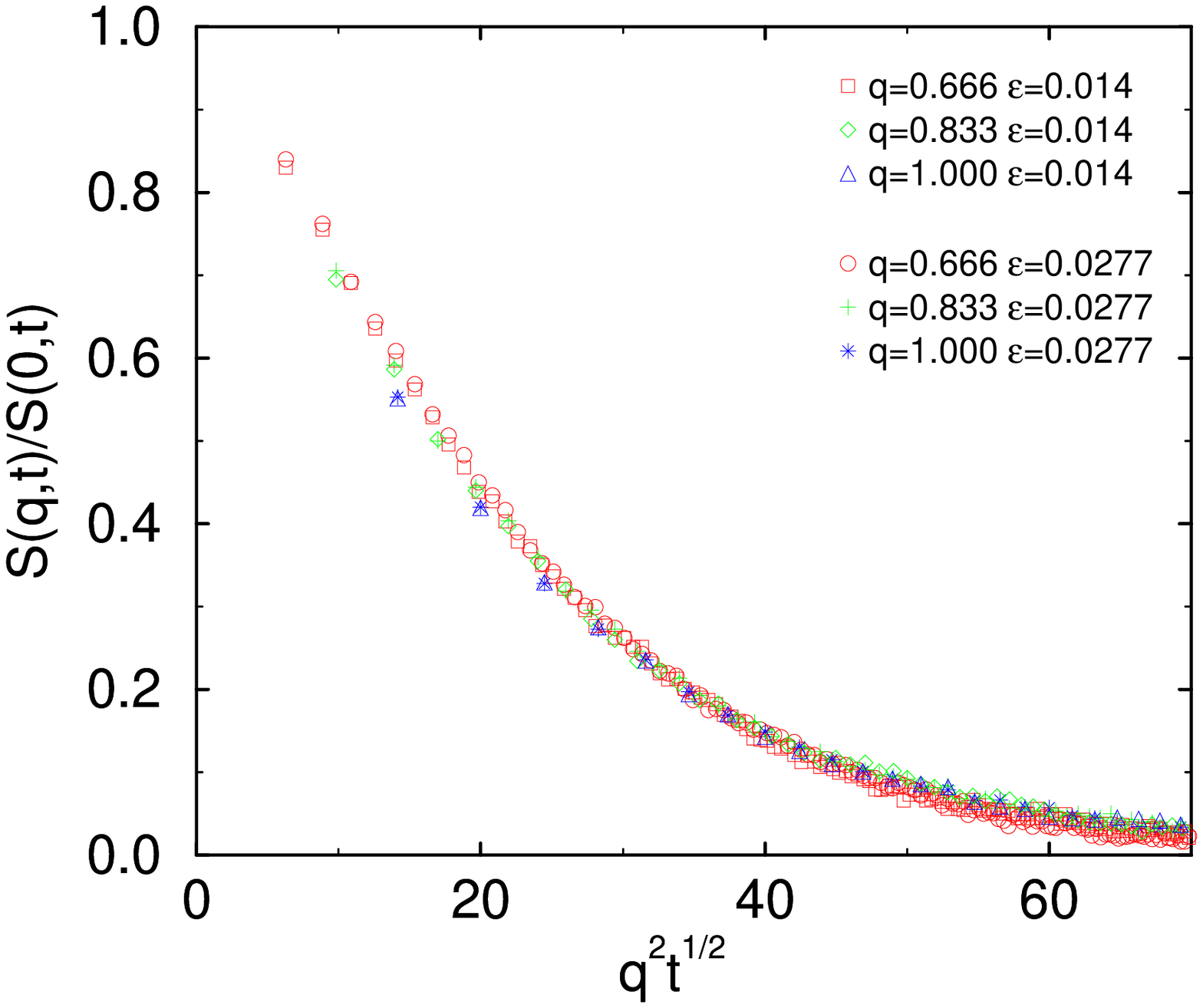}}
\end{minipage}%
\hfill%
\begin{minipage}[c]{54mm}%
\caption{\label{fig:dyn}}
	 Single chain dynamic structure factor above the critical point $T=2T_c$ ($\xi \approx R_g$) and
         at criticality for three different $q$ values. From ref.\cite{DYN}.
\end{minipage}%
\end{figure}

While collective composition fluctuations depend very strongly on the thermodynamic state it is interesting to explore the possible interplay between
temperature and single chain dynamics. Above the critical temperature the blend is homogeneous on large scales, but there is a clustering of like species
on a length scale of the correlation length $\xi$ of composition fluctuations. It has been suggested\cite{VIL} that the single chain dynamics is slowed down if
the correlation length is of the order of the radius of gyration. When we use the mean field prediction for the correlation length 
this condition is met at $\chi N = {\cal O}(1)$. This behavior was tested for a strictly symmetric blend of chain length $N=16$\cite{DYN}. For these short chain 
length no effects of entanglements are detectable in the single chain dynamics and non-random mixing effects are most pronounced. The temperature dependence 
of the correlation length has been presented in Fig.\ref{fig:xi}. 
Indeed, the mean field estimate describes the Monte Carlo results well above $T_c$. To investigate the relaxation of the internal chain structure, we present the 
simulation results for the coherent dynamic structure factor:
\begin{equation}
S({\bf q},t) = \frac{1}{N} \left\langle \sum_{i,j=1}^{N} \cos({\bf q}({\bf R}_i(t)-{\bf R}_j(0))\right\rangle
\end{equation}
where ${\bf R}_i(t)$ denotes the position of the $i$th monomer of a polymer at time $t$. We have chosen the wave vectors ${\bf q}$ in the plateau in the Kratky plot of the 
static single chain structure factor: $2\pi/R_g < {\bf q} <2\pi/b$. Hence, we investigate the relaxation of subelements of the chain. The simulation results are 
presented in Fig.\ref{fig:dyn} for three
different ${\bf q}$ values. $S({\bf q})$ is shown for the critical temperature and the temperature at which the previously determined correlation length matches the coil 
extension. The structure factor decays smoothly and does not show any signs of a spatial restriction of the internal chain motion due to the presence of 
composition fluctuations. Furthermore, $S({\bf q})$ is almost temperature independent and the data for the different wavevector ${\bf q}$ collapse onto a single master curve 
when plotted against the variable ${\bf q}^2\sqrt{t}$. Also the translational diffusion constant of a single chain and the mean square displacements of the monomers are 
virtually temperature independent. This suggests that in strictly symmetric blends of non-entangled chains there is only a very weak coupling between the
single chain dynamics and the thermodynamic state. The dynamics of single chains is dominated by the excluded volume interaction and packing effects. The
weak thermal interactions on the monomer scale $\chi \sim {\cal O}(1/N)$ do hardly influence the monomer dynamics. This is in accord with the approximate
decoupling of density and composition fluctuations exhibited in static properties.

This observation is also consistent with experiments on non-entangled chains\cite{HOMO}. However, in the long chain length limit the chains entangle and experiments do
reveal a coupling of the thermodynamic state and the single chain dynamics in blends\cite{HOMO} and diblock-copolymer systems\cite{LODGE}. Interesting effects have been predicted in 
the framework of the polymer mode coupling theory by Schweizer and co-workers\cite{SCHWEIZER4}. Most experiments -- e.g., by Ewen et al.\cite{EWEN} -- are on strongly asymmetric systems
where an even richer interplay between miscibility and single chain dynamics might occur.

\begin{figure}[htbp]
\begin{minipage}[t]{102mm}%
\setlength{\epsfxsize}{9cm}
\mbox{\epsffile{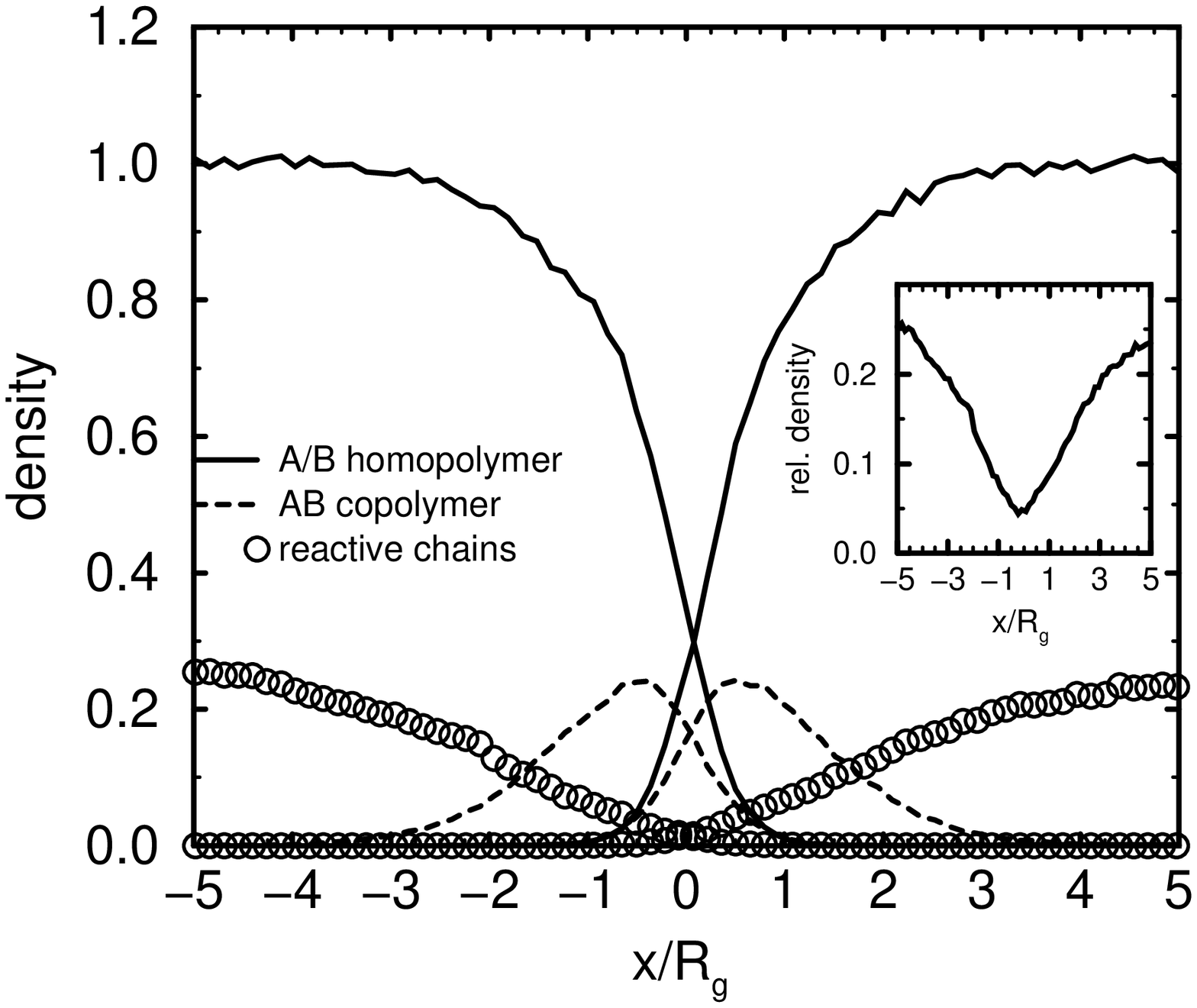}}
\setlength{\epsfxsize}{9cm}
\mbox{\epsffile{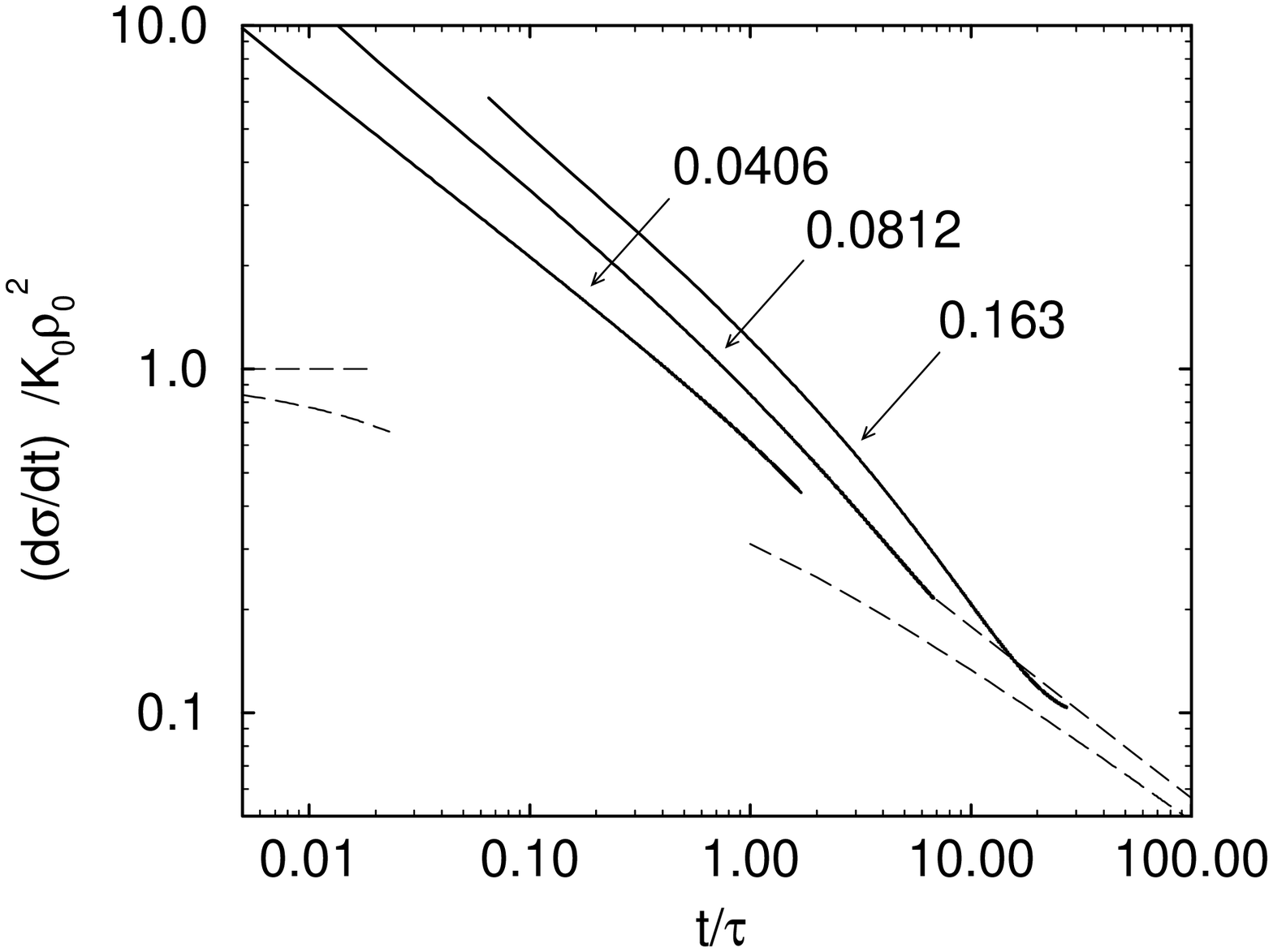}}
\end{minipage}%
\hfill%
\begin{minipage}[c]{54mm}%
\caption{\label{fig:reac}}
({\bf a}) Non-equilibrium density profiles for $\Phi_0R_g^3=0.163$ and $t/\tau=24.4$.
In this intermediate regime the reaction rate is determined by the flux of the reactive
chain to the interface. The inset presents the ratio of the monomer density of reactive chains to the total
monomer density of homopolymers.
({\bf b}) Comparison of the reaction rate with theoretical predictions. The solid lines correspond to Monte Carlo
simulations for various concentrations of reactive chains ($\Phi_0R_g^3$ as indicated in the figure), while the 
dashed lines are analytical estimes by Fredrickson and Milner\cite{F2} for $\Phi_0 R_g^3 \ll 1$. From ref.\cite{REAC}.
\end{minipage}%
\end{figure}

Another example where the single chain dynamics plays an important role are reactions in polymer blends. Interest stems from extensive commercial 
application of the reactive polymer blending technique\cite{EX1,EX2,EX3} and from recent progress in experimental studies\cite{EX4,EX5,EX6}. Since high molecular weight polymers are 
rather seldom completely miscible, they phase separate and the composite material can often be described as an assembly of interfaces. Block copolymers
containing both types of monomers can be used to tailor the mechanical properties of these interfaces. They reduce the interfacial tension and the droplet
coalescence rate, and mechanically strengthen the composite material due to entanglements across the interface. An effective method of creating copolymers at 
the interface is {\em in-situ} production\cite{EX1}, e.g., chemical reactions of end-functionalized homopolymers. We investigate an extremely idealized situation:
We consider a strictly symmetric blend well below the critical temperature. The almost pure phases $A$ and $B$ host a very small fraction $\Phi_0$ of 
chains with a terminal reactive group. The $A$ reactive ends react instantaneously and irreversibly with $B$ reactive ends whenever their mutual distance 
is less than $\sqrt{6}$. The detailed choice of this microscopic capture radius is not important for the dynamics. $AB$ diblock copolymers are formed at the interface between the bulk phases. We use chain length $N=32$ and rather strong incompatibility 
$\epsilon=0.1$. At the time $t=0$ the reaction 
starts. This corresponds to an experimental situation in which a (fast diffusive) reactive agent/catalyst is added to the blend or the reaction is initiated by 
radiation. The interfacial density $\sigma$ of copolymers at the interface grows and we define a growth rate $K(t)$ according to the rate equation:
\begin{equation}
\frac{{\rm d}\sigma}{{\rm d}t} = K(t) \Phi_0
\end{equation}
In the limit of very small number densities of reactive chains ($\Phi_0 R_g^3 \ll 1$) the following scenario emerges from the studies of O'Shaughnessy and Sawheney\cite{S1,S2}, and
Fredrickson and Milner\cite{F1,F2}. In the initial stage the density of reactive chains at the interface remains close to its bulk value. The reaction rate within the framework 
of the Rouse model\cite{ROUSE} is time independent $K_0 \approx 100.6 D_s R_g^2/\ln N$, where $D_s=1.17(5) \times 10^{-4}$ denotes the translational diffusion constant (in units
of the lattice spacing and the number of local monomer hopping attempts per monomer). As the reaction proceeds a depletion hole of the reactive chains emerges. Its
spatial extent grows like $\sqrt{D_st}$. 
After a time $\tau$ the concentration of the reactive chains at the interface is much smaller than the bulk value and the reaction rate is determined 
by the flux of reactive chains to the interface. In this regime the reaction rate decays with time as $K=\sqrt{D_s/\pi\Phi_0^2t}$; i.e., the interface acts much like an 
absorbing boundary. 
The crossover time $\tau$ can be estimated via the condition that the number of copolymers $K_0 \Phi_0^2 \tau$, which have 
been produced up to $\tau$ (per area), equals the number of reactive chain in the depletion zone $\Phi_0\sqrt{D_s\tau}$. This yields $\tau = D_s/K_0^2\Phi_0^2$ for 
the crossover time.  
The non-equilibrium density profiles for $\Phi_0R_g^3=0.163$ and $t/\tau=24.4$ are presented in Fig.\ref{fig:reac}({\bf a}).
Clearly, there is a pronounced depletion of the reactive chains at the interface, however, the copolymers do not strongly overlap at the interface.
In the final stage, the copolymers accumulate at the interface and form a brush. This prevents reactive chains from reaching the interface, and the 
reaction rate becomes extremely small. 

The time dependence of the reaction rate is presented in Fig.\ref{fig:reac}({\bf b}) for various densities $\Phi_0$ of reactive chains.  They are compared to the
predictions of Fredrickson and Milner\cite{F2}. The two sets of lines correspond to the asymptotic results for small and large ratios $t/\tau$. The higher ones represent 
the asymptotic behavior, the lower ones first--order corrections. The intermediate stage, where the interface acts like an absorbing boundary is well described
by the theory, while there are pronounced deviations in the initial stage. The theory underestimates the
reaction rate observed in the simulations by at least an order of magnitude. There are several possible reasons for these discrepancies: The theory neglects the local equilibrium structure of 
the homopolymer interface. Simulations indicate a pronounced enrichment of chain ends at the interface. Hence, the initial reaction rate is expected to be higher than 
for a uniform initial distribution of reactive segments. This effect has been used to rationalize the enhanced reaction rate at PS-PMMA interfaces.\cite{EX2}
Another effect is related to the initial concentration $\Phi_0$ of the reactive chains. The concentration in the simulations might be still too high to meet the assumptions in the
analytic theory, which assumes that the Rouse time $\tau_R = 6R_g^2/\pi^2 D_s$ is smaller than the crossover time $\tau$ to observe the time independent reaction rate 
in the initial rate. The theory might become appropriate for extremely small concentrations of reactive chains. In any case, the situation investigated in the Monte 
Carlo simulations corresponds to weight fractions of $25\%$, $12.5\%$, and $6.25\%$. These weight fractions are within the experimentally relevant range. Moreover, if one increases 
the chain length at fixed weight fraction the parameter $\Phi_0R_g^3$ increases as $\sqrt{N}$, taking the initial condition even further from the assumptions of the theory. 
Therefore we expect that experiments will also observe a decay of the reaction rate over the entire regime, unless extremely small volume fractions of reactive chains are used.

\section{Summary and outlook}
We have reviewed extensive Monte Carlo simulations on the miscibility behavior and single chain properties in binary polymer blends. 
For strictly symmetric blends -- i.e., symmetric species of polymers with identical architecture and an enthalpic repulsion between unlike species -- 
the Monte Carlo results agree with the mean field theory in the limit of long chain lengths. The $\chi$ parameter has been identified via the intermolecular 
paircorrelation function in the athermal limit. This parameterizes the local fluid-like packing structure on the monomer scale and accounts for the fact that 
the intramolecular energy does not contribute to the miscibility behavior. This correlation hole behavior imparts a $1/\sqrt{N}$ correction to the
$\chi$ parameter. The local fluid structure approximatively decouples from the composition.

The Monte Carlo simulations reveal, however, deviations from the mean field theory for short and moderate chain lengths: ({\bf A}) The chains in the minority 
phase shrink as to exchange energetically unfavorable intermolecular interactions against favorable intramolecular interactions. A quantitative comparison between
our Monte Carlo data, SCF calculations and heuristic scaling arguments show that this exchange is the dominant mechanism in strictly symmetric blends.
The relative reduction of the chain extension in the minority phase is of the order $\chi N/\sqrt{N}$. At constant $\chi N$ the relative perturbation decreases like
$1/\sqrt{N}$, and the chain conformations become composition independent in the long chain length limit. ({\bf B}) For finite chain lengths
composition fluctuations are important. They give rise to 3D Ising critical behavior with flat binodals and a strong divergence of collective scattering 
function $S({\bf q}\to 0)$ around the critical temperature. This results in an upward  parabolic composition dependence of the $\chi$ parameter, when 
extracted from the collective scattering function close to $T_c$. The region around the critical point where the mean field theory fails is determined by the 
Ginzburg criterion\cite{GINZBURG}.  Composition fluctuations become negligible, when $|\chi-\chi_c|/\chi_c \gg \chi_c/\rho^2\bar{b}^6$. These composition fluctuations result in an
overestimation of the critical temperature\cite{MON} by the mean field theory of the form $1-\chi_c^{\rm MF}/\chi_c \sim \sqrt{\chi_c}/\rho \bar{b}^3$. The data of
strictly symmetric blends of linear chains with two different interaction ranges, strictly symmetric blends of ring polymers as well as the simulation 
data of blends with a modest chain length asymmetry corroborate this scaling.  For high molecular weight the mean field theory yields a quantitative description
except for the ultimate vicinity of the critical point\cite{M0}. However, the relative size of this non-mean field region decreases as $1/N$. While the leading scaling
behavior for the critical temperature with chain lengths is observed in the Monte Carlo simulations by Deutsch and Binder\cite{HPD} already for rather short chain lengths 
$N \geq 16$; the asymptotic form of the finite $N$ corrections to the mean field theory are detectable only for larger chains lengths\cite{M0}.

In view of these findings the strictly symmetric blend in the bond fluctuation model is a suitable starting point for further investigations:
Two aspects of the dynamics in binary blends have been explored, but there remain many fascinating open questions of great practical importance.
Schweizer and co-workers\cite{SCHWEIZER4} have explored the dynamics of polymers in the framework of the polymer mode coupling theory, suggesting a rich interplay
between the dynamics and statics of binary polymer blends. The effect of shear on the miscibility behavior has also attracted much interest\cite{SHEAR}.

Moreover, a great deal of effort has been directed towards understanding interfacial properties in binary blends and self-assembly of diblock copolymers.
Various aspects of interfaces have been studied in the framework of the bond fluctuation model encompassing the detailed chain conformations at interfaces\cite{MBO},
the effect of capillary waves\cite{COP} on the phase diagram of ternary homopolymer-copolymer blends and interfacial profiles\cite{FREDDI,STIFF2,AW}, and the wetting 
behavior in symmetric blends\cite{WET}.
Interfacial properties between polymers of different stiffness -- e.g., the dependence of the interfacial tension on stiffness disparity -- have been explored
both in simulations and in self-consistent field calculations, which take due account of the chain architecture\cite{STIFF2}. The knowledge of the miscibility behavior and the 
range of validity of the mean field theory is a sound basis for comparing the results of these simulations to self-consistent field predictions. These application of the
bond fluctuation model have been recently reviewed in ref.\cite{FREV}.

Computer simulations have explored the consequences of various asymmetries. In the Monte Carlo simulations one can study the consequences 
of a single, well-defined modification of the architecture. Moreover, static properties on very different length scales -- e.g., the intermolecular paircorrelation
function, the spatial extension of single polymers, the correlation length  -- and thermodynamic properties -- e.g., the phase diagram or the $\chi$ parameter extracted 
from the inverse collective scattering function -- are simultaneously accessible. Deutsch and Binder\cite{HPD2} investigated the effect of asymmetric adhesive forces. 
The phase diagram remains symmetric and only the $\chi$ parameter depends on the asymmetry in a form suggested by the mean field theory.
The effect of chain length asymmetry, i.e., $N_A \neq N_B$ leads to truly asymmetric phase diagrams. In the limit of long chain lengths and not extremely large
asymmetry the critical temperature and composition are well describable in the framework of the mean field theory. For very strong asymmetries,
however, the assumption of composition independent chain conformations might break down for the long polymers ($N_A \ll \sqrt{N_B}$).
Under these extremely asymmetric conditions the short polymers do not screen the excluded volume along the long polymers, and the latter are swollen.
Our heuristic derivation of the mean field theory suggests that a composition dependence of the conformational entropy gives rise to an entropic contribution to
the $\chi$ parameter. This has been demonstrated for polymers consisting of monomers species which do not pack additively, i.e.\ monomers of the
same species can approach each other closer than monomers of different species. This results in  a large positive entropic contributions
which manifests itself in a lower critical solution temperature (LCST) for long chain lengths. The temperatures approach a limiting value from above when we increase the 
chain length. Similar effects can be expected for a large size disparity between the monomeric units. Simulations of hard sphere mixtures of a bimodal
size distribution show phase separation for high densities and sufficiently large size disparity\cite{FRENKEL2}. These effects are expected to be even more pronounced 
in polymer blends because the translational entropy per monomer is only of the order $1/N$. Moreover a disparity in the statistical segment length or stiffness
has been modeled. Stiffness disparity gives rise to a positive entropic contribution to the $\chi$ parameter. This entropic contribution is rather small 
$\Delta \chi \sim 10^{-3}$ but the simulation data suggest that it is chain length independent. A chain length independent entropic contribution
leads to a LCST behavior for long chain lengths. Apart from the entropic contribution, the enthalpic contribution increases too: the stiffer the chains are,
the more intermolecular contacts are formed, and the larger is the enthalpic contribution to the $\chi$ parameter. For the parameters investigated both effects
are of similar magnitude.

These Monte Carlo investigations in the framework of a coarse grained polymer model do not yield a quantitative prediction of the $\chi$ parameter on an atomistic basis. 
Nevertheless, they help in identifying relevant parameters for the miscibility on a coarse grained scale. The simulations indicate that effects of very different length 
scales which range from the microscopic monomer structure to the correlation hole effect on the size of an entire polymer coil are relevant for the miscibility behavior. 
While some of the effects -- e.g., the reduction of the critical temperature compared to the mean field estimate or the dependence of the $\chi$ parameter on the large
scale conformations via the correlation hole effect -- are universal, other observations -- e.g., the influence of the molecular architecture on the intermolecular packing --
depend strongly on the details of the modeling. Therefore a comparison of models with various degree
of coarse graining is appropriate for unraveling the correlation between the microscopic structure and the thermodynamic properties.  We just briefly mention two
recent approaches with demonstrate the wide span of length scales pertinent to the miscibility behavior.

On the most coarse grained side, Murat and Kremer\cite{MURAT} have developed an ultra coarse grained model which represents the entire chain only via its ellipsoidal
spatial density distribution. This might offer promising routes for exploring the behavior on very large length scales -- e.g., the morphology of a phase separated blend
and its consequences for the mechanical stability of the composite material.

On the microscopic side, recent simulations\cite{GREST} have explored the consequences of chain branching in the framework of a united atom model. 
The simulations demonstrate that the competition between packing and space filling results in a qualitative dependence of the intermolecular
pair correlation function (at short distances) on the degree of branching. Though the simulations suggest that the miscibility behavior is primarily
driven by enthalpic interactions, there are deviations from the regular mixing theory in 11 from 12 homopolymer mixtures. A comparison between simulation 
and experiment for homopolymer blends tentatively suggests that the segmental entropies of the components change upon mixing and give rise to an entropic
contribution to the $\chi$ parameter.

\subsection*{Acknowledgment}
It is a great pleasure to thank K. Binder, M. Schick, M.E. Cates, N.B. Wilding, F. Schmid, J.P. Wittmer, T. Geisinger, P. Janert, and A. Werner for fruitful and enjoyable 
collaborations. Special thanks go to K. Binder and F. Schmid for carefully reading the manuscript and their valuable comments. Financial support by the DFG under
grant No. Bi 314/17 and the BMBF under grant N$^0$ 03N8008C, and generous access to CRAY T3Es at the HLRZ Juelich, HLR Stuttgart, EPCC Edinburgh, SDSC San Diego are 
gratefully acknowledged.

\begin{thebibliography}{99}
\bibitem{COMMENT}    
                     The fact that the number of intermolecular contacts per chain $Nz^{\rm intra}$ scales with $N$, but its derivative with respect to the chain dimension $R$ is 
		     independent of the chain length can be rationalized 
		     in the framework of the Gaussian chain model. Let $R_0^2=bN^2$ denote the unperturbed end-to-end distance. For convenience, we replace the square well contacts
		     by Gaussian contacts of widths $\sigma_e$. Then, the number of intramolecular contacts can be written in the form:
{\tiny{
		     \begin{eqnarray*}
		     Nz^{\rm intra}(R) &\sim & \frac{1}{2} \int {\rm d}s\;{\rm d}s'\;{\rm d}^3\xi 
		        \left( \frac{3}{2\pi \sigma_e^2} \right)^{3/2}
		        \left( \frac{3}{2\pi b^2|s-s'|} \right)^{3/2}
		        \left( \frac{3}{2\pi b^2(N-|s-s'|)} \right)^{3/2}
		        \left( \frac{3}{2\pi b^2N} \right)^{-3/2} \\
			&& \exp\left( -\frac{3\xi^2}{2\sigma_e^2}\right)
			\exp\left( -\frac{3\xi^2}{2b^2|s-s'|}\right)
			\exp\left( -\frac{3(R-\xi)^2}{2b^2(N-|s-s'|)}\right)
                     \end{eqnarray*}
}}
		     in the limit $\sigma_2 \ll b^2N$ this expression can be simplyfied to:
{\tiny{
		     \begin{eqnarray*}
		     Nz^{\rm intra}(R) &\sim & \int_0^N {\rm d}\triangle s\; (N-\triangle s)
		        \left( \frac{3}{2\pi (\sigma_e^2+b^2\triangle s)} \right)^{3/2}
		        \left( \frac{N}{N-\triangle s} \right)^{3/2} \\
			&& \exp\left( -\frac{3R^2}{2b^2N}\; \frac{\triangle s}{N-\triangle s}\right)
                     \end{eqnarray*}
}}
		     The major contribution to this integral comes from $\triangle s \approx 0$, i.e., the neighbors along the chain dominate
		     the number of intramolecular contacts. $Nz^{\rm intra}$ scales like $N$. The derivative with respect to $R$ is given by:
{\tiny{
		     \begin{eqnarray*}
		     \frac{{\rm d}Nz^{\rm intra}}{{\rm d}R} &\sim & \int_0^N {\rm d}\triangle s\; (N-\triangle s)
		        \left( \frac{3}{2\pi (\sigma_e^2+b^2\triangle s)} \right)^{3/2}
		        \left( \frac{N}{N-\triangle s} \right)^{3/2} \\
			&& \frac{3R}{b^2N} \frac{\triangle s}{N-\triangle s}
			\exp\left( -\frac{3R^2}{2b^2N}\; \frac{\triangle s}{N-\triangle s}\right)
                     \end{eqnarray*}
}}
		     The integral converges in the limit $\sigma_e \to 0$ and scales like $N^0$ at $R=R_0$.
\end{thebibliography}
\end{document}